\documentclass[12pt,a4paper]{article}

\usepackage[a4paper,rmargin=1.7cm,lmargin=1.7cm,tmargin=1.0cm,bmargin=1.0cm,includefoot]{geometry}

\usepackage{booktabs}

\expandafter\let\csname equation*\endcsname\relax

\expandafter\let\csname endequation*\endcsname\relax

\usepackage{array}
\usepackage{epsfig}

\usepackage{cite}

\usepackage[namelimits]{amsmath}
\usepackage{amssymb}
\usepackage{amsfonts}
\usepackage{bbm}

\usepackage{amsmath}

\usepackage{epsfig}

\usepackage{comment}

\usepackage{multirow}
\usepackage{color}

\usepackage[T1]{fontenc}      % LÃª a codificaÃ§Ã£o de fonte T1 (font encoding default Ã© 0T1).
\usepackage{ae}               % Fonte "Almost European"
\usepackage{indentfirst}

\usepackage[affil-it]{authblk}

\usepackage{amssymb}

\usepackage{multirow}

\date{}

\newcommand{\tcs}{$\sigma_{\mathrm{tot}}$}
\newcommand{\tcss}{$\sigma_{\mathrm{tot}}(s)$}
\newcommand{\ro}{$\rho$}
\newcommand{\ros}{$\rho(s)$}

\begin{document}

\title{\bf Leading components in forward elastic hadron scattering: Derivative dispersion relations
and asymptotic uniqueness}

\author[1]{D.A. Fagundes}
\author[2]{M.J. Menon}
\author[2]{P.V.R.G. Silva}
\affil[1]{\small Departamento de Ci\^encias Exatas e Educa\c c\~ao, Universidade Federal de Santa Catarina - 
Campus Blumenau, 89065-300 Blumenau, SC, Brazil}
\affil[2]{\small Instituto de F\'{\i}sica Gleb Wataghin, Universidade Estadual de Campinas - UNICAMP \\
13083-859 Campinas, SP, Brazil}

\maketitle

\begin{abstract} 
Forward amplitude analyses constitute an important approach in the investigation
of the energy dependence of  the total hadronic cross-section \tcs\ and the $\rho$ parameter. 
The standard picture indicates for \tcs\
a leading log-squared dependence at the highest  c.m. energies, in accordance with the Froissart-Lukaszuk-Martin
bound and as predicted by the COMPETE Collaboration in 2002.
Beyond this log-squared (L2) leading dependence,
other amplitude analyses have considered
a log-raised-to-gamma
form (L$\gamma$), with $\gamma$ as a real free fit parameter. 
In this case, \textit{analytic} connections with \ro\ can be obtained either through dispersion relations (derivative forms),
or asymptotic uniqueness  (Phragm\'en-Lindel\"off theorems).
In this work we present a detailed discussion on the similarities and mainly
the \textit{differences} between the Derivative Dispersion Relation (DDR) and
Asymptotic Uniqueness (AU) approaches and results, with focus on the L$\gamma$ and L2 leading terms.
We also develop new Regge-Gribov fits with updated dataset on \tcs\ and $\rho$ from $pp$ and $\bar{p}p$ scattering,
including all available data in the region 5 GeV - 8 TeV. The recent tension between the TOTEM and ATLAS results 
at 7 TeV and mainly 8 TeV is discussed and considered in the data reductions.
Our main conclusions are the following: 
(1) all fit results present agreement with the experimental data analyzed 
and the goodness-of-fit is slightly better in case of the DDR approach;
(2) by considering only the TOTEM data at the LHC region, the fits with 
L$\gamma$ indicate $\gamma \sim 2.0 \pm 0.2$ (AU approach) and 
 $\gamma \sim 2.3 \pm 0.1$ (DDR approach);
(3) by including the ATLAS data the fits provide   
$\gamma \sim 1.9 \pm 0.1$ (AU) and 
 $\gamma \sim 2.2 \pm 0.2$ (DDR);
(4) in the formal and practical contexts, the DDR approach is more adequate for
the energy interval investigated than the AU approach.
A pedagogical and detailed review on the analytic results for \tcs\ and \ro\ from the Regge-Gribov, DDR and AU
approaches is presented.
Formal and practical aspects related to forward amplitude analyses
are also critically discussed.
\end{abstract}

\vspace{0.5cm}



\vspace{1.5cm}

\centerline{ Published in {\it International Journal of Modern Physics A} {\bf 32} (2017) 1750184}

\newpage

%..............................................................

\textbf{Table of Contents}

\vspace{0.3cm}

1. Introduction

\vspace{0.1cm}

2. Experimental Data

\ \ \ \ 2.1 Dataset

\ \ \ \ 2.2 Ensembles

\vspace{0.1cm}

3.  Analytic Models

\ \ \ \ 3.1 Notation

\ \ \ \ 3.2 Derivative Dispersion Relation Approach

\ \ \ \ \ \ \ \ \ 3.2.1 FMS-L$\gamma$ Model

\ \ \ \ \ \ \ \ \ 3.2.2 FMS-L2 Model

\ \ \ \ 3.3 Asymptotic Uniqueness Approach

\ \ \ \ \ \ \ \ \ 3.3.1 AU-L2 Model

\ \ \ \ \ \ \ \ \ 3.3.2 AU-L$\gamma$ Model

\ \ \ \ \ \ \ \ \ 3.3.2 AU-L$\gamma$=2 Model

\vspace{0.1cm}

4. Data Reductions and Results

\ \ \ \ 4.1 Fit Procedures

\ \ \ \ 4.2 FMS-L2 and FMS-L$\gamma$ Models

\ \ \ \ 4.3 AU-L$\gamma$=2 and AU-L$\gamma$ Models

\vspace{0.1cm}

5. General Discussion and Comments

\ \ \ \ 5.1 Analytic and Conceptual Differences

\ \ \ \ \ \ \ \ \ 5.1.1 DDR Approach

\ \ \ \ \ \ \ \ \ 5.1.2 AU Approach

\ \ \ \ 5.2 Fit Results

\ \ \ \ \ \ \ \ \ 5.2.1 Ensembles T and T+A

\ \ \ \ \ \ \ \ \ 5.2.2 FMS-L2 and FMS-L$\gamma$ Models

\ \ \ \ \ \ \ \ \ 5.2.3 FMS-L$\gamma$ and AU-L$\gamma$ Models 

\ \ \ \ \ \ \ \ \ 5.2.4 PDG 2016 and COMPETE

\ \ \ \ 5.3 Partial Conclusions

\ \ \ \ 5.4 Further Comments on the Log-raised-to-$\gamma$ Law

\vspace{0.1cm}

6. Conclusions and Final Remarks

\vspace{0.2cm}

Appendix A Comments on the Experimental Data Presently Available

\ \ \ \ A.1 Statistical and Systematic Uncertainties

\ \ \ \ A.2  Experimental Data at the Highest Energies

\vspace{0.1cm}

Appendix B Regge-Gribov Formalism

\ \ \ \ B.1 Historical Aspects

\ \ \ \ B.2 Simple Poles - Power Laws

\ \ \ \ \ \ \ \ \ B.2.1 Reggeon Contributions

\ \ \ \ \ \ \ \ \ B.2.2 Simple Pole Pomeron Contribution

\ \ \ \ \ \ \ \ \ B.2.3 Analytic Result

\ \ \ \ B.3 Double and Triple Poles - Logarithm Laws

\ \ \ \ B.4 L2 Models

\vspace{0.1cm}

Appendix C Dispersion Relations and the Effective Subtraction Constant

\ \ \ \ C.1 Integral Dispersion Relations and the High-Energy Approximation

\ \ \ \ C.2 Derivative Dispersion Relations with the Effective Subtraction Constant

\ \ \ \ \ \ \ \ \ C.2.1 Basic Concepts and Results

\ \ \ \ \ \ \ \ \ C.2.2 DDR Approach

\vspace{0.1cm}

Appendix D Asymptotic Uniqueness and the Phragm\'en-Lindel\"off Theorems

\ \ \ \ D.1 Basic Concepts 

\ \ \ \ D.2 Power Law (Simple Poles)

\ \ \ \ D.3 Log-squared Law (Triple Pole)

\newpage

\ \ \ \ D.4 Log-raised-to-gamma Law

\ \ \ \ \ \ \ \ \ D.4.1 Phase of the Amplitude

\ \ \ \ \ \ \ \ \ \ \ \ \ \ \ \ \ \ a) Exact Result

\ \ \ \ \ \ \ \ \ \ \ \ \ \ \ \ \ \ b) High-Energy Approximate Result

\ \ \ \ \ \ \ \ \ \ \ \ \ \ \ \ \ \ c) Relations with AU-L2 and FMS-L$\gamma$ Models

\ \ \ \ \ \ \ \ \ D.4.2 Binomial Expansion

\ \ \ \ \ \ \ \ \ \ \ \ \ \ \ \ \ \ a) General Result

\ \ \ \ \ \ \ \ \ \ \ \ \ \ \ \ \ \ b) Relations with AU-L2 and FMS-L$\gamma$ Models

%\end{document}

\newpage

\section{Introduction}
\label{s1}

The total cross-section, \tcs, is one of the most important physical quantity
in any particle collision process. In hadronic interactions,
although the rise of \tcs\ at high energies is an experimental fact,
the theoretical (QCD) explanation/description of this increase and, most importantly,
the exact energy dependence involved have been a long-standing
problem. The total cross section is connected with the imaginary part of the forward
\textit{elastic} scattering amplitude through the optical theorem, which at high energies reads \cite{pred}
\begin{eqnarray}
\sigma_{\mathrm{tot}}(s) = \frac{\mathrm{Im}\,A(s,t=0)}{s},
\label{ot}
\end{eqnarray}
where $s$ and $t$ are the Mandelstam variables. Therefore, the determination of \tcs($s$) demands
a theoretical result for the elastic amplitude in terms of the energy
(at least at $t=0$),
valid in all region above the physical threshold. As a soft scattering
state (the simplest one in the kinematic context), the dynamics involved in the
\textit{elastic scattering} is intrinsically 
nonperturbative and the crucial point concerns
the absence of a nonperturbative framework able to provide \textit{from the first principles} of QCD
a \textit{global description} of all physical quantities related to soft scattering states,
in particular, the elastic hadron scattering.

A general and formal result on the rise of $\sigma_{\mathrm{tot}}(s)$ at the \textit{asymptotic energy region} 
($s \rightarrow \infty$) is the upper bound derived by Froissart, Lukaszuk and Martin 
\cite{froissart,martin1,martin2,lukamartin}
\begin{eqnarray}
\sigma_{\mathrm{tot}}(s) < c \ln^2(s/s_0),
\label{fmb}
\end{eqnarray}
where $s_0$ is an unspecified energy scale and the pre-factor on the right-hand side is bounded by
\begin{eqnarray}
c \leq  \frac{\pi}{m_{\pi}^2} \approx\ \mathrm{60\ mb},
\label{cb}
\end{eqnarray}
where $m_{\pi}$ is the pion mass.

In the nonperturbative QCD context, the functional integral approach to
soft high-energy scattering (proposed by Nachtmann \cite{nach} and developed by
several authors \cite{land}), seems to be the closest formal approach to
derivations from first principles of QCD. In this context and under specific 
assumptions and selected scenarios, 
Giordano and Meggiolaro 
have recently predicted an \textit{asymptotic} ($s \rightarrow \infty$) behavior for
\tcs\ in the form $B\ln^2{s}$, with the pre-factor $B$ universal (independent
of the colliding hadron) and connected with the QCD spectrum
\cite{gm1,gm2}.

Beyond specific phenomenological models (for recent reviews see, for instance, \cite{giulia,dremin,kaspar,fiore}), 
the behavior of $\sigma_{\mathrm{tot}}(s)$ 
is usually investigated through \textit{forward amplitude analyses}. The approach consists of analytic
parameterizations for the total cross section, connected with the ratio $\rho$ between
the real and imaginary parts of the forward amplitude,
\begin{eqnarray}
\rho(s) = \frac{\mathrm{Re}\,A(s,t=0)}{\mathrm{Im}\,A(s,t=0)},
\label{rho}
\end{eqnarray}
by means of analytic or numerical methods
and simultaneous fits to the experimental data available on these two quantities
\cite{amaldi,ua42,bv,ckk,compete1,compete2,lm,ii1,ii2,bh1,bh2}. In the context of the Regge-Gribov formalism,
the COMPETE collaboration developed in 2002 a detailed and comparative analysis of different
parameterizations for $\sigma_{\mathrm{tot}}(s)$ selecting as the best model the one with
the \textit{log-squared leading contribution} at the highest energies \cite{compete1,compete2}.
The predictions for $\sigma_{\mathrm{tot}}$ from this analysis are in plenty agreement with the
data obtained at the LHC in 2012 (ten years later) and subsequent measurements
by the TOTEM Collaboration.
This parametrization, to be discussed in detail later, became a standard
result in forward amplitude analyses and has been used and updated
by the COMPAS group (IHEP, Protvino) in the
successive editions of the Review of Particle Physics (RPP), by the Particle Data Group (PDG)
\cite{pdg12,pdg14,pdg16}.

On the other hand, in 1977 Amaldi et al. introduced a parametrization for
\tcs\ with the leading contribution in the form of log-raised-to-$\gamma$, 
where the exponent $\gamma$ was treated as a real free fit parameter. The simultaneous
fit to \tcs\ and $\rho$ data, from $pp$ and $\bar{p}p$ scattering in the interval 5 GeV 
$< \sqrt{s} \leq$ 62.5 GeV, resulted in the value \cite{amaldi}
\begin{eqnarray} 
\gamma = 2.10 \pm 0.10. \nonumber
\end{eqnarray} 
In 1993, the UA4/2 Collaboration extended the analysis up to 546 GeV 
($\bar{p}p$ scattering) obtaining \cite{ua42}
\begin{eqnarray} 
\gamma = 2.24_{-0.31}^{+0.35}.
\nonumber
\end{eqnarray} 
This result has been confirmed by Bueno and Velasco in 1996, who develop also fits
to only \tcs\ data, obtaining \cite{bv}
\begin{eqnarray} 
\gamma = 2.42_{-0.6}^{+0.4},
\nonumber
\end{eqnarray} 
for cutoff at 5 GeV and
\begin{eqnarray} 
\gamma = 2.64_{-0.32}^{+0.50},
\nonumber
\end{eqnarray} 
for cutoff at 10 GeV.

After the first measurement of the total cross section from LHC7 by the TOTEM 
Collaboration, Fagundes, Menon and Silva (hereafter FMS) developed in 2012 an amplitude analysis
with basis on the parametrization introduced by Amaldi et al. for the total cross section
\cite{fms1}.
As in the above works, the data reductions were limited to $pp$ and $\bar{p}p$ scattering.
This analysis
was then developed and extended in \cite{fms2} and further updated  and discussed by Menon and Silva \cite{ms1,ms2},
leading to values of $\gamma$ in the interval 2.2 - 2.4 \cite{ms1}.
Including the first LHC8 data on \tcs\ by the TOTEM Collaboration, the simultaneous fit to
\tcs\ and $\rho$ resulted in \cite{ms2}
\begin{eqnarray} 
\gamma = 2.23 \pm 0.11.
\nonumber
\end{eqnarray}
A recent updated analyses by FMS, with the TOTEM and ATLAS data, has been presented in \cite{fms17a},
indicating $\gamma \sim 2.3 \pm 0.1$ (TOTEM data) and $\gamma \sim 2.2 \pm 0.2$
(TOTEM and ATLAS data). We shall return to these results along the paper.

The formal (theoretical) possibility of a rise of the total cross section faster than the 
log-squared bound, without violating unitarity, was discussed by Azimov in 2011 \cite{azimov}.
It should be also noted that the Froissart, Lukaszuk and Martin bound, Eq. (\ref{fmb}), is intended for $s \rightarrow \infty$ and presently,
experimental data from accelerators have been obtained up to $\sqrt{s} =$ 8 TeV. 
To a certain extent contrasting with the above $\gamma$ values, the COMPAS group (PDG) has quoted the value
1.98 $\pm$ 0.01, however, without reference to the method employed \cite{pdg14,pdg16}.

In the analyses by Amaldi et al., UA4/2 Collaboration and Bueno-Velasco, numerical methods (integral dispersion relations) 
have been used to connect \tcs\ and \ro. As regards \textit{analytic methods}, specially in case of the real exponent
in the logarithmic term, two methods can be used, either \textit{Derivative Dispersion Relations}
(DDR) or \textit{Asymptotic Uniqueness} (AU), which is based on the Phragm\'en-Lindel\"off theorems
(references properly given in what follows).
In what concerns the parametrizations with the log-raised-to-$\gamma$ leding term, DDR is the method employed in the 
FMS analyses \cite{fms1,fms2,ms1,ms2,fms17a} and AU, for example, is the method advocated by Block in \cite{block} (page 103).
 
The main goal of this work is a global and detailed analysis on the similarities
and mainly the differences between the DDR and AU approaches, from both formal and practical points of view,
with focus on the log-raised-to-$\gamma$ leading term.
Beyond discussing analytic and conceptual aspects involved, we also develop new Regge-Gribov simultaneous fits 
to \tcs\ and $\rho$ data from $pp$ and $\bar{p}p$ scattering above 5 GeV. The two methods (DDR, AU) are employed,
treating parameterizations with both the leading log-squared (L2) and the log-raised-to-$\gamma$ (L$\gamma$)
components.
The results are compared with those present in the 2016 edition of the PDG.
The dataset comprises all the experimental information presently available from the LHC
(RUN 1 and RUN 2), including the measurements at 8 TeV by the TOTEM Collaboration (5 points) and ATLAS 
Collaboration (1 point).
The tension between the TOTEM/ATLAS results (at 7 TeV and mainly at 8 TeV) is discussed and taken into account 
in the data reductions.

Our main conclusions are the following: 

\begin{enumerate}

\item 
 all fit results present agreement with the experimental data analyzed 
and the goodness-of-fit is slightly better in case of the DDR approach;

\item
by considering only the TOTEM data at the LHC region, the fits with L$\gamma$ indicate $\gamma = 1.99 \pm 0.17$ (AU approach) and 
 $\gamma \sim 2.301 \pm 0.098$ (DDR approach);

\item
 by including the ATLAS data, the fits provide   
$\gamma = 1.85 \pm 0.13$ (AU approach) and 
 $\gamma = 2.16 \pm 0.16$ (DDR approach);

\item
in the formal context and in what concerns practical applicability, the DDR approach 
is more consistent and adequate for
the energy interval investigated than the AU approach.

\end{enumerate}

The article is  organized as follows. The experimental data and ensembles to be employed in the data reductions 
are displayed in Sect. 2. The analytic models (parameterizations for \tcs\ and $\rho$),
based on the DDR and AU approaches, involving either the leading log-squared or log-raised-to-$\gamma$ contributions,
are presented in Sect. 3. The data reductions to forward $pp$ and $\bar{p}p$ elastic
scattering with both DDR and AU approaches are developed in Sect. 4 and all the results are discussed, in 
detail, in Sect. 5.
Our conclusions and final remarks are the contents of Sect. 6.
In Appendix A we discuss the experimental data presently available at the highest energies
and in three other appendices we present a pedagogical
review on some results and concepts referred to along the text, related to:
Regge-Gribov formalism (Appendix B), integral and derivative dispersion relations
(Appendix C), asymptotic uniqueness and the Phragm\'en-Lindel\"off theorems (Appendix D).

\section{Experimental Data}

In this section we first justify and display the experimental data on \tcs\ and $\rho$
used in our data reductions, limited to $pp$ and $\bar{p}p$ in the interval 5 GeV - 8 TeV (Sect. 2.1). 
After that, we discuss and define two ensembles for data reductions: one with only
the TOTEM data from the LHC and another one by including the ATLAS data (Sect. 2.2).
Important aspects related to the experimental data presently available
at the highest energies and with our selection of two ensembles are
presented in Appendix A.

\subsection{Dataset}

In order to investigate the behavior of \tcs($s$) and \ro($s$) mainly at the highest
and asymptotic energy region, we limit our analysis to particle-particle and
particle-antiparticle collisions corresponding only to the \textit{largest
energy interval with available data}, namely $pp$ and $\bar{p}p$ scattering.
As commented in our previous works \cite{fms1,fms2,ms1,ms2}, the main point is
that this choice \textit{allows the investigation of possible high-energy
effects that may be unrelated to the trends of the lower energy data
on other reactions}. With this focus, we attempt to extract the most complete
information on the free parameters involved,
independently of additional constraints, as for example, tests on universal
behaviors or factorization.

However, it is important to stress that both COMPETE Collaboration and COMPAS 
Group (PDG) consider in their analysis not only $pp$ and $\bar{p}p$, 
but also mesons-$p$, $\gamma p$ and $\gamma \gamma$ scattering data
(among others), available
up to $\sqrt{s} \sim$ 100 GeV. That is certainly important for tests on universality.
We shall return to this subject (difference in the datasets) after presenting
our fit results.

Our dataset on \tcs\ and \ro\ comprises all the accelerator data from $pp$ and $\bar{p}p$
elastic scattering above 5 GeV \cite{pdgdata} (same cutoff used in the COMPETE and PDG analyses),
\textit{without any kind of data selection or sieve procedure}, and
including all published results from LHC7 and LHC8 by the TOTEM and ATLAS Collaborations.
The data on \tcs\ at the highest energy region are displayed in Table 1, together with other information
discussed in Appendix A. The recent measurement of \ro\ at 8 TeV by the TOTEM
Collaboration \cite{totem6} is also included in the dataset. 
All these data and information on
\tcs\ and \ro\ are shown in Figure 1. The legend specifying the symbols of each point 
is omitted in the other figures of the text.
Although not taking part in the data reductions,
we  display in the figures, as illustration, some estimations of the \tcs\ from cosmic-ray experiments:
ARGO-YBJ results at $~\,$ 100 - 400 GeV \cite{argo}, Auger result at 57 TeV \cite{auger} 
and Telescope Array (TA) result at 95 TeV \cite{ta}. 

\noindent
%\begin{center}

\begin{table}[ht]
\centering
\caption{\label{t1} Experimental information on measurements of the $\bar{p}p$ and $pp$ total cross section at
the highest energies from collider experiments (CERN-SPS, Fermilab-Tevatron and CERN-LHC): central value (\tcs), statistical uncertainties
($\Delta\sigma_{\mathrm{tot}}^{\mathrm{stat.}}$), systematic uncertainties
($\Delta\sigma_{\mathrm{tot}}^{\mathrm{syst.}}$), total uncertainty from quadrature
($\Delta$\tcs) and total relative uncertainty ($\Delta$\tcs/\tcs).\vspace{0.2cm}}
\begin{tabular}{c c c c  c c c c }
\hline
 reaction    &$\sqrt{s}$&\tcs &$\Delta\sigma_{\mathrm{tot}}^{\mathrm{stat.}}$&
$\Delta\sigma_{\mathrm{tot}}^{\mathrm{syst.}}$&$\Delta$\tcs& $\Delta$\tcs/\tcs & Collaboration\\
(collider) & (TeV)& (mb) & (mb) & (mb) & (mb) & ($\times$ 100) \% & [reference] \\
\hline
$\bar{p}p$ (SPS)& 0.546& 61.9 & 1.5 & 1.0  &1.8 &2.9 & UA4 \cite{ua4} \\
$\bar{p}p$ (Tevatron)&0.546 & 61.26 & 0.93 & -  & 0.93& 1.5 & CDF \cite{cdf} \\
$\bar{p}p$ (SPS)& 0.900& 65.3 & 0.7 & 1.55  &1.66 & 2.5& UA5 \cite{ua5}\\
$\bar{p}p$ (Tevatron)& 1.80& 72.8 & 3.1 & -  &3.1 & 4.3& E710 \cite{e710}\\
 $\bar{p}p$ (Tevatron)& 1.80& 80.03& 2.24& -  &2.24& 2.8& CDF \cite{cdf}\\
$\bar{p}p$ (Tevatron)& 1.80& 71.42& 1.55& 2.6&3.03& 4.2& E811 \cite{e811}\\ 
\hline
 $pp$ (LHC) & 7.0 & 98.3  & 0.2   & 2.8  & 2.8& 2.9 & TOTEM \cite{totem1}\\
 $pp$ (LHC) & 7.0 & 98.6  &  -    & 2.2  & 2.2& 2.2 & TOTEM \cite{totem2}\\
 $pp$ (LHC) & 7.0 & 98.0  &  -    & 2.5  & 2.5& 2.6 & TOTEM \cite{totem3}\\
 $pp$ (LHC) & 7.0 & 99.1  &  -    &  4.3 & 4.3& 4.3& TOTEM \cite{totem3}\\
 $pp$ (LHC) & 7.0 & 95.35 & 0.38  &1.304 & 1.36& 1.4& ATLAS \cite{atlas7}\\
 $pp$ (LHC) & 8.0 & 101.7 & -     & 2.9  & 2.9 & 2.9& TOTEM \cite{totem4}\\
 $pp$ (LHC) & 8.0 & 101.5 & -     & 2.1  & 2.1  & 2.1 & TOTEM \cite{totem5}\\
 $pp$ (LHC) & 8.0 & 101.9 & -     & 2.1  & 2.1  & 2.1 & TOTEM \cite{totem5}\\
 $pp$ (LHC) & 8.0 & 102.9 & -     & 2.3  & 2.3  & 2.2 & TOTEM \cite{totem6}\\
 $pp$ (LHC) & 8.0 & 103.0 & -     & 2.3  & 2.3  & 2.2 & TOTEM \cite{totem6}\\
 $pp$ (LHC) & 8.0 & 96.07 & 0.18  & 0.85 $\pm$ 0.31  & 0.92 & 1.0 & ATLAS \cite{atlas8}\\
\hline
\end{tabular}
\end{table}

\begin{figure}[ht]
\begin{center}
 \includegraphics[scale=0.4]{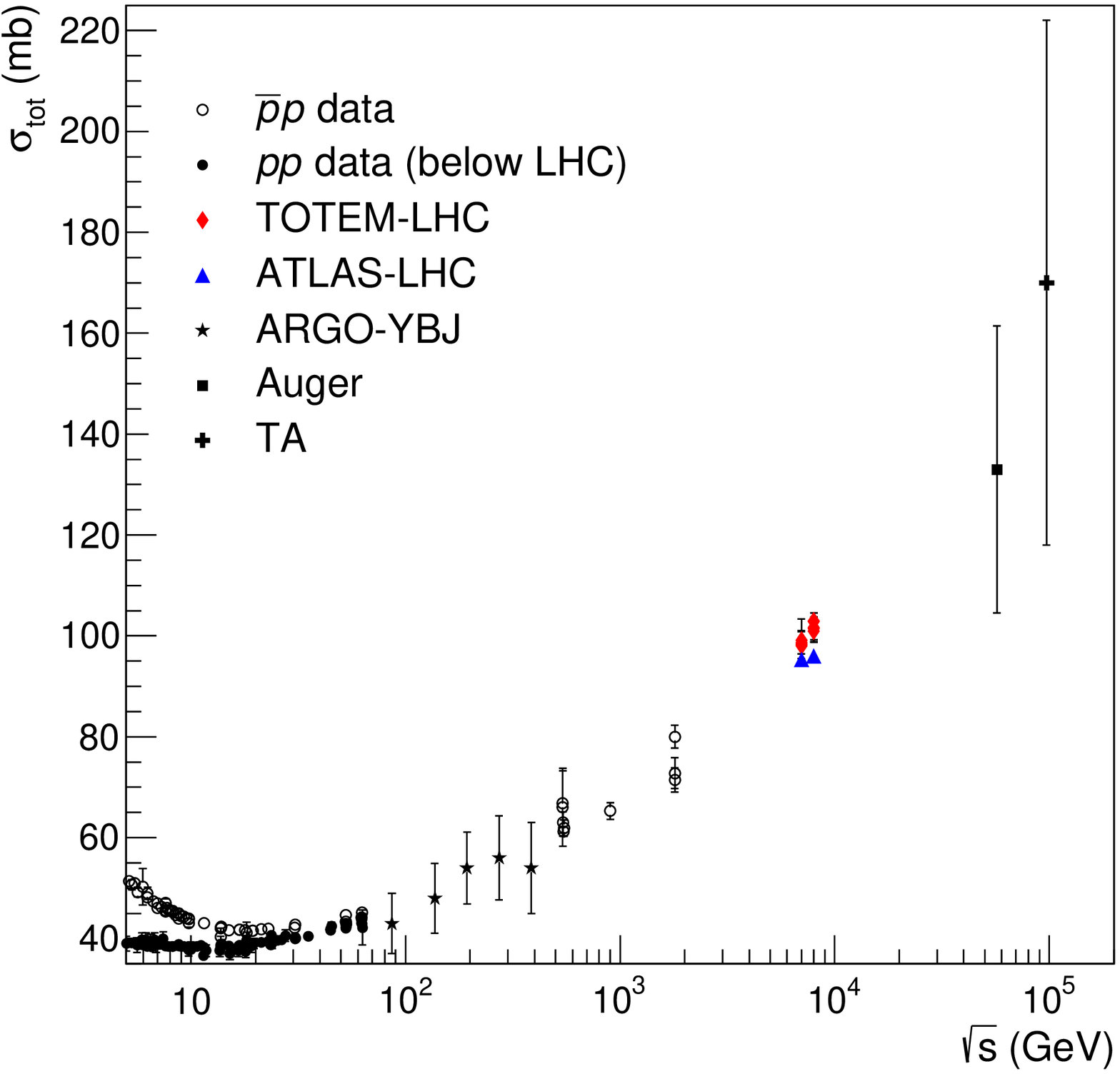}
 \includegraphics[scale=0.4]{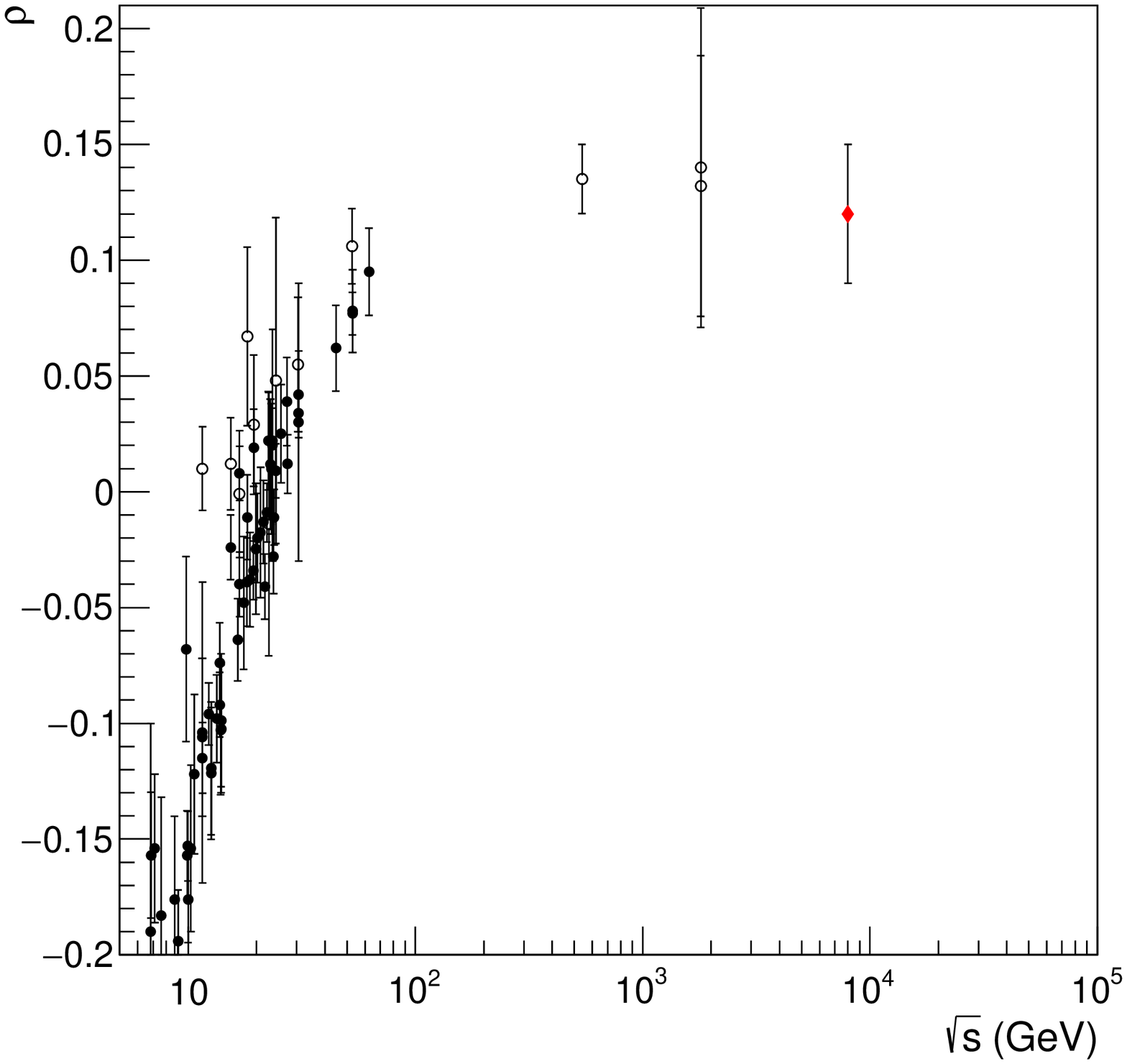}
 \caption{Accelerator data on \tcs\ and \ro\ from $pp$ and $\bar{p}p$ 
scattering used in this analysis. It is also displayed some estimations for the
$pp$ total cross section
from cosmic-ray experiments (ARGO-YBJ, Auger and TA).
The symbols here defined are assumed in all figures.}
\label{f1}
\end{center}
\end{figure}

\subsection{Ensembles}

As stated in Sect. 1,
our focus here concerns a study on the similarities and differences between two
\textit{approaches or methods}: DDR and AU. Based on our comments
in Appendix A (Sect. A.2) and also for the reasons discussed below,
we shall adopt as an ansatz two ensembles for data reductions, defined and denoted by

\vspace{0.3cm}

\noindent
{\bf Ensemble T}: accelerator data above 5 GeV and below the LHC plus all the TOTEM data.

\vspace{0.2cm}

\noindent
{\bf Ensemble T+A}: ensemble T plus the ATLAS data.

\vspace{0.3cm}

At first sight, ensemble T may sounds as a kind of biased choice
since only ensemble T+A does comprise all
the experimental data presently available. 
However, as we shall show, based on the important
role played by the $\gamma$ exponent as a free fit parameter,
the use of ensemble T provides a useful way to make explicit
the differences between the two methods, as well as
the correlations of $\gamma$ with other parameters involved in the data reductions. 
Moreover,
this choice is somehow supported by the consistence among several TOTEM results obtained through different
techniques and methods (independent of luminosity, or independent of $\rho$ and
based exclusively on the measurement of elastic scattering), comprising 4 points at 7 TeV and 5 points at 8 TeV, 
as well as the first
measurement of the $\rho$ parameter at the LHC8 (with the Coulomb-nuclear interference effects treated explicitly). 
In the ATLAS data (1 point at 7 TeV and
1 point at 8 TeV)  the measurement of \tcs\ used as input the phenomenological $\rho$ value 
extrapolated from the COMPETE analysis (7 TeV) or the COMPAS (PDG) fits (8 TeV) \cite{atlas7,atlas8}.

We stress that, in the present analysis, the use of ensembles T and T+A is strictly instrumental,
in the sense that they provide useful information on the practical differences
between the DDR and AU approaches, in case of models with the L$\gamma$ component. 
As commented in Appendix A.2, we have recently developed a global analysis through 
the DDR approach and including also an ensemble with the ATLAS data in place of the 
TOTEM data (denoted ensemble A) \cite{fms17a}). The goal there was to discuss bounds on the
rise of the total cross section, dictated by the experimental data available and using the
DDR approach.
Here our goal is to compare the two methods, DDR and AU, using two instrumental
ensembles of data, T and T+A.

As we shall show, given the small uncertainties in the
ATLAS results, the data reductions with ensembles T and T+A lead
to different scenarios for the rise of the total cross section at the highest energies.
Unfortunately, the situation for amplitude analyses is very similar to the pre-LHC era
and it seems currently useful the
statement by the COMPETE in 2002 \cite{compete2} ``Concerning the contradictory data, we are forced to use them
in our fits until the discrepancy is resolved by further experiments".

\section{Analytic Models}

In this section we introduce the analytic parameterizations for
\tcss\ and \ros\ of interest in this comparative study.
In order to further develop a detailed discussion, contrasting 
the DDR and AU approaches, it is instructive and important
to review the essential ideas and results related to three
correlated subjects:  Regge-Gribov theory, Dispersion Relation approach
and the Asymptotic Uniqueness approach.
These three subjects are treated, in some detail, in Appendices B, C and D, respectively,
where all the parameterizations presented in what follows are derived. 

Here,
after the
definition of a useful notation for models (Sect. 3.1), we display the analytic 
parameterizations for \tcss\ and \ros\ constructed through
the DDR approach (Sect. 3.2) and through the AU approach (Sect. 3.3).
The data reductions with these models are the subject of Sect. 4.

\subsection{Notation}

In what regards a suitable notation for our comparative study (DDR and AU), 
we have the comments and definitions that follows.
The parameterizations of interest have an structure
similar to that selected by COMPETE and following their notation,
could be represented by RRPL2 and RRPL$\gamma$.
Now, since the Regge contributions (RR) and the constant Pomeranchuk term (P)
have the same structure, they can be excluded in a short notation.
On the other hand, as we shall show, in case of L$\gamma$, the way to construct the parameterizations  
for \tcss\ and mainly the connections with \ros, are different in the DDR and AU
approaches and the analytic expressions and relations between 
the L2 and the L$\gamma$ terms are also not
the same in these two methods. 

Based on these details, we consider two aspects related to the DDR and
AU approaches for the definition of a notation:

1. Since our previous FMS analyses with the L$\gamma$ term
\cite{fms1,fms2,ms1,ms2,fms17a}
are based on DDR  with some specific assumptions, we shall adopt
here these FMS parameterizations as \textit{representative} of the DDR approach (in what concerns the leading L$\gamma$). 
For that reason we shall denote the DDR parameterizations as FMS-L$\gamma$ model and FMS-L2 model.

2. As we shall show, through the DDR approach, the FMS-L2 model is nothing more than
the FMS-L$\gamma$ model for $\gamma = 2$, namely a particular case. That, however, is not the
case within the AU approach, because the L$\gamma$ results for $\gamma = 2$ does not
correspond to an usual L2 parametrization. Therefore, in order to refer to a specific
model in the AU approach we need to distinguish these three cases, namely
L$\gamma$, L$\gamma$ for $\gamma = 2$ and L2. To this end
we adopt the following short notation: AU-L$\gamma$ model, AU-L$\gamma$=2 model and AU-L2
model, respectively.

Summarizing, we shall use the following notation for the analytic parameterizations for
\tcs($s$) and \ro($s$):
$$
\mbox{DDR Approach} \begin{cases}
 \mbox{FMS-L$\gamma$ Model} \\ 
 \mbox{FMS-L2 Model}
\end{cases}
\qquad
\mbox{AU\ Approach} \begin{cases}
 \mbox{AU-L2 Model} \\
 \mbox{AU-L$\gamma$ Model} \\
 \mbox{AU-L$\gamma$=2 Model}
\end{cases}
$$

In what follows we only display the corresponding formulas. As already noted,
all these analytic parameterizations are derived in detail in Appendices B, C and D
and are discussed in Sect. 5.1.
The symbols for the free parameters are mainly based on our previous notations
\cite{fms1,fms2}.

\subsection{Derivative Dispersion Relation Approach}

\subsubsection{FMS-L$\gamma$ Model}
The model is based on  two assumptions related separately to
\tcs\ and $\rho$ as follows \cite{fms1,fms2}.

\begin{description}

\item{1.}
The total cross section is given by the
parametrization introduced by Amaldi et al. in the 1970s \cite{amaldi}:

\begin{equation}
\sigma_{\mathrm{tot}}(s) = a_1\, \left[\frac{s}{s_0}\right]^{-b_1} + 
\tau\, a_2\, \left[\frac{s}{s_0}\right]^{-b_2}
+  \alpha + \beta \ln^{\gamma}\left(\frac{s}{s_0}\right),
\label{fmsstlg}
\end{equation}
where $\tau$ = -1 (+1) for $pp$ ($\bar{p}p$) scattering, 
while $a_1$, $b_1$, $a_2$, $b_2$, $\alpha$, $\beta$, $\gamma$ are real free fit parameters
and here, the energy scale is fixed at the physical threshold for scatterring states (see \cite{ms1},
Sect. 4.3
for discussions on this choice):

\begin{equation}
s_0 = 4m_p^2,
\label{escale}
\end{equation}
with $m_p$ the proton mass.

\item{2.} The $\rho(s)$ dependence is analytically determined through singly subtracted
derivative dispersion relations and using the Kang and Nicolescu representation \cite{kn} (see Appendix C
for details):

\begin{eqnarray}
\rho(s) &=& \frac{1}{\sigma_{\mathrm{tot}}(s)}
\left\{ \frac{K_{eff}}{s} + T^{R}(s) + T^{P}(s) \right\},
\label{fmsrholg}
\end{eqnarray}
where $K_{eff}$ is an \textit{effective subtraction constant} (discussed in Appendix C)
and the terms associated with Reggeon (R) and Pomeron (P) contributions read
\begin{eqnarray}
T^{R}(s) =
- a_1\,\tan \left( \frac{\pi\, b_1}{2}\right) \left[\frac{s}{s_0}\right]^{-b_1} +
\tau \, a_2\, \cot \left(\frac{\pi\, b_2}{2}\right) \left[\frac{s}{s_0}\right]^{-b_2} 
\label{fmsrhorlg}
\end{eqnarray}

\begin{eqnarray}
T^{P}(s) =
\mathcal{A}\,\ln^{\gamma - 1} \left(\frac{s}{s_0}\right) +
\mathcal{B}\,\ln^{\gamma - 3} \left(\frac{s}{s_0}\right) +
\mathcal{C}\,\ln^{\gamma - 5} \left(\frac{s}{s_0}\right),
\label{fmsrhoplg}
\end{eqnarray}
where
\begin{eqnarray} 
\mathcal{A} = \frac{\pi}{2} \, \beta\, \gamma,  
\quad 
\mathcal{B} = \frac{1}{3} \left[\frac{\pi}{2}\right]^3 \, \beta\, \gamma\, [\gamma - 1][ \gamma - 2], 
 \nonumber \\
\mathcal{C} = \frac{2}{15} \left[\frac{\pi}{2}\right]^5 \, \beta\, \gamma\, [\gamma - 1][ \gamma - 2]
[\gamma - 3][ \gamma - 4].
\label{fmsrhocoeflg}
\end{eqnarray}

\end{description}

\subsubsection{FMS-L2 Model}
In the particular case of $\gamma = 2$, from Eq. (\ref{fmsrhocoeflg}), 
we have $\mathcal{A} = \pi \beta$, $\mathcal{B} = \mathcal{C} = 0$
and through Eqs. (\ref{fmsstlg})-(\ref{fmsrhoplg}):

\begin{equation}
\sigma_{\mathrm{tot}}(s) = 
a_1\, \left[\frac{s}{s_0}\right]^{-b_1} + 
\tau\, a_2\, \left[\frac{s}{s_0}\right]^{-b_2}
+ \alpha + \beta \ln^{2}\left(\frac{s}{s_0}\right),
\label{fmsstl2}
\end{equation}

\begin{equation}
\rho(s) = \frac{1}{\sigma_{\mathrm{tot}}(s)}
\left\{\frac{K_{eff}}{s}
- a_1\,\tan \left( \frac{\pi\, b_1}{2}\right) \left[\frac{s}{s_0}\right]^{-b_1} +
\tau \, a_2\, \cot \left(\frac{\pi\, b_2}{2}\right) \left[\frac{s}{s_0}\right]^{-b_2}  
+ \pi \beta \ln\left(\frac{s}{s_0}\right)\right\}.
\label{fmsrhol2}
\end{equation}

These analytic expressions for \tcs($s$) and \ro($s$), Eqs. (\ref{fmsstl2}) and (\ref{fmsrhol2}), have solid basis on the Regge-Gribov
formalism (Appendix B) and (\ref{fmsrhol2}) can be directly obtained from (\ref{fmsstl2}) using DDR (Appendix C).
In the above two models, the reason to denote FMS instead of a general DDR is also related to the presence
and concept of an \textit{effective subtraction} (discussed in detail in Appendix C).

\noindent
$\bullet$ PDG-L2 Model

Eqs. (\ref{fmsstl2}) and (\ref{fmsrhol2}) have also
the same analytic structure as those selected by
the COMPETE Collaboration and used in the sucessive editions by the PDG, except for the presence here of the 
effective subtraction constant and the fixed energy scale, Eq. (\ref{escale}). 
We recall that in the PDG case the energy scale is given by
\begin{eqnarray}
s_0 = (2m_p + M)^2
\nonumber 
\end{eqnarray}
and $M$ and $\beta$ are constrained by
\begin{eqnarray}
\beta = \pi/M^2,
\nonumber 
\end{eqnarray}
with $M$ a free fit parameter \cite{pdg14,pdg16}.
In what follows we shall refer to this PDG version as PDG-L2 model
(namely without the subtraction constant and with the above two
constraints).

\subsection{Asymptotic Uniqueness Approach}

\subsubsection{AU-L2 Model} From Appendices D.2 and D.3, this model has the
same structure obtained with the Regge-Gribov (Appendix B) and DDR (Appendix C)
formalisms (except, in the last case, for the subtraction constant).
Although the formulas correspond to Eqs. (\ref{fmsstl2}) and (\ref{fmsrhol2}), without $K_{eff}$, for future reference and discussion we 
display the results:

\begin{equation}
\sigma_{\mathrm{tot}}(s) = 
a_1\, \left[\frac{s}{s_0}\right]^{-b_1} + 
\tau\, a_2\, \left[\frac{s}{s_0}\right]^{-b_2}
+ \alpha + \beta \ln^{2}\left(\frac{s}{s_0}\right),
\label{austl2}
\end{equation}

\begin{equation}
\rho(s) = \frac{1}{\sigma_{\mathrm{tot}}(s)}
\left\{
- a_1\,\tan \left( \frac{\pi\, b_1}{2}\right) \left[\frac{s}{s_0}\right]^{-b_1} +
\tau \, a_2\, \cot \left(\frac{\pi\, b_2}{2}\right) \left[\frac{s}{s_0}\right]^{-b_2}  
+ \pi \beta \ln\left(\frac{s}{s_0}\right)\right\},
\label{aurhol2}
\end{equation}
where, as before, $\tau$ = -1 (+1) for $pp$ ($\bar{p}p$) scattering. The energy scale
$s_0$ is a parameter to be treat in our data reductions (Sect. 4).

\subsubsection{AU-L$\gamma$ Model} From Appendices D.2 and D.4 (subsection D.4.1)
the AU approach leads to the following analytic results:

\begin{eqnarray}
\!\!\!\!\!\!\!\!\!\!\!\!\!\!\!\!\!\!\!\!\!\!\!\!
\sigma_{\mathrm{tot}}(s) = 
a_1\, \left[\frac{s}{s_0}\right]^{-b_1} + 
\tau\, a_2\, \left[\frac{s}{s_0}\right]^{-b_2} 
+ \alpha + 
\beta \cos \left(\frac{\gamma \phi}{2}\right) \ln^{\gamma/2}\left(\frac{s}{s_0}\right)
\left[\ln^2\left(\frac{s}{s_0}\right) + \pi^2\right]^{\gamma/4},
\label{austlg}
\end{eqnarray}

\begin{eqnarray}
\rho(s) &=& \frac{1}{\sigma_{\mathrm{tot}}(s)}
\left\{
- a_1\,\tan \left( \frac{\pi\, b_1}{2}\right) \left[\frac{s}{s_0}\right]^{-b_1} +
\tau \, a_2\, \cot \left(\frac{\pi\, b_2}{2}\right) \left[\frac{s}{s_0}\right]^{-b_2}   
\right.  \nonumber \\
&+& \left. \beta \sin \left(\frac{\gamma \phi}{2}\right)
\ln^{\gamma/2}\left(\frac{s}{s_0}\right)
\left[\ln^2\left(\frac{s}{s_0}\right) + \pi^2\right]^{\gamma/4}\right\},
\label{aurholg}
\end{eqnarray}
where 

\begin{eqnarray}
\phi = \phi(s) = \tan^{-1}\left(\frac{\pi}{\ln(s/s_0)}\right).
\label{auphi}
\end{eqnarray}

\subsubsection{AU-L$\gamma$=2 Model} For $\gamma = 2$, Eqs. (\ref{austlg}) and (\ref{aurholg}) read

\begin{eqnarray}
\!\!\!\!\!\!\!\!\!\!\!\!\!\!\!\!
\sigma_{\mathrm{tot}}(s) = 
a_1\, \left[\frac{s}{s_0}\right]^{-b_1} + 
\tau\, a_2\, \left[\frac{s}{s_0}\right]^{-b_2} 
+ \alpha + 
\beta \cos \left(\phi\right) \ln\left(\frac{s}{s_0}\right)
\left[\ln^2\left(\frac{s}{s_0}\right) + \pi^2\right]^{1/2},
\label{austg=2}
\end{eqnarray}

\begin{eqnarray}
\rho(s) &=& \frac{1}{\sigma_{\mathrm{tot}}(s)}
\left\{
- a_1\,\tan \left( \frac{\pi\, b_1}{2}\right) \left[\frac{s}{s_0}\right]^{-b_1} +
\tau \, a_2\, \cot \left(\frac{\pi\, b_2}{2}\right) \left[\frac{s}{s_0}\right]^{-b_2}   
\right.  \nonumber \\
&+& \left. \beta \sin \left(\phi\right)
\ln\left(\frac{s}{s_0}\right)
\left[\ln^2\left(\frac{s}{s_0}\right) + \pi^2\right]^{1/2}\right\},
\label{aurhog=2}
\end{eqnarray}
with $\phi = \phi(s)$ given by Eq. (\ref{auphi}).

We note that, differently from the FMS-L$\gamma$ and FMS-L2 models,
this AU-L$\gamma$=2 model does not correspond to the AU-L2 model,
given by Eqs. (\ref{austl2})-(\ref{aurhol2}).
The \textit{analytic} similarities and mainly differences between the AU and DDR approaches,
related to L2 and L$\gamma$ models, are discussed in Sect. 5.1.

\section{Data Reductions and Results}

We develop here several fits to $pp$ and $\bar{p}p$ scattering
in the interval 5 GeV - 8 TeV. As explained in Sect. 2.2 we consider two ensembles,
consisting of the data below the LHC region and including only the TOTEM data (ensemble T)
or including also the ATLAS data (ensemble T+A).

In order to confront the DDR and AU approaches with L2 and L$\gamma$ models,
four parameterizations for \tcs\ and \ro\ are considered: 
FMS-L2 and FMS-L$\gamma$ models, Eqs. (\ref{fmsstlg})-(\ref{fmsrhocoeflg}) (since L2 is the particular case of L$\gamma$
for $\gamma$ = 2), AU-L$\gamma$=2 model, Eqs. (\ref{austg=2})-(\ref{aurhog=2}) and AU-L$\gamma$ model, 
Eqs. (\ref{austlg})-(\ref{auphi}) 
(since the L2 is not a particular of L$\gamma$ for $\gamma$ = 2).

\subsection{Fit Procedures}

The nonlinearity of the fits demands a methodology for the choice of the
initial values of the free parameters (feedback) and that is an important point.
As in our previous analyses, we initialize our parametric set
with the central values  already obtained by the PDG in their most recent
data reductions. Here we use the values of the parameters published
in the 2016 edition, which for the $pp$ and $\bar{p}p$
scattering read \cite{pdg16}:
\begin{eqnarray}
a_1 &=& 13.07 \pm 0.17\ \mathrm{mb}, \quad b_1 = 0.4473 \pm 0.0077, \quad 
a_2 = 7.394 \pm 0.081\ \mathrm{mb},  \nonumber \\
b_2 &=& 0.5486 \pm 0.0049, \quad 
\alpha = 34.41 \pm 0.13\ \mathrm{mb}, \quad \beta = 0.2720 \pm 0.0024\ \mathrm{mb}, \nonumber  \\
s_0 &=& 15.977 \pm 0.075\ \mathrm{GeV}^2 
\label{ivpdg}
\end{eqnarray}

Recall that these results were obtained with the PDG-L2 model (Sect. 3.2.2) from fits to data comprising several reactions
and not only to $pp$ and $\bar{p}p$ data. Also, the dataset did not include 
the latest TOTEM measurements and the ATLAS result at 8 TeV.
For future discussion, we display in Figure 2 the 
corresponding results for \tcs $(s)$ and $\rho(s)$ with the PDG-L2 model (2016), 
for $pp$ and $\bar{p}p$ scattering \cite{pdg16}, together with the experimental
data used here.
\begin{figure}[ht]
 \includegraphics[scale=0.4]{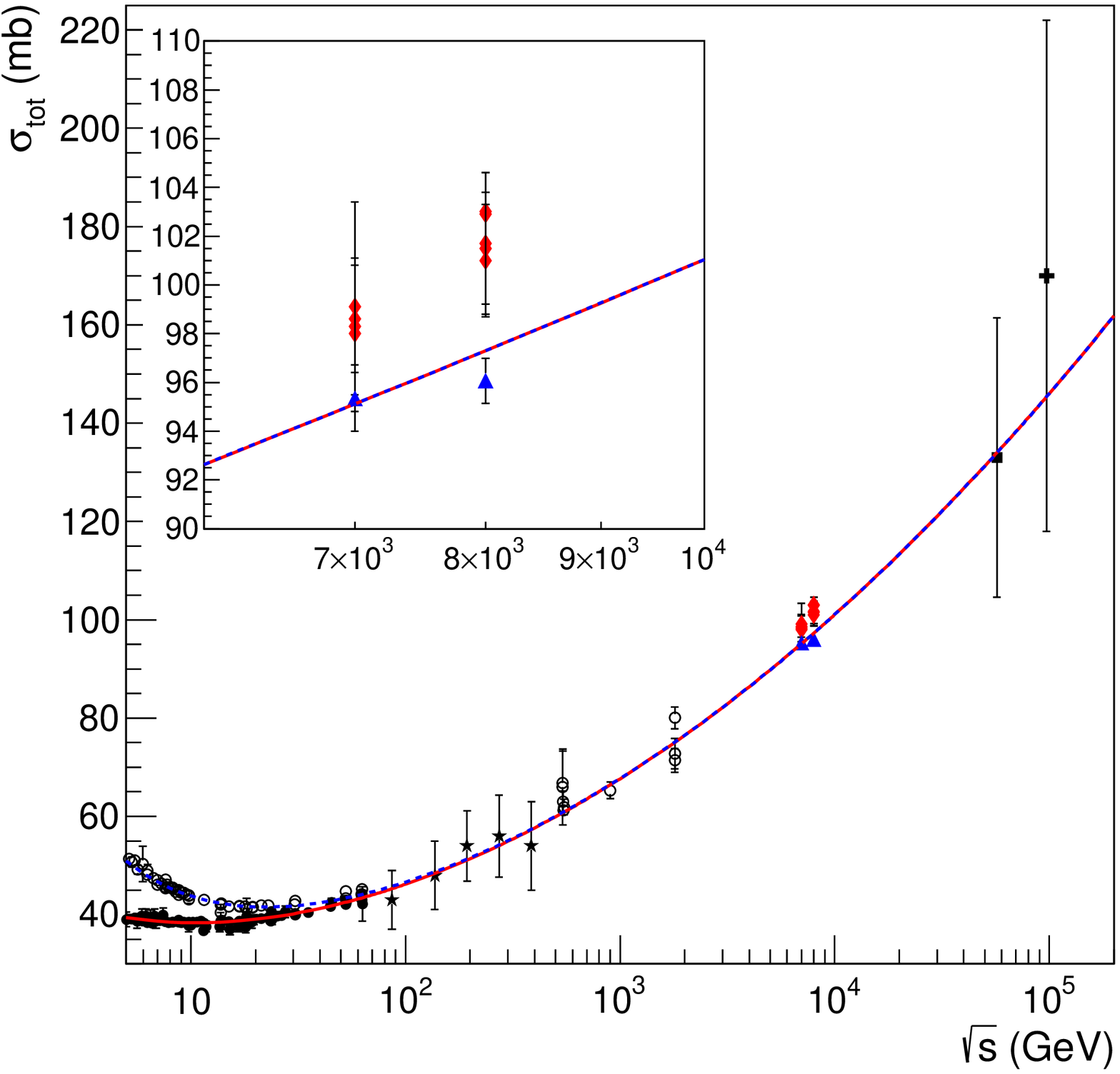}
 \includegraphics[scale=0.4]{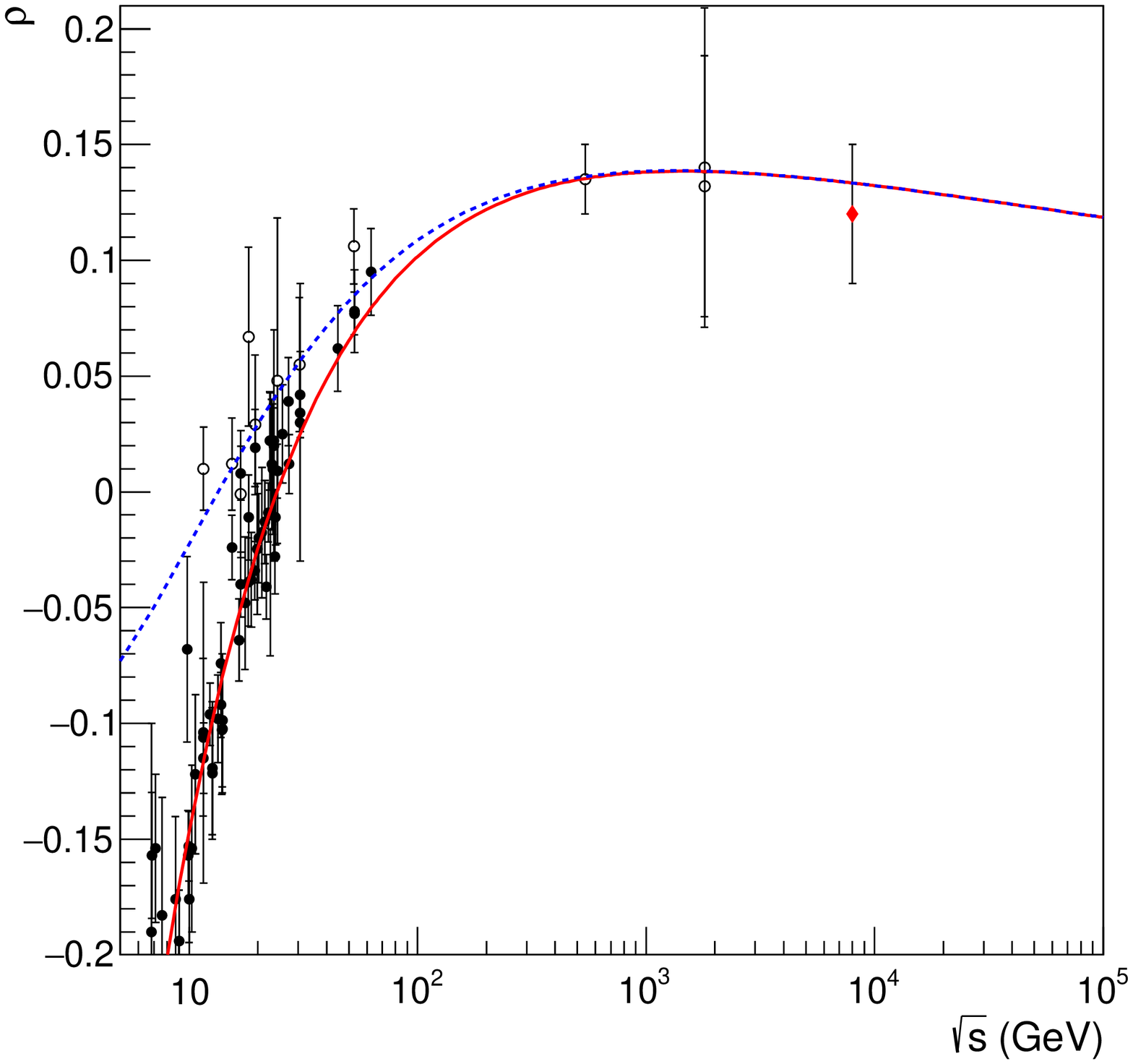}
 \caption{PDG-L2 results from the PDG 
2016 \cite{pdg16}, parameters from Eq. (\ref{ivpdg}), compared with 
$pp$ and $\bar{p}p$ data. The insert shows the \tcs\ data and curve
in the region 6 - 10 TeV.}
\label{f2}
\end{figure}

The fit procedure is as follows. With these central values as feedback we develop the fits
with the FMS-L2 model and AU-L$\gamma$=2 model and, in each case,
we use the resulting central values of the free parameters
as initial values for data reductions with the corresponding L$\gamma$ models.
These fit procedures can be summarized by the scheme that follows, where
I.V. stands for initial values of the free parameters.
$$
 \mbox{PDG-L2, Eq. (\ref{ivpdg})}\ 
\xrightarrow{\text{I.V.}}
\quad
\begin{cases}
 \mbox{fit\ FMS-L2 model}\ \xrightarrow{\text{I.V.}}\ \mbox{fit\ FMS-L$\gamma$ model} \\
 \mbox{fit\ AU-L$\gamma$=2 model}\ \xrightarrow{\text{I.V.}}\ \mbox{fit\ AU-L$\gamma$ model}
\end{cases}
$$
These procedures have been used with each one of the two ensemble: T and T+A.

As in our previous analysis, 
the data reductions have been performed with the objects of the class TMinuit of ROOT Framework 
\cite{root}. We have employed the default MINUIT error analysis \cite{minuit}
with the \textit{selective criteria} explained in \cite{ms1} (section 2.2.4).
The error matrix provides the variances and covariances associated with each free parameter,
which are used in the analytic evaluation of the uncertainty regions 
associated with the fitted and predicted
quantities (through standard error propagation procedures \cite{bev}).
As tests of goodness-of-fit we shall consider the chi-square per degree of freedom
($\chi^2/\nu$) and
the corresponding integrated probability, $P(\chi^2)$ \cite{bev}. 

\vspace{0.3cm}

\noindent
$\bullet$ Energy Scale.

Here, as a first step in our study and following the discussions
in \cite{ms1} (Sects. 4.2 and 4.3),
we shall consider, in all fits, the energy scale fixed
at the physical threshold, Eq. (\ref{escale}),
\begin{eqnarray}
s_0 = 4m_p^2 \sim \mathrm{3.521\ GeV}^2. 
\nonumber
\end{eqnarray}
Detailed analyses with $s_0$ as a free
fit parameter are in progress.

\vspace{0.3cm}

In what follows, we present the fit results with
the DDR Approach (Sect. 4.1) followed by those with the AU approach.
(Sec. 4.2). A detailed discussion on all these results is the content
of Sect. 5.2.

\

\subsection{FMS-L2 and FMS-L$\gamma$ models}

In applying the above fit procedure with Eqs. (\ref{fmsstlg})-(\ref{fmsrhocoeflg}),
since the subtraction constant is absent in the PDG parametrization,
we consider the initial value 0 for $K_{eff}$.  
The fit results, already obtained in our recent analysis \cite{fms17a}, are displayed in Table 2, including statistical information
on the quality of the fit. The corresponding results for \tcs $(s)$ and \ro $(s)$ with ensembles T and T+A,
together with the experimental data analyzed,
are shown in Fig. 3 with the FMS-L2 model and in Fig. 4 with
the FMS-L$\gamma$ model.

\begin{table}[ht]
\centering
\caption{\label{t2} Fit results with the FMS-L2 and FMS-L$\gamma$ models, Eqs. (\ref{fmsstlg})-(\ref{fmsrhocoeflg}), 
to ensembles T and T+A.
Energy scale fixed, $s_0 = 4m_p^2 =$ 3.521 GeV$^2$.
It is also displayed statistical information on the
quality of the fits (chi squared per degree of freedom and integrated probability).
The corresponding figures are referred to in the last line.
Results from \cite{fms17a}.}
\begin{tabular}{c c c c c}
\hline
Ensemble:              & \multicolumn{2}{c}{T}       &        \multicolumn{2}{c}{T+A}                              \\
\cmidrule(lr){2-3} \cmidrule(lr){4-5}
Model:                 &     FMS-L2          & FMS-L$\gamma$          & FMS-L2              & FMS-L$\gamma$        \\
\hline
$a_1$ (mb)             & 32.11 $\pm$ 0.60    & 31.5 $\pm$ 1.3         & 32.16 $\pm$ 0.67    & 31.60 $\pm$ 0.98     \\
$b_1$                  & 0.381 $\pm$ 0.017   & 0.528 $\pm$ 0.057      & 0.406 $\pm$ 0.016   & 0.484 $\pm$ 0.084    \\
$a_2$  (mb)            & 16.98 $\pm$ 0.72    & 17.10 $\pm$ 0.74       & 17.01 $\pm$ 0.72    & 17.07 $\pm$ 0.73     \\
$b_2$                  & 0.545 $\pm$ 0.013   & 0.546 $\pm$ 0.013      & 0.545 $\pm$ 0.013   & 0.546 $\pm$ 0.013    \\
$\alpha$ (mb)          & 29.25 $\pm$ 0.44    & 34.0 $\pm$ 1.1         & 30.06 $\pm$ 0.34    & 32.8 $\pm$ 2.2       \\
$\beta$  (mb)          & 0.2546 $\pm$ 0.0039 & 0.103 $\pm$ 0.029      & 0.2451 $\pm$ 0.0028 & 0.151 $\pm$ 0.071    \\
$\gamma$               & 2 (fixed)           & \bf{2.301 $\pm$ 0.098} & 2 (fixed)           & \bf{2.16 $\pm$ 0.16} \\
$K_{eff}$ (mbGeV$^2$)  & 50 $\pm$ 17         & 109 $\pm$ 36           & 61 $\pm$ 17         & 90 $\pm$ 42          \\
\hline
$\nu$                  &  242                & 241                    & 244                 & 243                  \\
$\chi^2/\nu$           & 1.09                & 1.07                   & 1.15                & 1.14                 \\
$P(\chi^2)$            & 0.150               & 0.213                  & 0.059               & 0.063                \\
\hline
Figure:                &        3            &        4               &         3           &    4                 \\
\hline
\end{tabular}
\end{table}
\begin{figure}[ht]
 \includegraphics[scale=0.4]{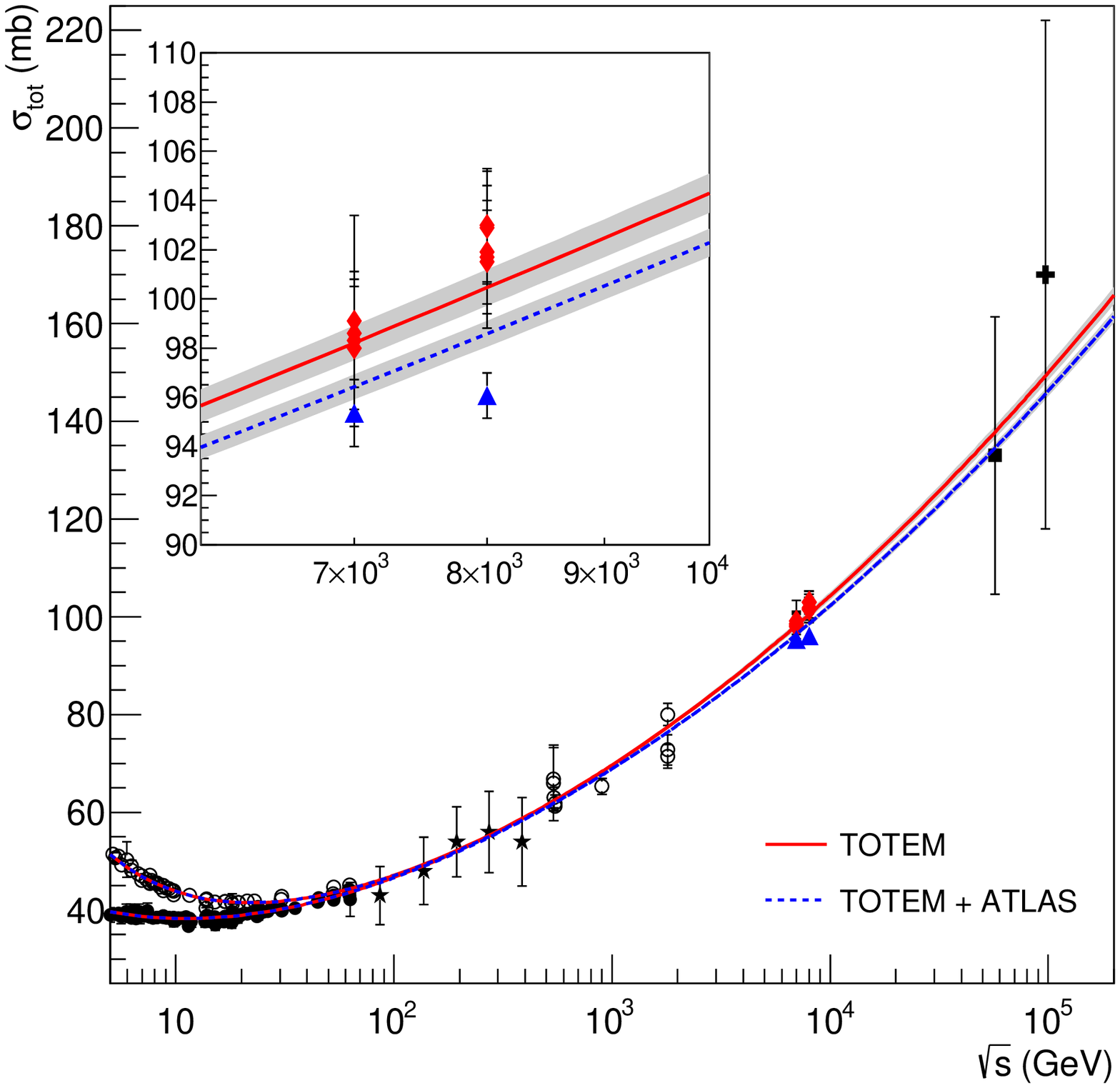}
 \includegraphics[scale=0.4]{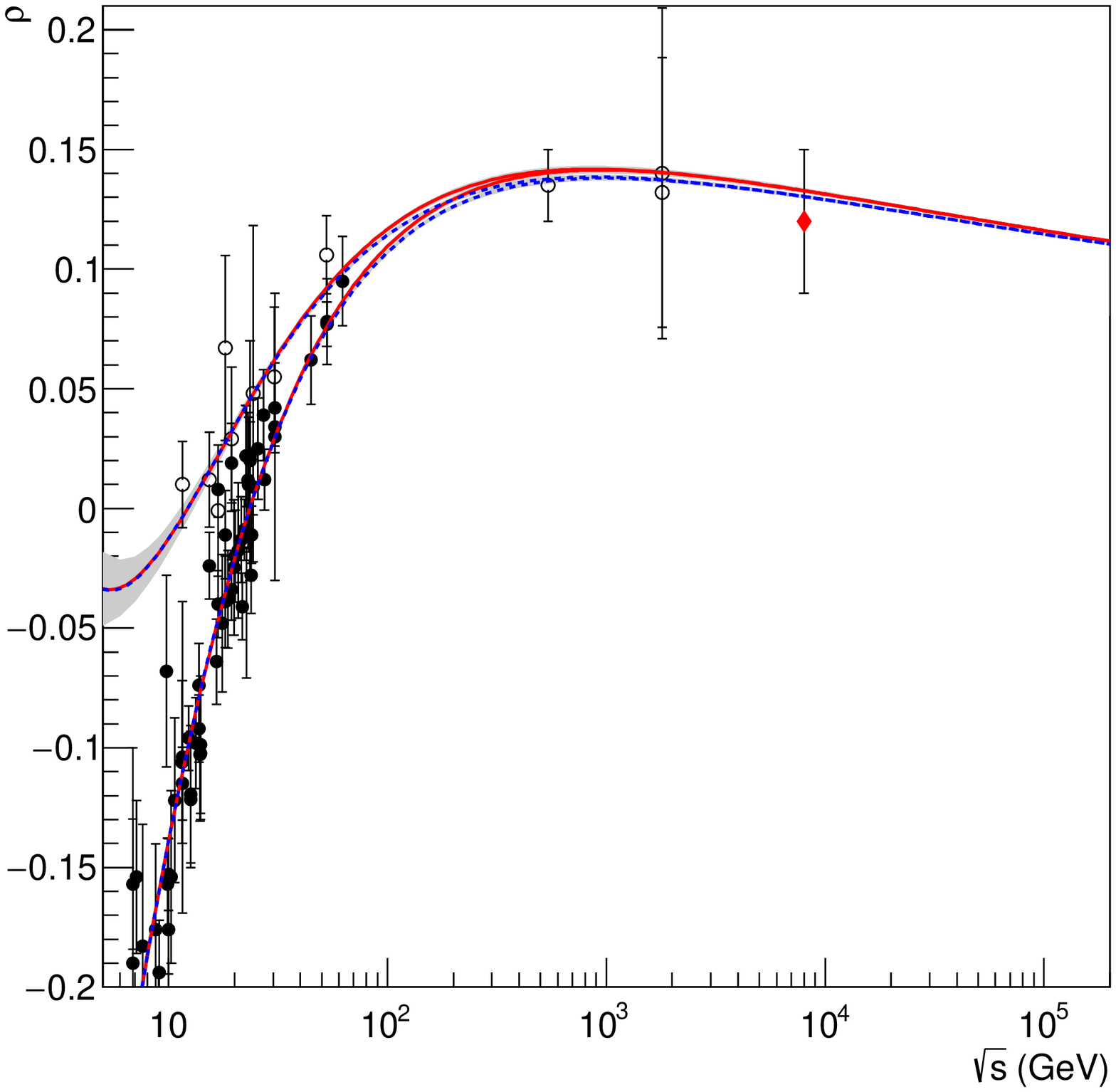}
 \caption{Fit results with the FMS-L2 model to ensembles T and T+A,
Eqs. (\ref{fmsstlg})-(\ref{fmsrhocoeflg}), Table 2.}
\label{f3}
\end{figure}
\begin{figure}[ht]
 \includegraphics[scale=0.4]{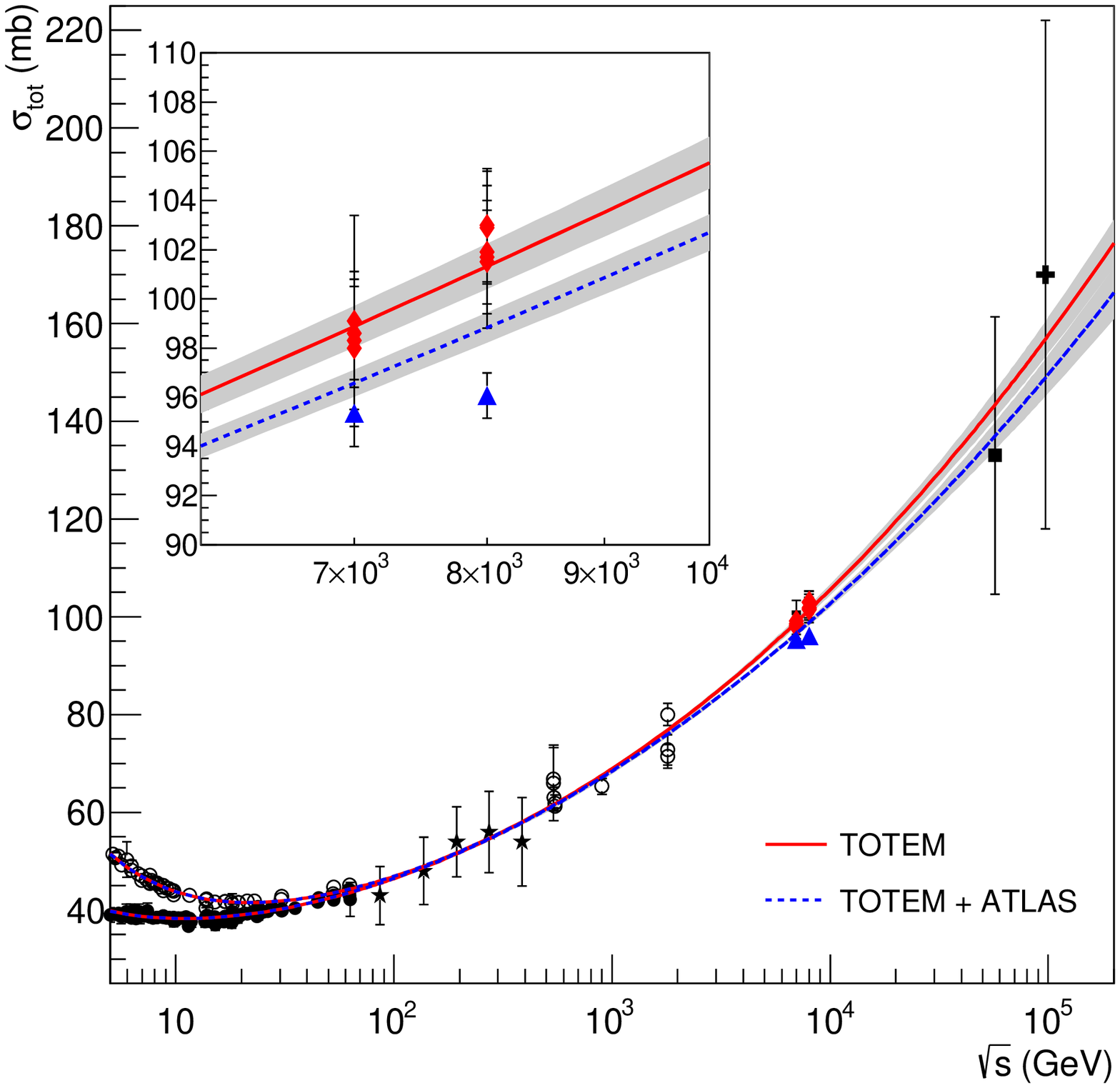}
 \includegraphics[scale=0.4]{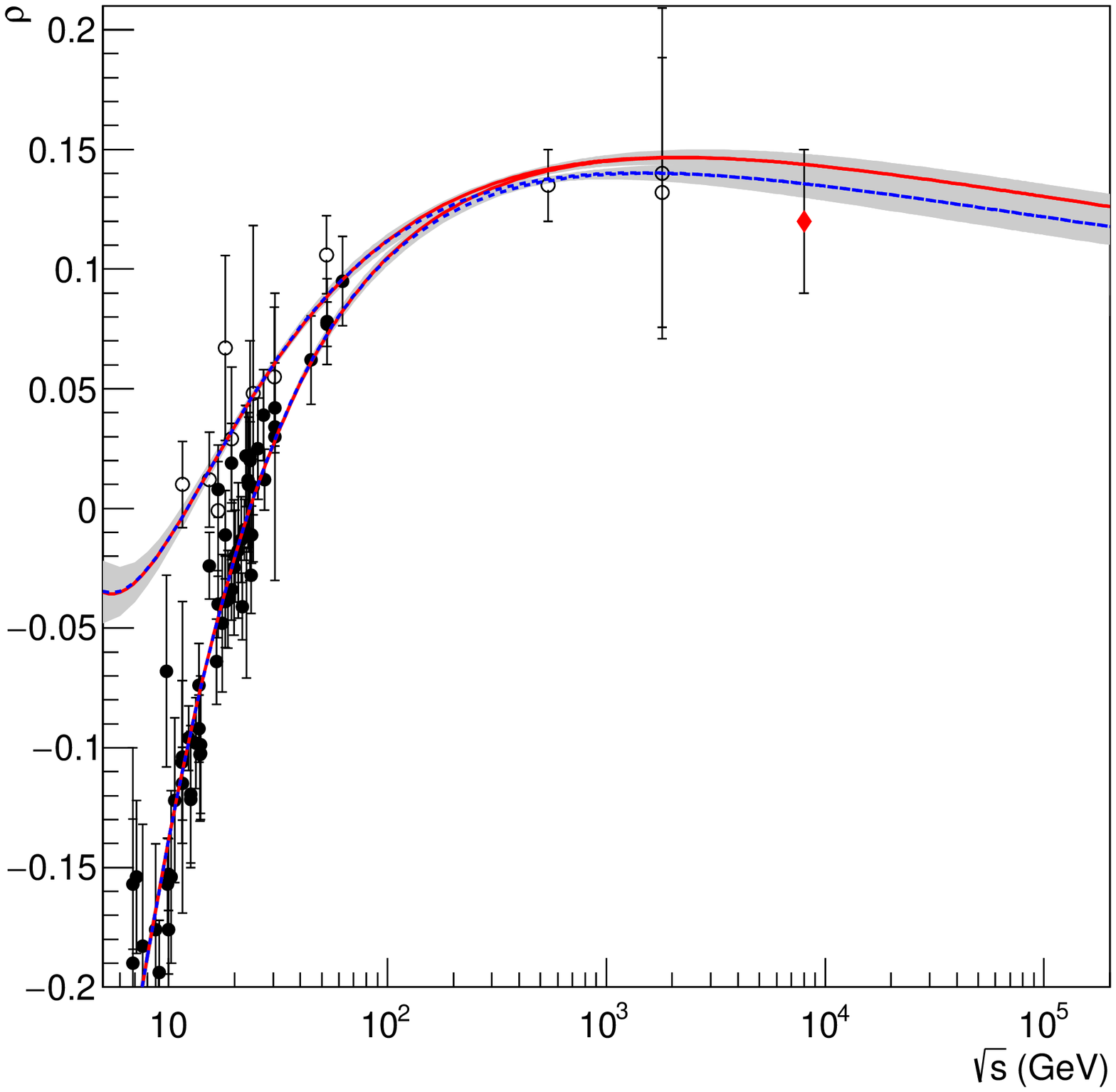}
 \caption{Fit results with the FMS-L$\gamma$ model to ensembles T and T+A,
Eqs. (\ref{fmsstlg})-(\ref{fmsrhocoeflg}), Table 2.}
\label{f4}
\end{figure}

\subsection{AU-L2 and AU-L$\gamma$ Models}

The fit results are displayed in Table 3 and the corresponding results for \tcs $(s)$ and \ro $(s)$ with ensembles T and T+A,
together with the experimental data analyzed,
are shown in Fig. 5 with the AU-L$\gamma$=2 model and in Fig. 6 with
the AU-L$\gamma$ model.

\begin{table}[ht]
\centering
\caption{\label{t3} Data reductions with the AU-L$\gamma$=2 model, Eqs. (\ref{austg=2})-(\ref{aurhog=2}) and AU-L$\gamma$ model, 
Eqs. (\ref{austlg})-(\ref{auphi}), 
to ensembles T and T+A.
Energy scale fixed, $s_0 = 4m_p^2 =$ 3.521 GeV$^2$.}
\begin{tabular}{c c c c c}
\hline
Ensemble:       & \multicolumn{2}{c}{T}                        & \multicolumn{2}{c}{T + A}                \\
\cmidrule(lr){2-3} \cmidrule(lr){4-5}
Model:          &    AU-L$\gamma$=2    & AU-L$\gamma$      & AU-L$\gamma$=2      & AU-L$\gamma$          \\
\hline
$a_1$ (mb)      &  31.42 $\pm$ 0.47    & 31.5 $\pm$ 3.1    & 31.10 $\pm$ 0.50    & 34.0 $\pm$ 3.1     \\
$b_1$           &  0.355 $\pm$ 0.014   & 0.353 $\pm$ 0.060 & 0.376 $\pm$ 0.013   & 0.326 $\pm$ 0.044  \\
$a_2$ (mb)      &  17.30 $\pm$ 0.72    & 17.30 $\pm$ 0.73  & 17.38 $\pm$ 0.72    & 17.28 $\pm$ 0.72   \\
$b_2$           &  0.553 $\pm$ 0.013   & 0.553 $\pm$ 0.013 & 0.555 $\pm$ 0.013   & 0.553 $\pm$ 0.013  \\
$\alpha$ (mb)   &  28.61 $\pm$ 0.43    & 28.5 $\pm$ 3.9    & 29.46 $\pm$ 0.32    & 25.9 $\pm$ 3.7     \\
$\beta$  (mb)   &  0.2584 $\pm$ 0.0038 & 0.26 $\pm$ 0.14   & 0.2483 $\pm$ 0.0027 & 0.39 $\pm$ 0.16    \\
$\gamma$        &  2 (fixed)           & {\bf1.99 $\pm$ 0.17}   & 2 (fixed)           & {\bf 1.85 $\pm$ 0.13}    \\
\hline
$\nu$           &  243                 & 242               & 245                 & 244                \\
$\chi^2/\nu$    & 1.125         & 1.130             & 1.191               & 1.188              \\
$P(\chi^2)$     & 0.0878        & 0.0809            & 0.0217              & 0.0232             \\
\hline
Figure:         &    5          &        6          &        5            &        6    \\
\hline
\end{tabular}
\end{table}
\begin{figure}[ht]
 \includegraphics[scale=0.4]{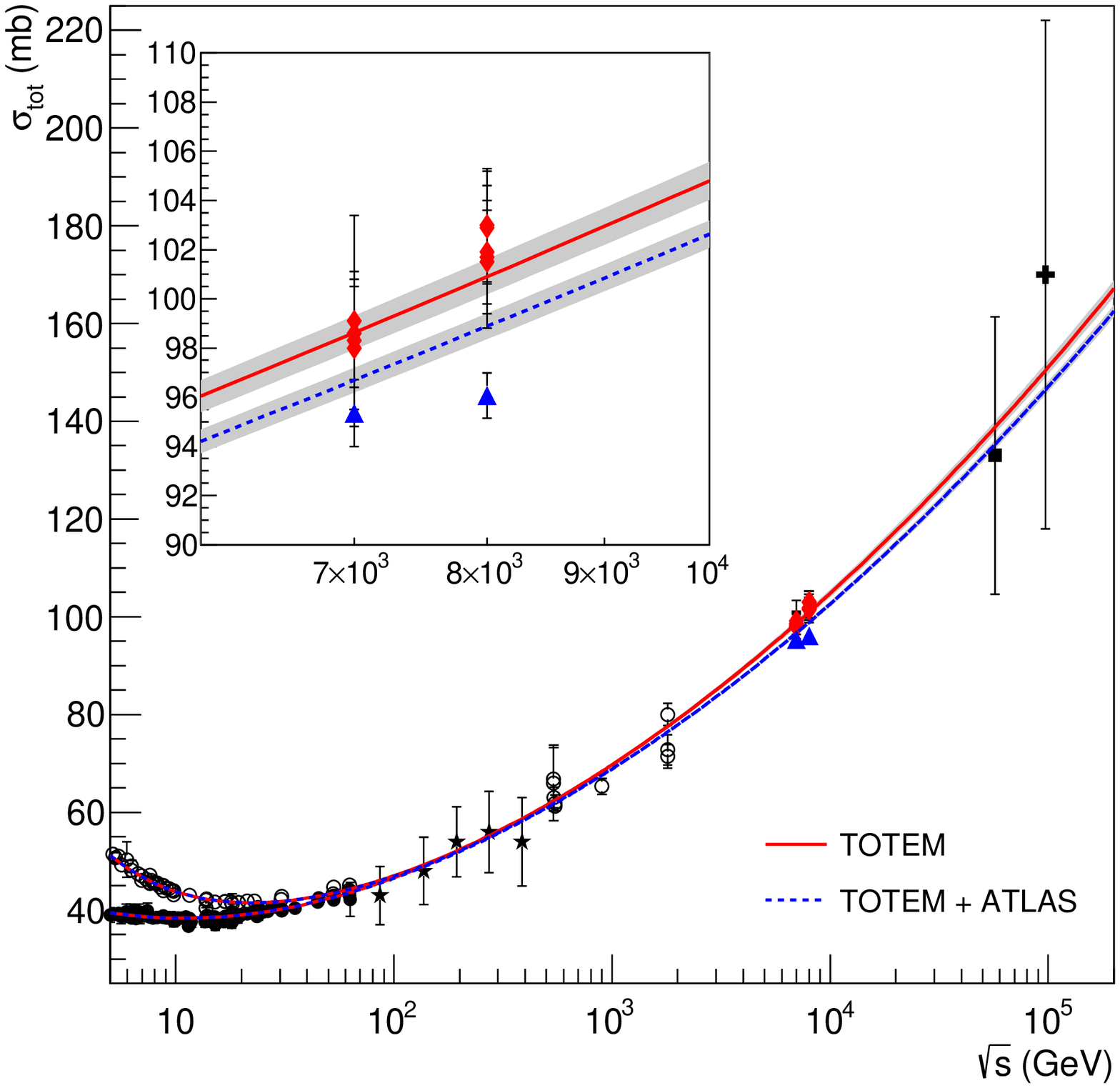}
 \includegraphics[scale=0.4]{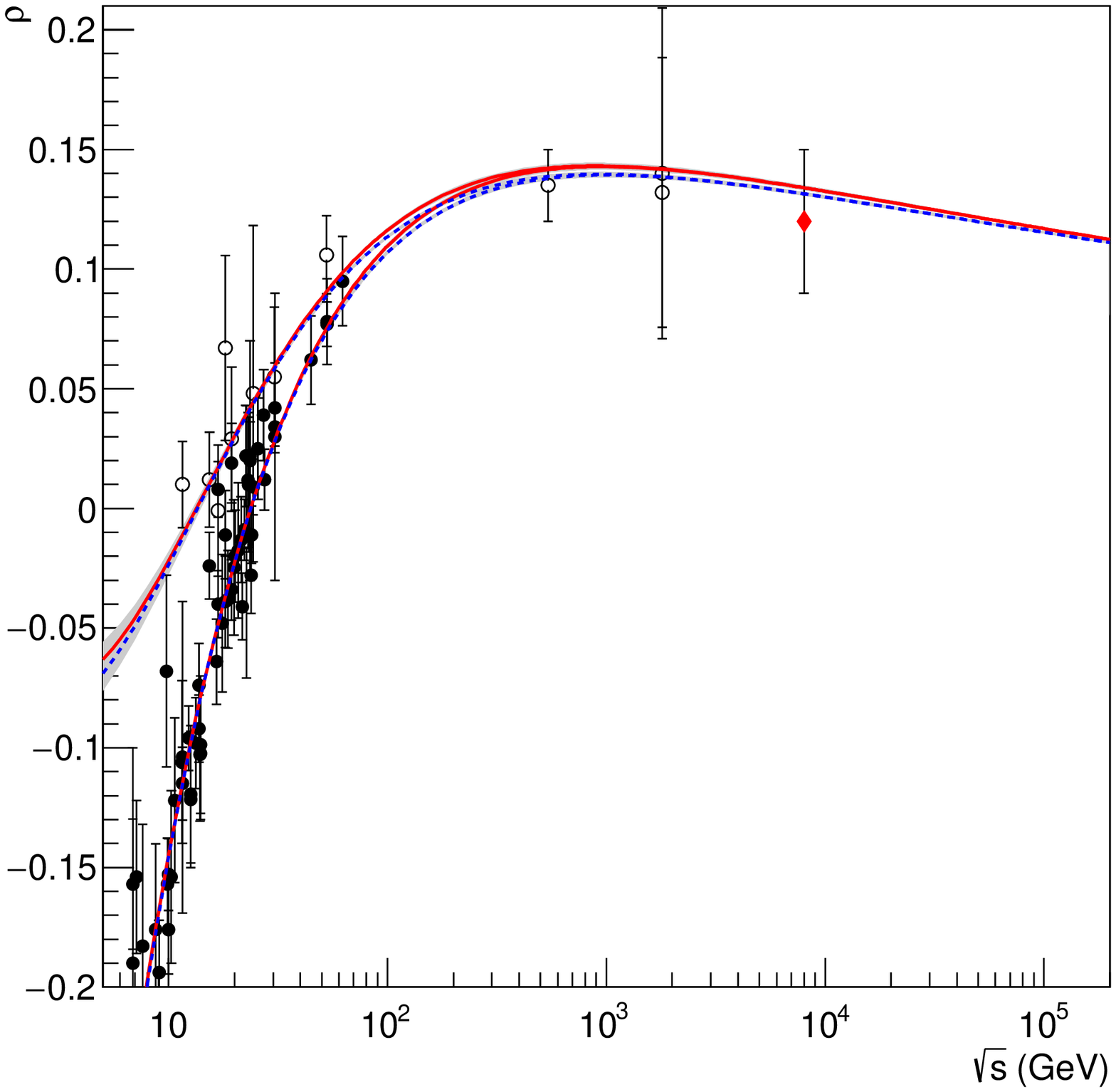}
 \caption{Fit results with the AU-L$\gamma$=2 model to ensembles T and T+A.
Eqs. (\ref{austg=2})-(\ref{aurhog=2}), Table 3.}
\label{f5}
\end{figure}
\begin{figure}[ht]
 \includegraphics[scale=0.4]{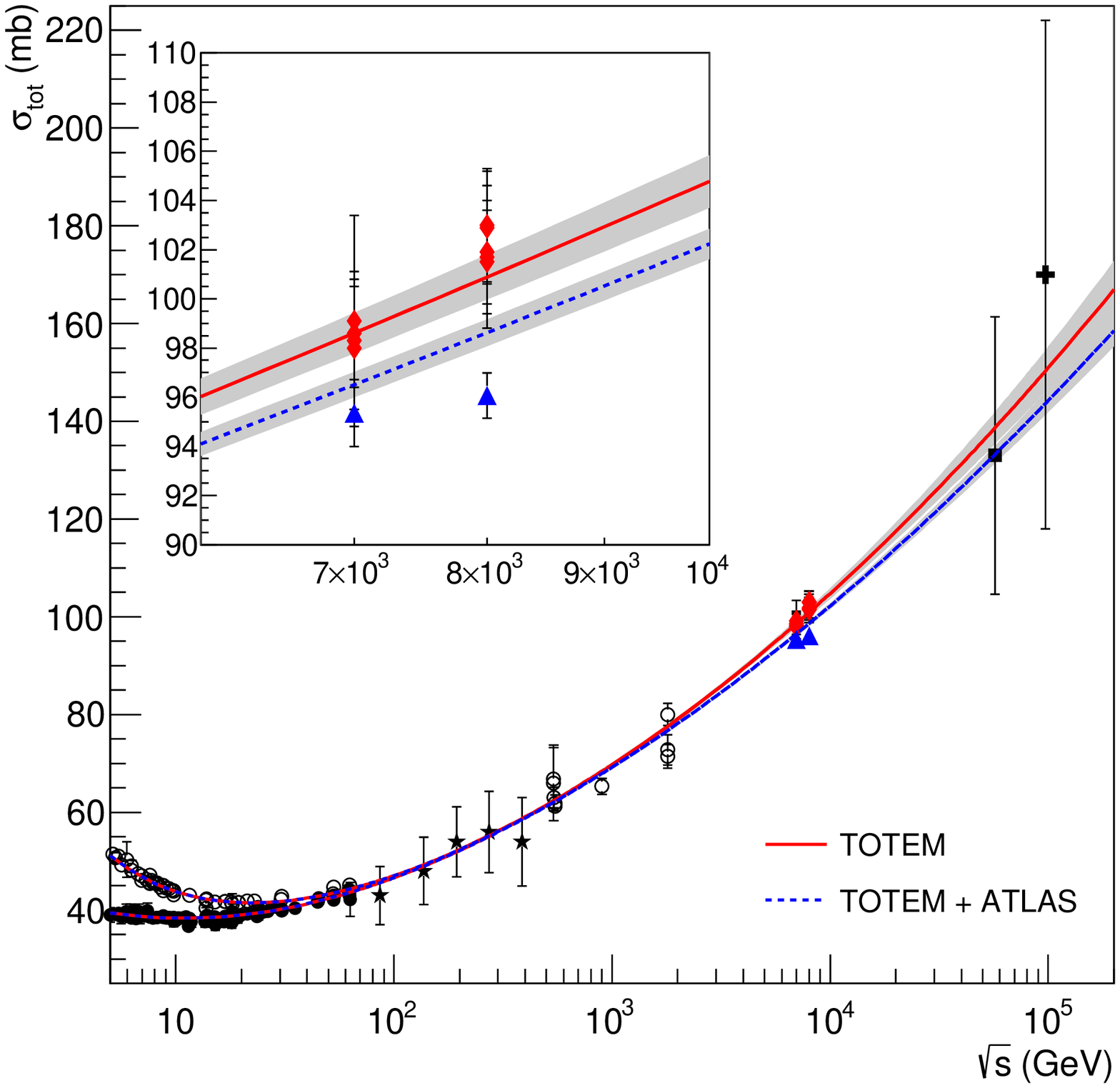}
 \includegraphics[scale=0.4]{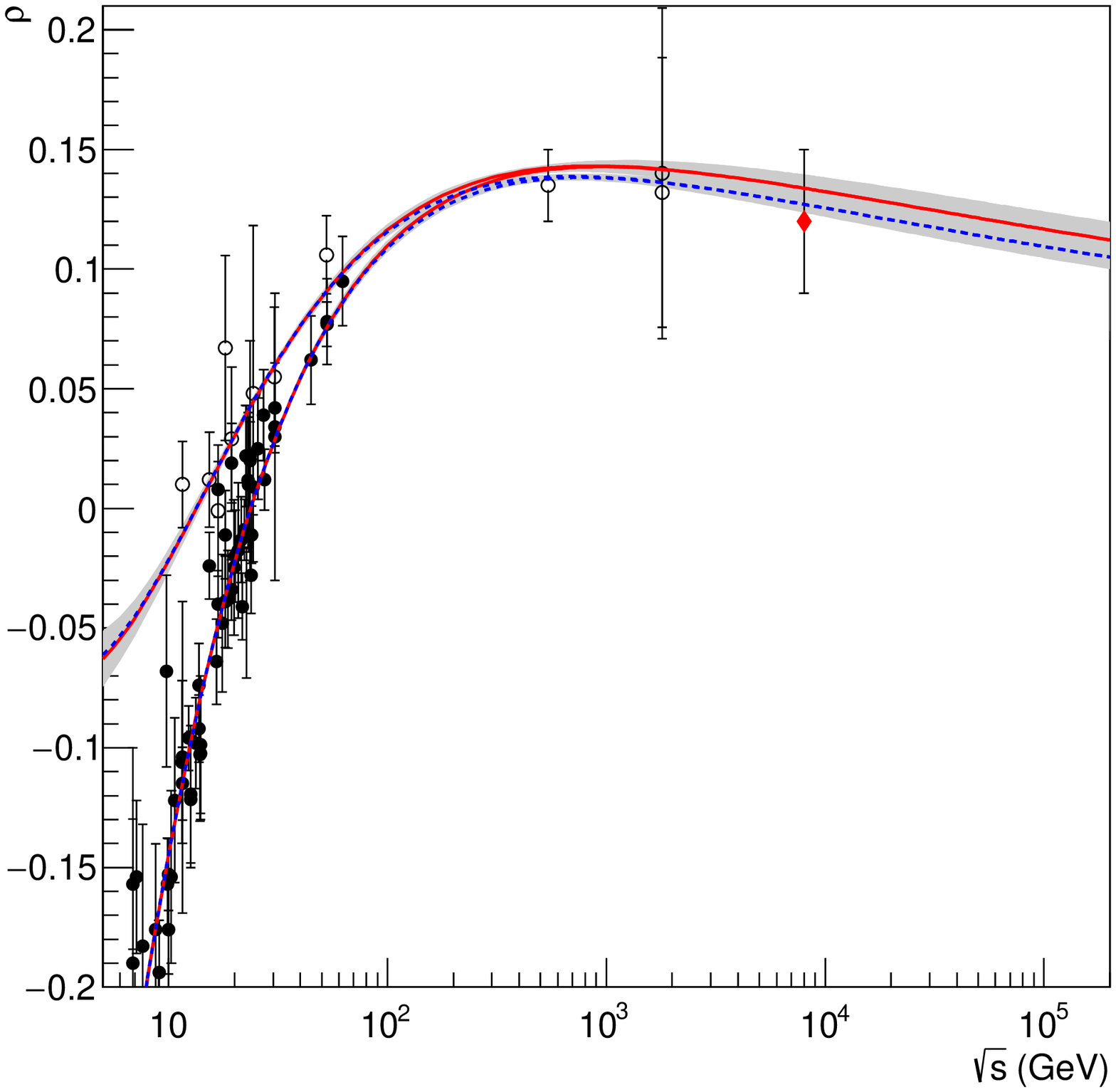}
 \caption{Fit results with the AU-L$\gamma$ model to ensembles T and T+A.
Eqs. (\ref{austlg})-(\ref{auphi}), Table 3.}
\label{f6}
\end{figure}

\newpage

\section{General Discussion and Comments} 
\label{s5}

In this section we develop a critical and detailed discussion on all the models
and data reductions treated in Sects. 3 and 4: the analytic and conceptual differences between
the DDR and AU approaches related to L2 and L$\gamma$ models (Sect. 5.1);
the corresponding fit results with ensembles
T and T+A (Sect. 5.2). We proceed presenting our
partial conclusions (Sect. 5.3) and some further comments
on the log-raised-to-$\gamma$ law (Sect. 5.4).

\subsection{Analytic and Conceptual Differences}

Let us confront the analytic results presented in Sect. 3, namely
the FMS-L2, FMS-L$\gamma$ models, Eqs. (\ref{fmsstlg})-(\ref{fmsrhocoeflg}) and the
AU-L$\gamma$=2, Eqs. (\ref{austg=2})-(\ref{aurhog=2}), AU-L$\gamma$, Eqs. (\ref{austlg})-(\ref{auphi}) models.
Our focus here concerns the analytic and conceptual differences
among them. 

First we note that all models present the same Reggeon contributions
(related to the parameters $a_1$, $b_1$ and $a_2$, $b_2$) and the same critical
Pomeron contribution ($\alpha$). 
Beyond the presence of the effective subtraction constant in the FMS models (not in the
AU cases), a central analytic point in these parameterizations concerns the
log-raised-to-$\gamma$ term and the corresponding connection between \tcs\
and $\rho$.

\subsubsection{DDR Approach}

As discussed in Appendix C, this analytic approach is based on the use of derivative
dispersion relations \textit{without the high-energy approximation}. This important aspect
is taken into account (at least in first order) through the concept of an
\textit{effective subtraction constant} as a free fit parameter (see Appendix C for details).

The general analytic expressions of the FMS-L$\gamma$ model are given by Eqs. (\ref{fmsstlg})-(\ref{fmsrhocoeflg}).
As we showed in Sect. 3.2.2, the FMS-L2 model is a particular case for
$\gamma = 2$. This specific case has the same analytic structure as the
PDG-L2 model and the COMPETE parameterizations $RRPL2$ for \tcs\ and \ro,
except for the presence of the effective subtraction constant, the absence
of a constraint connecting $\beta$ and $M$ (PDG) and the fixed scale $s_0 = 4m_p^2$.

These L2 models (both PDG and FMS) are constructed in accordance with the
Regge-Gribov theory (as shown in Appendix B). In this context,
the parameters $a_1$, $a_2$, $\alpha$ and $\beta$ correspond to the strengths of the
Reggeons and of the Pomerons (critical and triple pole),
being, therefore, constant factors at $t=0$ (independent of the energy).
We shall return to this point in what follows.

\subsubsection{AU Approach}

The AU models (L2, L$\gamma$=2, L$\gamma$) are derived in detail in Appendix D.
Although the AU-L2 model can also be deduced in this context (Sect. D.3), a crucial
point is the fact that for $\gamma = 2$ the
L$\gamma$ model does not correspond to the L2. The essential distinction respect L2 concerns
the Pomeron contributions to \tcs\ and \ro, as we stress and discuss in what follows.

By omitting the arguments $s$ and $t=0$ of the amplitude, in the AU-L2 model
the Pomeron contributions ($P$) to the imaginary and real parts of the
amplitude are given by Eqs. (\ref{austl2})-(\ref{aurhol2}),

\begin{equation}
\frac{\mathrm{Im} A_{L2}^{P}}{s} =
\alpha + \beta \ln^{2}\left(\frac{s}{s_0}\right),
\label{iaul2}
\end{equation}
\begin{equation}
\frac{\mathrm{Re} A_{L2}^{P}}{s} =
\pi \beta \ln\left(\frac{s}{s_0}\right).
\label{raul2}
\end{equation}
With the AU-L$\gamma$ model for $\gamma = 2$ (L$\gamma$=2 model), the contributions
from Eqs. (\ref{austg=2})-(\ref{aurhog=2}) read

\begin{eqnarray}
\frac{\mathrm{Im} A_{L\gamma=2}^{P}}{s} =
\alpha + \beta \cos (\phi)
\ln\left(\frac{s}{s_0}\right)
\left[\ln^2\left(\frac{s}{s_0}\right) + \pi^2\right]^{1/2},
\label{iaug=2}
\end{eqnarray}
\begin{eqnarray}
\frac{\mathrm{Re} A_{L\gamma=2}^{P}}{s} =
\beta \sin (\phi)
\ln\left(\frac{s}{s_0}\right)
\left[\ln^2\left(\frac{s}{s_0}\right) 
+ \pi^2 \right]^{1/2}.
\label{raug=2}
\end{eqnarray}

The essential difference between (\ref{iaul2})-(\ref{raul2}) and (\ref{iaug=2})-(\ref{raug=2}) does not concern 
the additional factor $\pi^2$, but the presence in (\ref{iaug=2})-(\ref{raug=2}) of 
trigonometric functions, which \textit{depend on the energy} through
$\phi = \phi(s)$, as given by Eq. (\ref{auphi}). Although in the asymptotic
limit ($s \rightarrow \infty$),
 \begin{eqnarray}
\phi \rightarrow 0, \qquad \cos(\phi) \rightarrow 1, \qquad \sin(\phi) \rightarrow 0,
\nonumber
\end{eqnarray}
that is not the case in the \textit{finite energy-interval investigated} (5 GeV - 8 TeV), where,
even if limited to the interval $[-1, 1]$, both cosine and sine can, in principle, take on negative,
null and positive values (depending on the ratio $s/s_0$ in Eq. (\ref{auphi})). 

This analytic dependence on the energy seems extremely difficult to be interpreted or
justified in the
Regge-Gribov context. For example, if in (\ref{iaug=2}) we neglect the constant $\pi^2$
so that the term $\ln^2(s/s_0)$ may be associated
with a triple pole (Appendix B), how to interpret 
$\cos(\phi)\ln^2(s/s_0)$ in (\ref{iaug=2})? If, on the other hand,
we associate this cosine with the pre-factor $\beta$, it seems obscure
to relate the Pomeron strength $\beta$ in (\ref{iaul2}) with an energy dependent
strength $\beta \cos{\phi}$ in (\ref{iaug=2}). Similar considerations apply to the
real part of the amplitude (note also the $\ln{s}$ dependence in (\ref{raul2})
contrasting with the $\ln^2{s}$ in (\ref{raug=2})), as well as to the more general
AU-L$\gamma$ model.

We illustrate this effect in Fig. 7, using the AU-L$\gamma$=2
and AU-L$\gamma$ models,
from the fits to ensemble T+A (Table 3). 
We plot three
dimensionless terms associated with the Pomeron component to \tcs.
Note that in this case, the energy scale is fixed at the
physical threshold $s_0 = 4m_p^2$, a relativelly small value which attenuates the
oscilation. That, however, is not the case for larger values of $s_0$,
as can be easily verified.
\begin{figure}[ht]
 \includegraphics[scale=0.4]{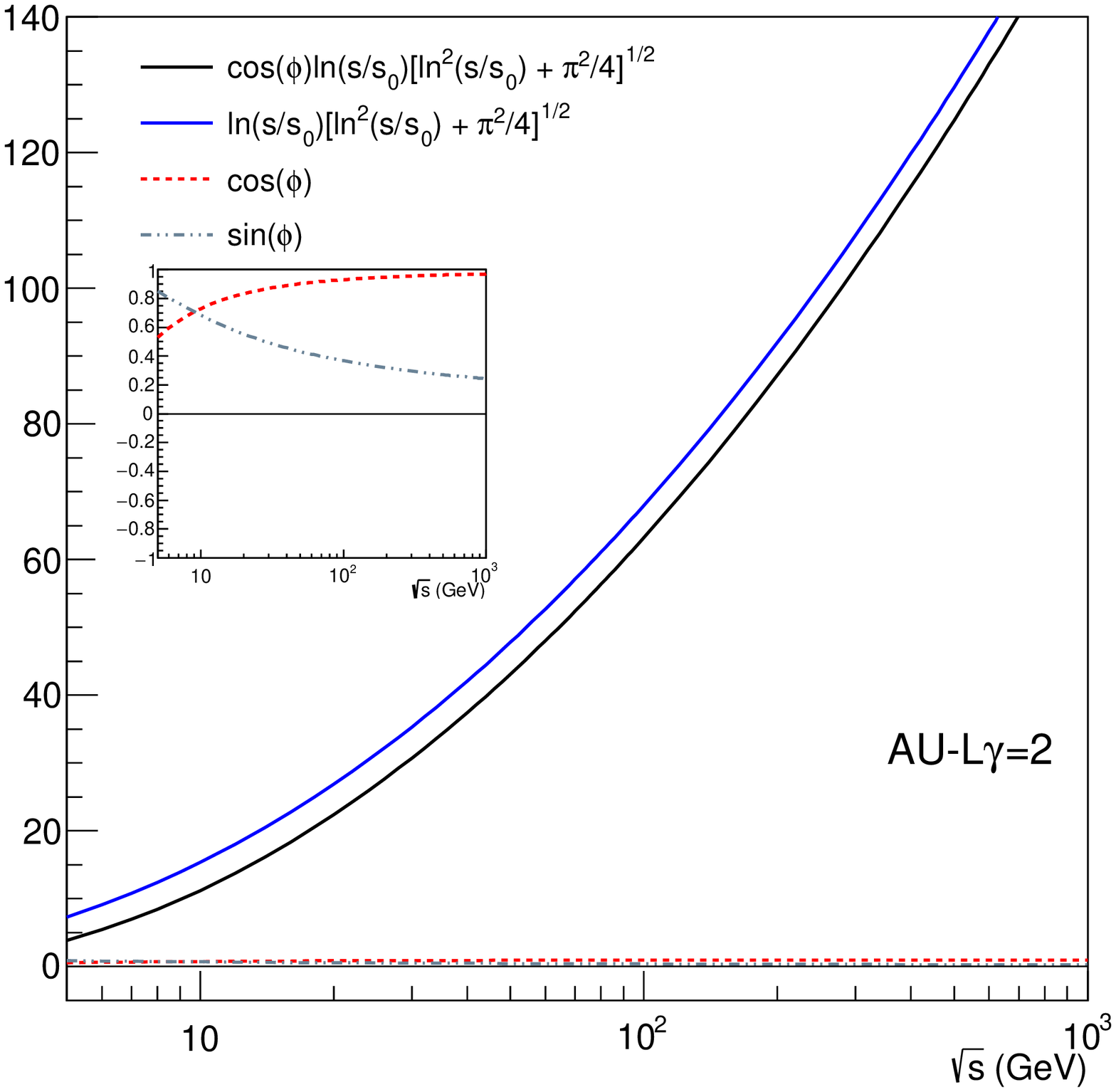}
 \includegraphics[scale=0.4]{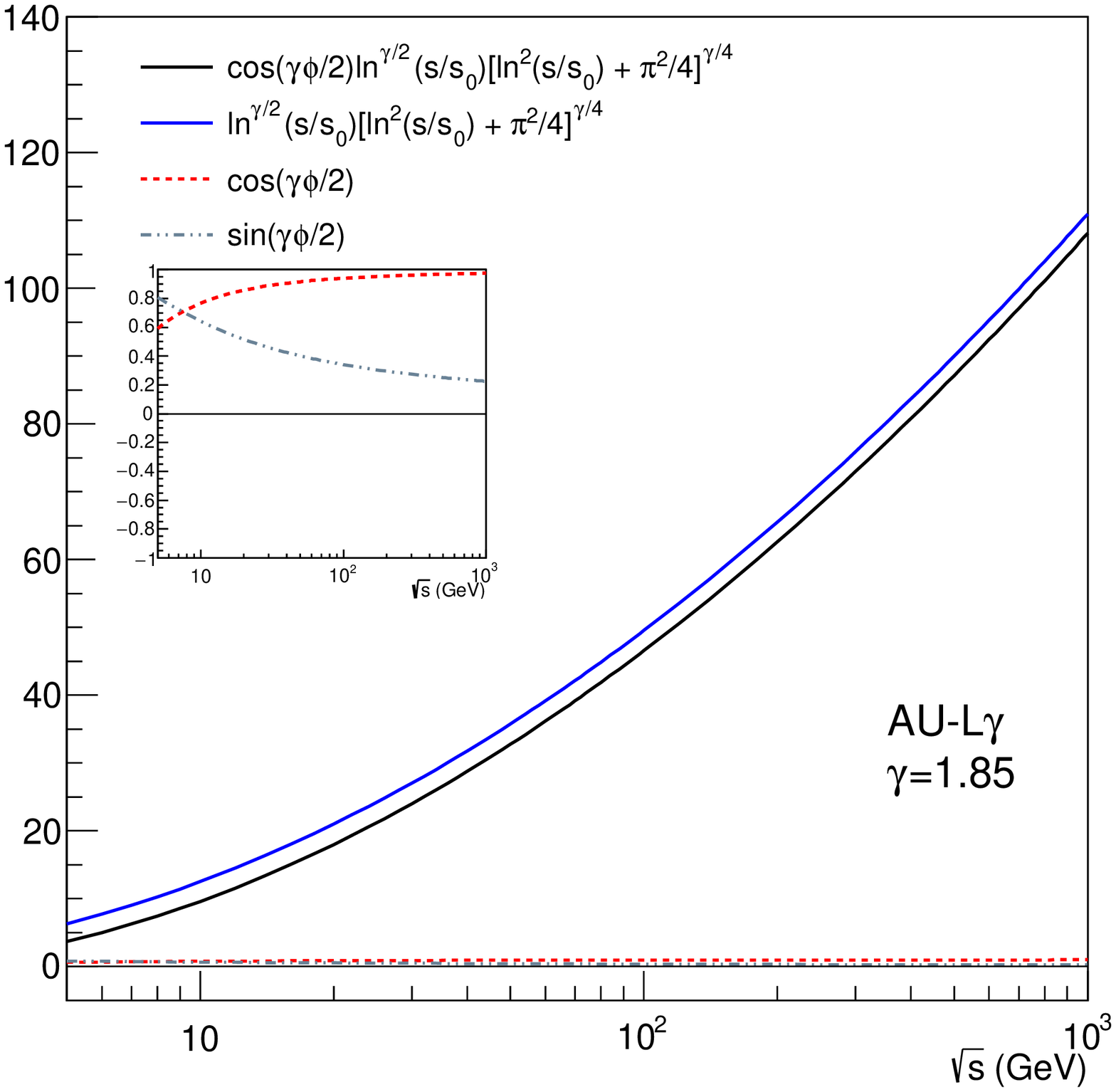}
\caption{Dimensionless contributions to the leading Pomeron component in
AU-L$\gamma$=2 and AU-L$\gamma$ models, from fits to ensemble T+A,
Table 3.}
\label{f7}
\end{figure}

Therefore, and as noted in Appendix C, the AU approach for L$\gamma$ and L$\gamma$=2 models
introduces energy dependent functions in the parametrization for \tcs($s$)
which are not present in the original input, Eq. (\ref{fmsstlg}).

\subsection{Fit Results}

We have developed eight data reductions, using four models
(FMS-L2, FMS-L$\gamma$, AU-L$\gamma$=2 and AU-L$\gamma$) and two
ensembles (T and T+A). The fit results with the FMS models
are presented in Table 2 and those with the AU models
in Table 3 (and quoted figures in the Tables).

Let us discuss all the \textit{fit results}, by confronting 
separately (two-by-two) the following aspects: 
ensembles T and T+A (Sect. 5.2.1), L2 and L$\gamma$ models
with the DDR approach (Sect. 5.2.2), L$\gamma$ models with the
DDR and AU approaches (Sect. 5.2.3). After that we present some
comments on the PDG 2016 and COMPETE results (Sect. 5.2.4).

\subsubsection{Ensembles T and T+A}

The results are displayed in Table 2, Fig. 3 (FMS-L2), Fig. 4 (FMS-L$\gamma$)
and Table 3, Fig. 5 (AU-L$\gamma$=2), Fig. 5 (AU-L$\gamma$).
We have the comments that follows.

\begin{itemize}

\item  
Within all models, the goodness-of-fit is slightly better with ensemble T
than with T+A: $\chi^2/\nu \sim 1.07 - 1.13$ (T) and 
$\chi^2/\nu \sim 1.14 - 1.19$ (T+A), or  
$P(\chi^2) \sim 0.08-0.2$ (T) and
$P(\chi^2) \sim 0.02-0.06$ (T+A).

\item
From the figures, all TOTEM data are quite well
described with ensemble T, but in case of ensemble T+A all curves lie between the data points, barely reaching
the extrema of the uncertainty bars.

\item
Ensemble T indicates a rise of the total cross section faster than
ensemble T+A, as shown by the extrapolated curves and, for example, by
the $\gamma$ values with the FMS-L$\gamma$ model:
$\gamma \sim 2.30 \pm 0.10$ (T) and  $\gamma \sim 2.16 \pm 0.16$ (T+A). 

\item
The ATLAS datum at 8 TeV is not described by any fit result:
all curves within the uncertainties lie above this point, even with ensemble
T+A.

\end{itemize}

\subsubsection{FMS-L2 and FMS-L$\gamma$ Models}

The fit results are displayed in Table 2 and Figs. 3 and 4.

\begin{itemize}
 
\item
Taking into account the distinct characteristics of ensembles T and T+A,
both models present agreement with the experimental data analyzed and
cannot be distinguished in statistical grounds: with ensemble T,
$\chi^2/\nu$ = 1.09 (L2) and 1.07 (L$\gamma$) and with ensemble T+A,
$\chi^2/\nu$ = 1.15 (L2) and 1.14 (L$\gamma$).

\item
With ensemble T, the L$\gamma$ results confirm our previous determination
of the parameter $\gamma$. The slight high value, $\gamma \sim 2.30 \pm 0.10$
(as compared with the previous $2.23 \pm 0.11$ \cite{ms2}) is a consequence
of the lastest TOTEM data at 8 TeV.

%\item
%With ensemble T+A the results with L2 and L$\gamma$ are very
%similar, since $\gamma \sim 2.2 \pm 0.2$.

\item
We note the anti-correlation between the parameters $\beta$ and $\gamma$:
$\beta \sim 0.25$ mb for $\gamma = 2$ and 
$\beta \sim 0.10 $ mb for $\gamma \sim 2.3$ (T). We shall return to this
point in the next Section.

\end{itemize}

\subsubsection{FMS-L$\gamma$ and AU-L$\gamma$ Models}

The fit results are presented in Table 2 and Figure 4 (FMS) and
in Table 3 and Figure 6 (AU).

\begin{itemize}

\item
Once more, taking into account the distinct characteristics of ensembles T and T+A,
both models show agreement with the experimental data analyzed, 
with a goodness-of-fit slightly better in case of FMS: with ensemble T,
$\chi^2/\nu \sim 1.08$, $P(\chi^2) \sim 0.2$ (FMS) and 
$\chi^2/\nu \sim 1.13$, $P(\chi^2) \sim 0.09$ (AU) and with ensemble T+A,
$\chi^2/\nu \sim 1.14$, $P(\chi^2) \sim 0.06$ (FMS) and 
$\chi^2/\nu \sim 1.19$, $P(\chi^2) \sim 0.02$ (AU).

\item
The resulting $\gamma$-values with FMS are greather than with AU:
within ensemble T, $\gamma \sim 2.3 \pm 0.1$ (FMS) and $\sim 2.0 \pm 0.2$ (AU)
and within ensemble T+A, $\gamma \sim 2.2 \pm 0.2$ (FMS) and $\sim 1.9 \pm 0.1$ (AU).

\item
Note that, the $\gamma$ values
obtained with the AU method are in agreement, within the uncertainties, with the result
quoted by the COMPAS Group in \cite{pdg14,pdg16}, namely 1.98 $\pm$ 0.01
(although based on distinct datasets, as we stressed in Sect. 2.1).

\end{itemize}

\subsubsection{PDG 2016 and COMPETE}

We discuss some results related to $pp$ and $\bar{p}p$ scattering,
already published by the PDG (2016)
\cite{pdg16} and by the COMPETE Collaboration (2002) \cite{compete1,compete2}.

\begin{itemize}

\item 
The results from the PDG 2016 with the PDG-L2 model \cite{pdg16}
are displayed in Fig. 2, Eq. (\ref{ivpdg}). As already commented,
the fit comprises several reactions and in the figure we plot
the results for $pp$ and $\bar{p}p$ scattering.
Given the systematic character of the uncertainties in the TOTEM
measurements (discussed in Appendix A.1), we understand that the PDG-L2
fit result does not describe the TOTEM data at 7 TeV
and mainly 8 TeV, since the curve lies below the (systematic) error bar.
This behavior  
is also present in the fits from 2014 \cite{pdg14} with cutoff at 5 GeV;
however, the description is consistent if the cutoff is raised to 7 GeV.

\item
Although the COMPETE prediction from 2002 is in agreement with
the TOTEM data, it should be recalled that the value of the energy scale
is greater that the energy cutoff. As a consequence the triple pole Pomeron contribution
to the total cross section decreases as the energy increases between these
two values, which seems to contradict the supercritical Pomeron
character as a rising contribution (see discussion by Menon and Silva:
Sect. 4.2 and Fig. 7 in \cite{ms1}).

\end{itemize}

\subsection{Partial Conclusions}

On the basis of the discussions in Sects. 5.1 and 5.2, we are led to the 
partial conclusions that follows.

\begin{enumerate}

\item
In what concerns the
\textit{L$\gamma$ models} (and the energy-interval investigated), the DDR approach
is consistent with the Regge-Gribov theory: (1) the derivative
dispersion relations\footnote{Introduced in first order by Gribov and
Migdal (Appendix B).} apply to any function of interest in
amplitude analyses (power or logarithmic laws); (2) the triple pole
contribution is nothing more than a particular case of L$\gamma$ for
$\gamma = 2$; (3) with the FMS models, given a parametrization for \tcs($s$), the determination of
$\rho(s)$ does not involve the high-energy approximation 
(due to the \textit{effective subtraction constant}) and therefore,
most importantly, it is not associated with the asymptotic condition
($s \rightarrow \infty$). Or, in other words, the approach is adequate
for the finite energy interval investigated.

On the other hand, the AU approach leads to analytic results for \textit{both} \tcs($s$) and \ro($s$)
with oscillatory terms, depending on the energy, which do not have justifications
in the Regge-Gribov context. Moreover, the AU-L$\gamma$ does not reproduce the
AU-L2 model for $\gamma$ = 2 and this model, AU-L$\gamma$=2, has also oscillatory
terms.

In conclusion, regarding L$\gamma$, \textit{the FMS models are more consistent in the formal context 
and adequate for the energy interval
investigated than the AU models}.

\item
Taking into account all the experimental data presently available (ensemble T+A),
the discrepancy between the TOTEM and ATLAS data does not allow a high-quality
fit on statistical grounds. In the strictly sense of goodness-of-fit, we could say that the fit
results favor the TOTEM data. Moreover, all fit results fail to describe the ATLAS datum at 8 TeV.

\item
The FMS-L2 and FMS-L$\gamma$ models present agreement with the experimental data
analyzed and cannot be discriminated on statistical grounds.

\item
The fit results indicate that
the TOTEM data and the ATLAS data favor different scenarios for
the asymptotic rise of the total cross section.

\end{enumerate}

Predictions 
of the FMS-L2 and FMS-L$\gamma$ models (ensembles T and T+A) for \tcs$(s)$
and $\rho(s)$
at 13, 14, 57 and 95 TeV
are presented in our recent analysis \cite{fms17a}.

\subsection{Further Comments on the Log-raised-to-$\gamma$ Law}

As introduced by Amaldi et al., the leading contribution to the total cross section
in the form 
$
\beta \ln^{\gamma} ({s}/{s_0})
$
seems to have had a strictly empirical origin, possibly, as a check on how close the rising of the total cross
section, dictated by the experimental data, could be in respect the \textit{analytic} bound by Froissart, Lukaszuk and Martin.
We are not yet sure to have a justification for a real exponent
in the logarithm, directly related to the \textit{standard} Regge-Gribov picture, namely if an specific 
singularity in the $t$-channel could imply
in this contribution to the amplitude of the crossed channel. Perhaps, if not speculative, 
the real exponent
could
represent a kind of effective contribution, similar to an effective exponent in the simple
pole super-critical Pomeron\footnote{Note that even double and triple poles are mathematical
possibilities (Appendix B.3), \cite{land}, sect. 2.3, \cite{edenland}.}.
In spite of this possible limitation, to treat the exponent $\gamma$ as a free fit parameter
leads to some interesting consequences and useful results which are worth noting.

\begin{itemize}

\item
In the general case, data reductions with this term involve three free parameters,
$\beta$, $\gamma$ and $s_0$, which are strongly correlated as demonstrated in the Appendix of Ref. \cite{fms2} and
also discussed in \cite{ms1} (see Sect. 4.2 and Table 6). For $s_0$ fixed, as assumed in this analysis,
$\beta$ and $\gamma$ are anti-correlated.
Typically, from Tables 2 and 3:
\begin{eqnarray}
& &\gamma \sim 2.3\ \mathrm{(FMS)}\ \iff \beta \sim 0.10\ \mathrm{mb}\ \mathrm{(T)}   \nonumber \\
& &\gamma \sim 2.2\ \mathrm{(FMS)}\ \iff \beta \sim 0.15\ \mathrm{mb}\ \mathrm{(T+A)}  \nonumber \\
& &\gamma = 2.0\ \mathrm{(FMS, AU)}\ \iff \beta \sim 0.26\ \mathrm{mb}\ \mathrm{(T)},
\beta \sim 0.25\ \mathrm{mb}\ \mathrm{(T+A)}  \nonumber \\
& &\gamma \sim 1.85\ \mathrm{(AU)}\ \iff \beta \sim 0.39\ \mathrm{mb}\ \mathrm{(T+A)}
\label{gammabeta}
\end{eqnarray}

These different values associated with $\gamma$ and $\beta$ may have some connections
with recent phenomenological and theoretical ideas and results, as
discussed in the next item.

\item
In the phenomenological context, a fractional (real) exponent in the interval
$1 < \gamma < 2$ is predicted in the QCD mini-jet model with soft gluon re-summation
\cite{giulia1,giulia2}. 
As commented in \cite{pdg14, pdg16}, a rise slower than L2 is also
predicted in the Cheng and Wu formalism \cite{chengwu}.

An effective real exponent may also be associated with the
presence of sub-leading contributions, beyond the leading log-squared component.
As commented in our introduction, in the nonperturbative approach to soft high-energy scattering
(related to first principles of QCD), important analytic developments have been recently
obtained, mainly in what concerns the rise of the total hadronic
cross section at the highest energies and its relation with the QCD spectrum.
Under specific assumptions and selected scenarios, Giordano and Meggiolaro \cite{gm1,gm2}
have predicted a leading contribution to $\sigma_{\mathrm{tot}}(s)$, as $s \rightarrow \infty$, 
of the form $B \log^2 s$ and a
sub-leading contribution proportional to\footnote{Recent results
on this sub-leading contribution are presented and discussed by Giordano, Meggiolaro
and Silva in \cite{gms}.} $\log s \log(\log s)$.
The pre-factor $B$ is universal (independent of the colliding hadron)
and can be entirely determined from the QCD spectrum.
Most importantly, considerations on the spectrum of stable particles in \textit{unquenched} (full) QCD 
yields
\begin{eqnarray}
B_{th} \simeq 0.22 \ \mathrm{mb},
\nonumber
\end{eqnarray}
and in the case of \textit{quenched} QCD (Q), the estimated lower value reads
\begin{eqnarray}
B_{th}^{Q} \simeq 0.42 \ \mathrm{mb}.
\nonumber
\end{eqnarray}
These two results indicate that the inclusion of dynamical
fermions (full QCD) has a pronounced influence in the (universal) value of 
$B$, a result in contrast with the
usual phenomenological models which consider that this asymptotic behavior is governed
by the gluonic sector of QCD.

From Tables 2 and 3 (and estimates in Eq. (\ref{gammabeta})), the $\beta$ results from models with $\gamma = 2$ (FMS and AU)
are in agreement with the full QCD 
prediction\footnote{As well as the PDG-L2 result, Eq. (\ref{ivpdg}).},
while the AU-L$\gamma$ result favors the quenched case. The FMS-L$\gamma$, with ensemble T, predicts
the smallest $\beta$ value (0.10 mb) corresponding to a mass
\begin{eqnarray}
M = \sqrt{\frac{\pi}{\beta}} \sim 3.5\ \mathrm{GeV}.
\nonumber
\end{eqnarray}

\item
There is an important and interesting aspect
related to a leading Pomeron component of \tcs\ from the DDR approach,
in the form
\begin{eqnarray}
\sigma^P(s) = \alpha + \beta \ln^{\gamma}\left(\frac{s}{s_0}\right),
\label{plg}
\end{eqnarray}
with $\gamma$ a real fit parameter. That concerns the slope ($B$) and curvature ($C$) of
\tcs\ \textit{in terms of the variable} $\ln{s}$. Indeed, in the particular case of a L2 model
(or in the asymptotic Froissart, Lukaszuk and Martin bound), we have
\begin{eqnarray}
B_{L2}(\ln s) = 2 \beta \ln \left(\frac{s}{s_0}\right)\ \  (\mathrm{linear}), 
\qquad
C_{L2}(\ln s) = 2 \beta \ \ (\mathrm{constant}),
\label{scl2}
\end{eqnarray}
while in the general case,
\begin{eqnarray}
B(\ln s) = \beta \gamma \ln^{\gamma - 1} \left(\frac{s}{s_0}\right),
\qquad
C(\ln s) = \beta\,\gamma (\gamma - 1) \ln^{\gamma - 2} \left(\frac{s}{s_0}\right),
\label{sclg}
\end{eqnarray}
both, therefore, being local relations. This means that,
any deviation from a linear rate of change of \tcs($\ln s$) and from a
constant curvature ($2 \beta$) can be ``detected"\ by a L$\gamma$ model\footnote{In case
of a AU approach, the analytic structure and oscillatory term do not
allow this simple interpretation.} 
as given by Eq. (\ref{plg}). Moreover, Eq. (\ref{sclg}) allows to quantify the differences in the
rate of rise of \tcs\ from the TOTEM and ATLAS data, as shown in \cite{fms17a}. 

\item
Another aspect related to parametrization (\ref{plg}) concerns its dependence with the
energy scale $s_0$ and the assumption, in this work, of fixing the scale
at the physical threshold for scattering states, namely $4m_p^2$. 
Since $\gamma$ is a real (not integer)
exponent and once Eq. (\ref{plg}) represents a cross section, this function is real valued
only in the interval $s \geq s_0$. Therefore, it ``starts"\ at $s=s_0$, where
 \begin{eqnarray}
\sigma^P(s_0) = \alpha
\nonumber
\end{eqnarray}
and from this point on $\sigma^P(s)$ increases as the energy increases,
in accordance with the concept of a super-critical Pomeron.

In case of the same energy scale for the Reggeon components,
as in the FMS-L$\gamma$ model, Eq. (\ref{fmsstlg}), at $s=s_0$ the contributions
to the $pp$ and $\bar{p}p$ total cross section read
\begin{eqnarray}
\sigma_{pp}^R(s_0) = a_1 - a_2
\qquad
\sigma_{\bar{p}p}^R(s_0) = a_1 + a_2
\nonumber
\end{eqnarray}
and from this point on $\sigma_{pp}^R(s)$ and  $\sigma_{\bar{p}p}^R(s)$
decreases as the energy increases.  

It is interesting to investigate these quantities at the threshold points
as well as their evolution with the energy. For example, from our fit
results with the FMS-L$\gamma$ model to ensemble T+A, the numerical results
are (Table 2):
\begin{eqnarray}
\sigma^P(s_0) \sim \mathrm{32.8\ mb},
\qquad
\sigma_{pp}^R(s_0) \sim \mathrm{14.5\ mb},
\qquad
\sigma_{\bar{p}p}^R(s_0) \sim \mathrm{48.7\ mb}.
\nonumber
\end{eqnarray}
Therefore at $s=s_0$, 
$\sigma_{pp}^R(s_0) < \sigma^P(s_0) < \sigma_{\bar{p}p}^R(s_0)$
and above $s_0$, the Reggeon contributions decrease and the Pomeron
increases as $s$ increases. That means the Pomeron component is not
only a leading contribution at higher energies but it is also a
significant component at intermediate and low energies. 
From Tables 2 and 3, this effect is also present in all fit results,
independently of model or ensemble.
This behavior is illustrated in Fig. 8 with the FMS-L$\gamma$ model
and ensemble T+A, where we plot the components
$\sigma_{pp}^R(s)$, $\sigma_{\bar{p}p}^R(s)$ and $\sigma^P(s)$.

\end{itemize}

\begin{figure}[ht]
\begin{center}
 \includegraphics[scale=0.5]{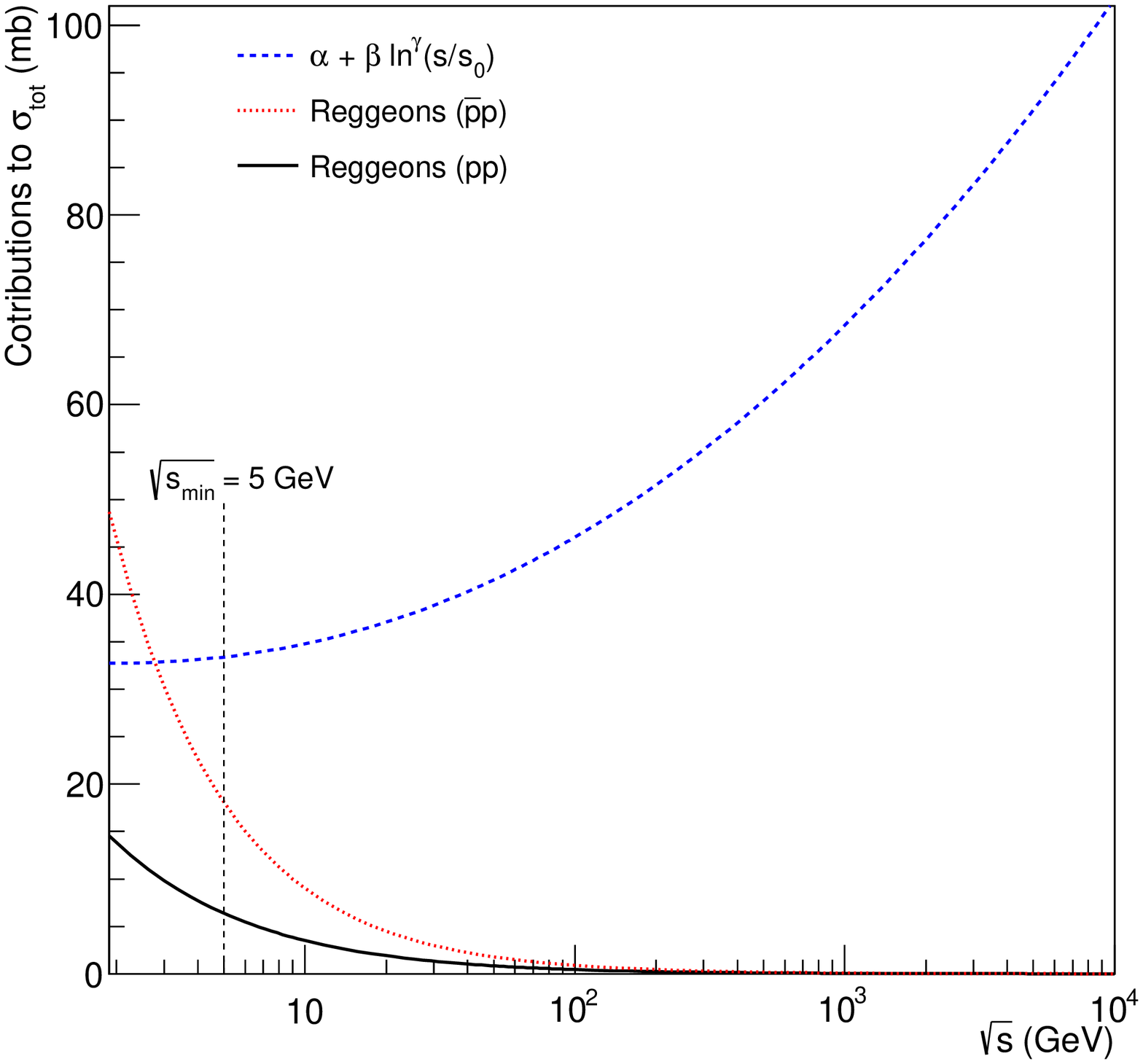}
\caption{Reggeon components, $\sigma_{pp}^R(s)$, $\sigma_{\bar{p}p}^R(s)$ and
Pomeron component,  $\sigma^P(s)$, of \tcs($s$) for $pp$ and $\bar{p}p$
scattering,
above the physical threshold $\sqrt{s_0} = 2m_p \sim $ 1.9 GeV$^2$. Results obtained with
the FMS-L$\gamma$ model
and ensemble T+A (energy cutoff at $\sqrt{s_{\mathrm{min}}}$ = 5 GeV).}
\label{f8}
\end{center}
\end{figure}

\section{Conclusions and Final Remarks}

We have presented a critical discussion on some analytic parameterizations
for the forward elastic amplitude, associated with $pp$ and $\bar{p}p$ scattering
in the energy region 5 GeV - 8 TeV. The focus has been in
the Derivative Dispersion Relations (DDR) and Asymptotic Uniqueness (AU) approaches 
related to the log-raised-to-$\gamma$ (L$\gamma$) leading component of \tcs($s$),
as well as in the particular cases of $\gamma$ = 2.
The models have been introduced in Sect. 3 and the detailed derivations
in Appendices B, C and D.

As regards the L$\gamma$ models, the main
difference between the DDR and AU approaches concerns the way to determine
$\rho(s)$ from analytic inputs for \tcs$(s)$. 

In the DDR approach, here represented by the FMS-L$\gamma$ and FMS-L2 models,
use is made of singly subtracted derivative dispersion relations
with the \textit{effective subtraction constant}, which avoids
the high-energy approximation, at least in first order (Appendix C).
With
this approach, the L2 model is a particular case of the L$\gamma$ model
for $\gamma = 2$ and the analytic results for \tcs $(s)$ and $\rho(s)$
are consistent with the Regge-Gribov formalism (Appendix B).
 
In the AU approach, the L2 model (also derived through DDR and in the
Regge-Gribov context)
does not correspond to the AU-L$\gamma$
model for $\gamma = 2$. That led us to define the AU-L$\gamma$=2 model.
As a consequence of the exponent $\gamma$
as a \textit{real} free parameter, the method introduces energy-dependent factors
in the parameterizations of both \tcs $(s)$ and $\rho(s)$, which are not present
in the Regge-Gribov forms (Sect. 5.1.2). Oscillatory terms appears
in both AU-L$\gamma$ and AU-L$\gamma$=2 models.

Based on the relatively large amount of experimental information from the
TOTEM experiment (9 points), as compared with ATLAS (2 points) and, once the
previous analyses with the FMS models were
based mainly in the TOTEM data, we have considered two ensembles of dataset: T and
T+A (Sect. 2.2). With each ensemble we have developed new data reductions
with four models: FMS-L2, FMS-L$\gamma$, AU-L$\gamma$=2 and AU-L$\gamma$.

In statistical grounds, the FMS-L2 and FMS-L$\gamma$ models present good and equivalent descriptions of the
experimental data analyzed (with both ensembles T and T+A). Besides consistent
with the Regge-Gribov formalism, they predict a rising Pomeron contribution
as the energy increases above the cutoff (in agreement with the super-critical
Pomeron concept). We cannot select a unique solution. However, extrapolations to higher
energies are slightly different (Figures 3 and 4).

In case of ensemble T we have obtained different
values for the $\gamma$ parameter:
$\gamma_{\mathrm{FMS}} \sim 2.3 \pm 0.1$ (Table 2) and
$\gamma_{\mathrm{AU}} \sim 2.0 \pm 0.2$ (Table 3). The former is consistent with
a rise of the total cross section faster than the Froissart-Lukaszuk-Martin bound
and the latter consistent with that bound.
In case of ensemble T+A:
$\gamma_{\mathrm{FMS}} = 2.16 \pm 0.16$ (Table 2) and
$\gamma_{\mathrm{AU}} = 1.85 \pm 0.13$ (Table 3). 
The former suggests a quantitative saturation of the bound and the
latter a rise below that bound. With both ensembles T and T+A, the $\gamma$-values
from the FMS-L$\gamma$ model are greather than those from the AU-L$\gamma$ model.
As already noted in \cite{fms17a}, the
above
$\gamma_{\mathrm{FMS}}$-value with ensemble T+A is in plenty agreement with the
result obtained by Amaldi et al. forty years ago (data up to 62.5 GeV), namely
$\gamma_A = 2.10 \pm 0.10$ \cite{amaldi}.

It is important to recall that the Froissart-Lukaszuk-Martin bound
is an \textit{asymptotic result}, namely intended for the regime $s \rightarrow \infty$.
In what concerns the energy dependence of \tcs\ in the experimental context, this
\textit{asymptotic concept} has been associated with distinct scenarios
along the last 60 years, depending on the available data. Indeed, as recalled in Appendix B.1, 
asymptotia was already identified
with: a decreasing cross section to zero in the beginning of 60s;
a decreasing cross section to a constant value (not zero) by the
end of 60s; a rising cross section along the 70s, followed by different
analytic representations in the form of logarithmic and/or powers laws.
It seems an appeal for several authors to be a contemporary of an
asymptotic regime. Nonetheless, it seems not obvious that
presently, with the LHC data, we should associate the rise of
\tcs\ with the so expected asymptotic regime.

Despite the different topics and subjects treated in this work,
several aspects of our study deserve further investigation.
Among them, the energy scale plays a central role. As commented
in Sect. 4.1, in this first step towards a detailed comparative study
between the DDR and AU methods, we have considered the energy scale
$s_0$ fixed at the physical threshold ($4m_p^2$). Data reductions,
with $s_0$ as a free fit parameter are very important,
for example, due the correlations among the parameters $\beta$, $s_0$
and $\gamma$ in L$\gamma$ models (and $\beta$ and $s_0$ in the
L2 case), as we have already discussed and demonstrated using
the DDR method (Sect. 4.2 and Table 6 in \cite{ms1}).
However, the implementation of data reductions may demand suitable
manipulation of some analytic formulas and additional variants,
which go beyond the objectives of the present analysis, as we shall show in
a subsequent work.
Another topic concerns analysis with the AU
result obtained through the binomial expansion (Appendix D.4.2).
Investigations along this and other lines are in progress.

At last, we stress that
the discrepancies between the total cross sections measurements at the
highest energies reached in accelerators, from both $\bar{p}p$ scattering
(CDF - E710/E811) and $pp$ scattering (TOTEM - ATLAS), as discussed
in Appendix A.2, turns out difficult any amplitude analysis.

We hope that this detailed review and the points raised here, together with
the new data from Run 2 at 13 TeV, can shed light on the subject.

\section*{Acknowledgments}

We are thankful to Professor Vladimir Ezhela (PDG, COMPAS Group, IHEP, Protvino)
for useful correspondence.
Work supported by 
S\~ao Paulo Research Foundation (FAPESP),
Contract 2013/27060-3 (P.V.R.G.S.)

\appendix

\section{Comments on the Experimental Data Presently Available}
\label{saa}

In this Appendix, we discuss some aspects related to the experimental
data presently available on \tcs\ from $pp$ and $\bar{p}p$ scattering
at the highest energies (Table 1). These aspects concern the 
statistical and systematic uncertainties involved 
(Sect. A.1) and some discrepancies among different measurements
(Sect. A.2).

\subsection{Statistical and Systematic Uncertainties}

Since our main focus concerns investigation on the leading contribution
to \tcs($s$), the experimental information at the highest energy region 
plays an important role. For discussions on goodness of fits throughout
the main text, we stress 
here some characteristics of the uncertainties associated
with the measurements of \tcs \
from $pp$ and $\bar{p}p$ presently available at the highest energies.

Uncertainties in the measurements are usually expressed as
statistical and systematic, which are then added in quadrature 
providing the total uncertainty.
Statistical uncertainties are associated with fluctuations and random errors
and can be treated through probability distributions (usually Poisson or Gaussian)
and eventually represented by the corresponding standard deviation. In particular, in data reductions,
the $\chi^2$-test for goodness of fit is based on the assumption of
Gaussian distribution associated with each experimental point. In that case
a measurement
$\sigma_{\mathrm{tot}} \pm \Delta \sigma_{\mathrm{tot}}^{stat}$, the uncertainty $\Delta \sigma_{\mathrm{tot}}^{stat}$
corresponds to one standard deviation, namely the confidence level for
the ``true value" to lie between $\sigma_{\mathrm{tot}} - \Delta \sigma_{\mathrm{tot}}^{stat}$ and
$\sigma_{\mathrm{tot}} + \Delta \sigma_{\mathrm{tot}}^{stat}$ is $\sim 68.3\%$.

On the other hand, systematic uncertainties are equally probable quantities (do not follow a
Gaussian distribution). That means the ``true value" is equally likely to lie in any
place limited by the systematic uncertainty (the ``error" bar). Explicitly,
a measurement represented by 
$\sigma_{\mathrm{tot}} \pm \Delta \sigma_{\mathrm{tot}}^{syst}$
means that the ``true value"  may lie in any place
between $\sigma_{\mathrm{tot}} - \Delta\sigma_{\mathrm{tot}}^{syst}$ and
$\sigma_{\mathrm{tot}} + \Delta\sigma_{\mathrm{tot}}^{syst}$, with equal probability.
Or, in other words, $\Delta\sigma_{\mathrm{tot}}^{syst}$ does not correspond
to one standard deviation.

Therefore, if data reductions are developed with statistical and systematic uncertainties
added in quadrature (as usual), 
care should be taken with strict interpretation of the $\chi^2$-test for goodness of fit
since it is based on the assumption of Gaussian error distribution. Moreover and most
importantly, the concept of agreement between a model prediction or fit result (curve)
and an experimental point is different if the uncertainty is statistical or systematic:
a curve that does not cross a systematic error bar can not be considered in agreement
with the experimental information.
 
In Table 1
we collect the statistical and systematic uncertainties associated with
the measurements presently available on the total cross section from $\bar{p}p$ and $pp$ scattering
at the highest energies: 6 points from $\bar{p}p$
(546 GeV, 900 GeV and 1.8 TeV) and 11 points from $pp$ (7 and 8 TeV). It is also displayed
the corresponding total uncertainty (quadrature) and the relative uncertainty.
We have included as systematic the uncertainties associated with 
error propagated from uncertainties in fit
parameters (TOTEM, 8 TeV \cite{totem6}) and
extrapolations (ATLAS, 8 TeV \cite{atlas8}).

An important point to notice in these results is the character 
eminently (or even strictly)  systematic of the uncertainties
in all the TOTEM measurements at 7 and 8 TeV\footnote{The absence, or comparatively small values, of 
the statistical uncertainties is
a consequence of the large number of experimental information at the LHC.}.
That, however, is not the case with all the
high energy 
$\bar{p}p$ measurements (546 GeV - 1.8 TeV) or the $pp$ measurements by the ATLAS Collaboration at both
7 and 8 TeV.
Therefore, based on the above discussion, it is important to take
account of these  differences when interpreting the data reductions, mainly in what concerns
the concept of agreement or not between a fit result (curve) and a experimental point
with only systematic uncertainty. We treat
this subject in Sect. 5.2, when discussing our fit results.

\subsection{Experimental Data at the Highest Energies}

A crucial point to note, mainly in forward amplitude analyses, is the differences between
the \tcs\ measurements  at just the highest energies in both $\bar{p}p$ and $pp$ scattering.
For $\bar{p}p$ at 1.8 TeV the CDF result contrasts with both the E710 and E811
measurements. Since the uncertainties in the CDF and E710 are statistics (Table 1), the difference
in the central values corresponds to 2.3 standard deviation:
\begin{eqnarray} 
\frac{\sigma_{\mathrm{CDF}} - \sigma_{\mathrm{E710}}}{\Delta \sigma_{\mathrm{E710}}} = 2.3.
\nonumber
\end{eqnarray}

For $pp$ scattering,  the ATLAS results at 7 and 8 TeV
(one point at each energy) contrasts also with those by the TOTEM Collaboration (4 points at 7 TeV and
5 points at 8 TeV). At 8 TeV, comparison of the ATLAS result with the latest TOTEM measurement \cite{totem6}
led to
\begin{eqnarray} 
\frac{\sigma_{\mathrm{TOTEM}} - \sigma_{\mathrm{ATLAS}}}{\Delta \sigma_{\mathrm{TOTEM}}} 
= \frac{103.0 - 96.07}{2.3}
= 3.
\nonumber
\end{eqnarray} 
If, instead, the ATLAS uncertainty is used, this ratio reads 7.5.

In an extreme possibility, it is obvious that combination of the CDF result for $\bar{p}p$ with the TOTEM results for $pp$
indicates a completely different scenario in respect the combination of E710/E811 for $\bar{p}p$
and ATLAS for $pp$. Certainly, these differences turns it difficult any amplitude analisys. 

The implications of the differences between the TOTEM and ATLAS data have been recently
investigated through our FMS amplitude analysis, by using both the L2
and L$\gamma$ leading contributions  \cite{fms17a}. In that analysis, the dataset also comprises all the accelerator data from 
$pp$ and $\bar{p}p$ above 5 GeV and below 7 TeV and three ensembles for data reductions have been
considered by adding either: only the TOTEM results, or only the ATLAS results,
or both sets. Of interest here, one of the conclusions of this study is that the TOTEM data
favor a rise of \tcs($s$) faster than the ATLAS data \cite{fms17a}.

\section{Regge-Gribov Formalism}
\label{sab}

From a strictly mathematical point of view, usual amplitude analyses are characterized
by a few number of functions of the energy $s$; they can be classified into
two classes: power laws ($s^a$, with $ a \ge  0$ or $ a< 0$)
and logarithmic laws ($\ln^{\gamma} s$, with $\gamma = 1$, or $\gamma = 2$ or $\gamma$
a \textit{real} free parameter in data reductions). Except for this last example
(discussed in Sect. 5.4), all the other
functions have solid physical basis on the $S$-matrix formalism and the Regge-Gribov 
theory \cite{pred,land,collins,eden}.

In this formal context,
we present here a review on the derivation of the analytic parameterizations for $\sigma_{\mathrm{tot}}(s)$ and $\rho(s)$, 
mainly related to the RRPL2 models introduced in Sects. 3.2 and 3.3.
After a short review on some historical aspects (Sect. B.1), 
we treat the Reggeons and the supercritical Pomeron associated with 
simple poles (Sect. B.2), followed by the Pomeron corresponding to double and triple poles (Sect.B.3). 
Some critical comments are also presented.
This appendix is useful for discussions throughout the main text,
as well as for comparison with connections between real and imaginary
parts of the amplitude, obtained by means of mathematical methods in Appendices C (dispersion relations) 
and D (asymptotic uniqueness).
Here, for simplicity, we omit in the beginning the energy scale $s_0$ in the
power and logarithmic functions (namely we set $s_0 =$ 1 GeV$^{2}$).

\subsection{Historical Aspects}

Analytic parameterizations for the total cross section 
are expressed as a sum of contributions associated with Reggeons ($R$)
and Pomerons ($P$) 

\begin{equation}
\sigma_{\mathrm{tot}}(s) = \sigma^{R}(s) +  \sigma^{P}(s).
\label{strp}
\end{equation}
For reference and discussions along the text, we shortly recall some 
historical aspects 
related to the construction
of these contributions \cite{pred,land,collins,eden}. Specific formulas
are derived and discussed in the sections that follows (B.2 and B.3).

The Reggeons were the first to be conceived in 1960s to account
for the decreasing of $\sigma_{\mathrm{tot}}(s)$ in the region
$\sqrt{s} \lesssim$ 20  GeV and also for the differences
between particle-particle and antiparticle-particle scattering.
The Reggeon exchanges are associated with the highest interpolated mesonic trajectories provided by
spectroscopic data ($t$-channel), relating Re $J$ with
the masses $M^2$ (the Chew-Frautschi plot).
The trajectories are approximately linear, defining an effective slope and intercept.
The functional form for the total cross-section associated with a simple pole
consists of a power law of $s$ (Sect. B.2) with \textit{negative} exponent (given by the intercept) 
and around 0.5, predicting, therefore, that $\sigma_{\mathrm{tot}}(s)$ decreases asymptotically to
zero. 

The possibility of an asymptotic non zero limit led to the introduction of the
first Pomeron concept, associated with a simple pole with intercept 1 (after named \textit{critical pomeron}). 
This situation, by the end
of 1960s,  is instructively discussed by Barger, Olsson and Reeder in the 1968 paper
``Asymptotic Projections of Scattering Models" \cite{bor}.

In the 1970s, experimental results by the IHEP-CERN Collaboration at Serpukhov
and at the CERN-ISR indicated the
rise of $\sigma_{\mathrm{tot}}$ above $\sim$ 20 GeV.
In the absence of a mesonic trajectory able to account for this rise,
an \textit{ad hoc} trajectory has been introduced, with intercept slightly greater than one,
namely an \textit{increasing contribution with the energy}. This $C=+1$ trajectory
(to account for an asymptotic equality between $pp$ and $\bar{p}p$ scattering)
has been associated with a simple pole in the amplitude, corresponding, therefore,
to a power law in $s$, with \textit{positive} exponent slightly greater than 0 (named
\textit{supercritical Pomeron} or \textit{soft Pomeron}).

In the mathematical context, this rise could also be explained by logarithmic
dependences in the energy. In this case, in principle, the Froissart-Lukaszuk-Martin bound Eq. (\ref{fmb}) 
limits the functional forms to
either linear or quadratic or both,  which in the Regge-Gribov context can be associated
with higher order poles in the amplitude, double pole for linear and triple pole for
quadratic dependences (Sect. B.3). 

As commented in our introduction, 
all these possibilities have been carefully analyzed and discussed by the COMPETE
Collaboration in their seminal papers of 2002 \cite{compete1,compete2}.
The dataset consisted of all the
forward data available at that time
on $pp$, $\bar{p}p$, mesons-$p$, $\gamma p$ and $\gamma \gamma$
scattering, in the interval 5 GeV - 1.8 TeV. 
The analysis involved tests on different
energy cutoffs and on the universality of the leading high-energy contribution.

The selected parametrization for \tcs($s$) includes
two Regge terms (simple poles at $J \approx 0.5$), associated with $C = +1$ and $C = -1$ mesonic trajectories, namely
($a$, $f$) and ($\rho$, $\omega$), respectively. The selected Pomeron
contribution consists of a constant Pomenranchuck contribution (a critical Pomeron
with pole at $J=1$)
plus a triple pole at $J=1$ contribution in the form of log-squared of $s$
(a triple pole Pomeron).
For comparison with other models, this parametrization was denoted 
RRPL2 by the COMPETE Collaboration, representing the two Reggeons (R, R), the constant
Pomenranchuck contribution (P) and the triple pole Pomeron (L2).
The connections with $\rho(s)$ was obtained through crossing and dispersion relations
\cite{compete1,compete2}.

This selected COMPETE parametrization 
became a standard reference in successive editions of the RPP by the PDG. 
Moreover, a remarkable result concerns the fact that ten years later \cite{compete1}, the COMPETE 
\textit{extrapolation}
for the $pp$ total cross-section at 7 TeV showed to be in complete agreement with the
first high-precision measurement by the TOTEM Collaboration, as well as with the subsequent
measurements by this group \cite{totem5,totem6}.

\subsection{Simple Poles - Power Laws}

\subsubsection{Reggeon Contributions}

\vspace{0.5cm}

In the $s$-channel and in the large energy limit, the contribution to the Lorentz invariant scattering amplitude from 
a simple pole in the complex $J$-plane ($t$-channel),
\begin{eqnarray}
\frac{1}{J - \alpha},
\label{absp}
\end{eqnarray}
namely a single reggeon exchange, can be expressed by (\cite{pred}, Eq. 5.76,
\cite{collins}, Eq. 2.8.10)
\begin{eqnarray}
A(s,t) =  \beta(t) \xi(t) s^{\alpha(t)},
\nonumber
\end{eqnarray}
where $\beta(t)$ is the residue, $\alpha(t)$ the exchanged trajectory and
$\xi(t)$ the signature factor, given by
\begin{eqnarray}
\xi(t) = - \frac{1 + \zeta e^{-i \pi\alpha(t)}}{\sin \pi \alpha(t)}, 
\nonumber
\end{eqnarray}
with $\zeta = + 1$ or  $\zeta = - 1$ for analytic continuations 
through even or odd integer values of the angular momenta, respectively. 

Let us consider non-degenerated exchanged trajectories, with even and odd angular
momenta from the $a_2/f_2$ and $\omega/\rho$ mesonic families,
associated with the corresponding even ($+$) and odd ($-$) signatures.
Taking the signatures separately,
\begin{eqnarray}
\xi_{+}(t) = - \cot \frac{\pi}{2} \alpha(t) + i,
\qquad
\xi_{-}(t) = - \tan \frac{\pi}{2} \alpha(t) - i,
\nonumber
\end{eqnarray}
the \textit{forward} ($t = 0$) complex even and odd \textit{amplitudes} read
\begin{eqnarray} 
A_{+}(s,0) &=& \beta_{+}(0) \left[i - \cot \frac{\pi}{2} \alpha_{+}(0)\right] s^{\alpha_{+}(0)},
\label{aba+}  
\end{eqnarray}
\begin{eqnarray}
A_{-}(s,0) &=& -\, \beta_{-}(0)\left[i + \tan \frac{\pi}{2} \alpha_{-}(0)\right] s^{\alpha_{-}(0)}.
\label{aba-}
\end{eqnarray}

These are crossing symmetric and antisymmetric functions of the energy,
\begin{eqnarray}
A_{+}(- s) = + A^{*}_{+} (s), 
\qquad
A_{-}(- s) = - A^{*}_{-} (s), 
\label{abcross}
\end{eqnarray}
where $*$ denotes complex conjugation and are connected with the normalized physical amplitudes
for the $pp$ and $\bar{p}p$ scattering by \cite{eden},
\begin{eqnarray}
\frac{A_{pp}}{s} = \frac{A_{+} + A_{-}}{s},
\qquad
\frac{A_{\bar{p}p}}{s} = \frac{A_{+} - A_{-}}{s}.
\label{abph}
\end{eqnarray}

Denoting the strengths and intercepts by
\begin{eqnarray}
\beta_{+}(0) = a_1, \qquad \alpha_{+}(0) - 1 = - b_1, \qquad 
\beta_{-}(0) = a_2, \qquad \alpha_{+}(0) - 1 = - b_2,
\nonumber
\end{eqnarray}
the \textit{full complex forward amplitudes} can be expressed by
\begin{eqnarray}
\frac{A(s, t=0)}{s} &=& \left[ - a_1 \tan \left(\frac{\pi b_1}{2}\right) s^{- b_1} + 
\tau a_2 \cot \left(\frac{\pi b_2}{2} \right) s^{-b_2} \right] \nonumber \\ 
&+& i \left[a_1 s^{-b_1} + \tau a_2 s^{-b_2} \right],
\nonumber
\end{eqnarray}
where $\tau = -1$ for $pp$ and $\tau = +1$ for $\bar{p}p$.

Therefore, the contributions from the Reggeons ($R$) to the total cross section
and the $\rho$ parameter, Eqs. (\ref{ot}) and (\ref{rho}),
are given by
\begin{eqnarray}
\sigma^{R}(s) &=& a_1 s^{-b_1} + \tau a_2 s^{-b_2}, 
\label{absigr}
\end{eqnarray}
\begin{eqnarray}
\rho^{R}(s) &=& \frac{1}{\sigma_{R}(s)} \left\{- a_1 \tan \left(\frac{\pi b_1}{2}\right) s^{- b_1} + 
\tau a_2 \cot \left(\frac{\pi b_2}{2} \right) s^{-b_2}\right\}
\label{abrhor}
\end{eqnarray}

Since from the Chew-Frautschi plot
for the $a_2/f_2$ and $\omega/\rho$ trajectories, $b_1 > 0$ and $b_2 >0$ (approximately $1/2$)
these cross sections are decreasing functions of the energy. From the optical theorem and by
inverting Eqs. (\ref{abph}),
\begin{eqnarray}
\mathrm{Im} A_+(s) = a_1s^{-b_1} > 0
\quad
\mathrm{and}
\quad
\mathrm{Im} A_-(s) = - a_2s^{-b_2} < 0.
\label{absigns}
\end{eqnarray}
With the associated $\tau$ values, $\sigma_{\bar{p}p} > \sigma_{pp}$ in the energy region
where these cross sections are not equal (in agreement with the experimental data). We shall 
return to this point in Appendix D.

\subsubsection{Simple Pole Pomeron Contribution}

In this context, a \textit{rising cross section} 
is obtained by considering a positive intercept slightly greater than $1$.
Once dominating at the highest energies, where 
$\sigma_{\bar{p}p} - \sigma_{pp} \rightarrow 0$, it is associated
with a symmetric amplitude.
Denoting the Pomeron ($P$) intercept $\alpha_P = 1 + \epsilon$
(with $\epsilon$ slightly greater than $0$) and $\delta$ the strength, 
we obtain for a simple pole ($S$) Pomeron:
\begin{eqnarray}
\sigma_{S}^{P}(s) &=& \delta s^{\epsilon}, \ \epsilon > 0, \nonumber \\ 
\rho_{S}^{P}(s) &=& \frac{1}{\sigma_{P}(s)} \left\{\delta \tan \left(\frac{\pi \epsilon}{2}\right) s^{\epsilon} \right\}.
\nonumber
\end{eqnarray}

\subsubsection{Analytic Result}

Therefore, including the energy scale $s_0$, with non degenerated Regge trajectories and the supercritical 
Pomeron associated with a simple pole in the amplitude, the
Regge-Gribov formalism provides:
\begin{eqnarray}
\! \! \! \! \! \! \! \! \! \! \! \! \! \! \! \! \! \! \! \! \! \! \! \! \! \! \! \! \! \! \! \!
\sigma_{\mathrm{tot}}(s) = a_1 \left[\frac{s}{s_0}\right]^{-b_1} + \tau a_2 \left[\frac{s}{s_0}\right]^{-b_2} +
\delta \left[\frac{s}{s_0}\right]^{\epsilon}, 
\label{absigrp}
\end{eqnarray}
\begin{eqnarray}
\! \! \! \! \! \! \! \! \! \! \! \! \! \! \! \! \! \! \! \! \! \! \! \! \! \! \! \! \! \! \! \!
\rho(s) = \frac{1}{\sigma_{\mathrm{tot}}(s)} \left\{- a_1 \tan \left(\frac{\pi b_1}{2}\right) \left[\frac{s}{s_0}\right]^{- b_1} + 
\tau a_2 \cot \left(\frac{\pi b_2}{2} \right) \left[\frac{s}{s_0}\right]^{-b_2}  + 
\delta \tan \left(\frac{\pi \epsilon}{2}\right) \left[\frac{s}{s_0}\right]^{\epsilon}    \right\},
\label{abrhorp}
\end{eqnarray}
with $\tau = -1$ for $pp$ and $\tau = +1$ for $\bar{p}p$.

Although not reaching the highest rank in the COMPETE analysis \cite{compete1},
these parameterizations for $\sigma_{\mathrm{tot}}(s)$ and $\rho(s)$ present consistent descriptions of
the LHC data, as recently discussed by Donnachie and Landshoff \cite{dl} and also Menon and Silva
\cite{ms2}.

A fundamental aspect of this simple pole Pomeron contribution
is the fact that the power-law with positive exponent ($\epsilon \sim 0.08 - 0.09$, for example, \cite{dl,ms2})
implies in a strictly rise of the associated cross section with the energy.
This behavior is directly related to
the standard or original concept of the super-critical (or soft) Pomeron,
namely a \textit{rising cross section}.
As discussed in the main text (Sect. 5.2.4), some amplitude analyses do
not reproduce this behavior in the whole range of energy investigated.

\subsection{Double and Triple Poles - Logarithmic Laws}

In the Regge context, the power law $s^{\alpha}$ is associated with a simple pole of the
amplitude in the complex $J$-plane, Eq. (\ref{absp}).
As discussed in \cite{land} (Sect. 2.3), higher order
poles can be generated by derivatives of the simple pole,
\begin{eqnarray}
\frac{d^n}{d \alpha^n}\left[\frac{1}{J - \alpha}\right] = 
\frac{n!}{[J - \alpha]^{n+1}},
\quad
n = 1, 2,...,
\nonumber
\end{eqnarray}
with $N = n +1$ the order of the pole. Translating this derivative
to the power-law,
\begin{eqnarray}
\frac{d^n}{d \alpha^n} s^{\alpha} = s^{\alpha} \ln^n s,
\quad
n = 1, 2,...
\nonumber
\end{eqnarray}

Therefore, associated with a pole of order $N$ ($t$-channel), the contribution to the amplitude
in the $s$-channel
is $s^{\alpha} \ln^{N-1}(s)$. In the case of the Pomeron ($P$), with a pole at $J = \alpha = 1$, 
the contribution to the
total cross section is 
 \begin{eqnarray}
\sigma^{P}(s) = \frac{\mathrm{Im} A(s)}{s} \propto \ln^{N-1} s.
\nonumber
\end{eqnarray}
Based on the Froissart-Lukaszuk-Martin bound the possible leading contributions 
are $\ln s$ (double pole) or $\ln^2 s$ (triple pole).

In these cases, and yet in the context of Regge-Gribov, the real part of the amplitude can
be evaluated through a representation of the Watson-Sommerfeld integral, introduced by
Gribov and Migdal in the end of 1960s \cite{grib1, grib2}. The result, in the forward direction, 
can be put in the form \cite{grib2} (Eq. 44.b in there)
\begin{eqnarray}
\frac{\mathrm{Re}A(s,0)}{s}  = 
\frac{\pi}{2} \frac{d}{d\ln s} \left[ \frac{\mathrm{ImA(s,0)}}{s} \right].
\label{abddrfo}
\end{eqnarray}

As shown in Appendix C, this formula corresponds to the first order series-expansion
of a derivative dispersion relation for even amplitudes. For the double pole ($D$) and triple pole
($T$) we obtain, respectively
\begin{eqnarray}
\sigma^P_{D} = \beta \ln s
\qquad
\rightarrow
\qquad
\rho^P_{D} = \frac{\pi/2}{\ln s},
\label{abdoublep}
\end{eqnarray}
\begin{eqnarray}
\sigma^P_{T} = \beta \ln^2 s
\qquad
\rightarrow
\qquad
\rho^P_{T} = \frac{\pi}{\ln s}.
\label{abtriplep}
\end{eqnarray}
In the case of a simple pole (power law), it is necessary to take into account the full series (Appendix C).

We note  that the logarithmic laws demand an energy scale $s_0$ for consistency,
implying in a null value at $s = s_0$. More importantly, if the energy
increases in the region $s < s_0$,
the double pole contribution increases through \textit{negative values} and the
triple pole contribution \textit{decreases} through positive values up to $s_0$.
Obviously, both a negative contribution and a decreasing contribution,
as the energy increases, are not consistent with the standard 
super-critical (soft) Pomeron concept.

\subsection{L2 Models} 

By adding the Reggeon contributions, Eqs. (\ref{absigr})-(\ref{abrhor}), the triple pole Pomeron contribution, Eq. (\ref{abtriplep}),
a critical Pomeron (constant contribution),
and introducing the corresponding energy scale, we obtain the analytic expressions for
$\sigma_{\mathrm{tot}}(s)$ and $\rho(s)$
in the FMS-L2 model, Eqs. (\ref{fmsstl2})-(\ref{fmsrhol2}) (except for the effective subtraction constant $K_{eff}$,
discussed in Appendix C), the PDG-L2 model, Sect. 3.2.2 (except for the constraints) and
the AU-L2 model, Eqs. (\ref{austl2})-(\ref{aurhol2}).

\section{Dispersion Relations and the Effective Subtraction Constant}
\label{sac}

Dispersion relations, in integral form (IDR) or derivative form (DDR), constitute an important analytic tool 
in the investigation of $\sigma_{\mathrm{tot}}(s)$ and $\rho(s)$: they connect the real and imaginary
parts of the forward crossing even ($+$) and odd ($-$) amplitudes, providing the hadronic
amplitudes for $pp$ and $\bar{p}p$ elastic scattering. In this appendix we review the determination of 
$\rho(s)$ from analytic parameterizations for \tcs $(s)$ by using dispersion relations.
The main focus are: (1) to discuss the important role of the subtraction constant in data reductions, 
as a practical way to take into account the finite lower limit (the physical threshold) in integral relations. 
That avoids the high-energy approximation and led us to
introduce the concept of an \textit{effective subtraction constant}, which applies as well
to derivative forms; (2) to review the DDR results for $\rho(s)$ in the FMS models with emphasis
on the log-raised-to-$\gamma$ law. The replacement of IDR by DDR is also discussed.

\subsection{Integral Dispersion Relations and the High-Energy Approximation}

The parameterizations usually employed for the total cross section in forward amplitude
analyses demand singly-subtracted integral dispersion relations, with the subtraction present
in the even part of the amplitude. In terms of the c.m. energy squared, the even ($+$) and
odd ($s$) amplitudes can be expressed by \cite{idr,idr2}
 
\begin{eqnarray} 
\frac{\textrm{Re}\,A_{+}(s)}{s}  =
\frac{K}{s} + \frac{2s}{\pi}\,P\int_{s_{th}}^{\infty} 
ds' \left[\frac{1}{s'^2-s^2}\right]  
\frac{\textrm{Im}\,A_{+}(s')}{s'} , 
\label{acidre}
\end{eqnarray}

\begin{eqnarray} 
\frac{\textrm{Re}\,A_{-}(s)}{s} =
  \frac{2}{\pi}\,P\int_{s_{th}}^{\infty} 
ds' \left[\frac{s'}{s'^2-s^2}\right]  
\frac{\textrm{Im}\,A_{-}(s')}{s'},
\label{acidro}
\end{eqnarray}
where $P$ denotes principal Cauchy value, $K$ is the \textit{subtraction constant} and $s_{th}$
denotes the physical threshold for scattering states. In the case of $pp$ and $\bar{p}p$,
from $s = 2 m_p(E + m_p)$, for $E_{th} = m_p$:
\begin{eqnarray}
s_{th} = 4 m_p^2 \approx 3.521 \mathrm{GeV}^2.
\nonumber
\end{eqnarray}

In the application of IDR and also in the replacement of IDR by DDR in amplitude
analyses, several authors
consider the \textit{high-energy approximation}, which consists in taking the
limit 
\begin{eqnarray}
s_{th} \rightarrow 0
\nonumber
\end{eqnarray}
in the above integrals. Although usual, this approximation
is not well justified, for several reasons, as discussed in what follows.

\begin{description}

\item{1.} 
Experimental data indicate that below $\sqrt{s} \sim$ 2 GeV the total cross section is characterized by narrow peaks, 
caused by the formation of resonances. As the energy increases, reaching the scattering region
($s_{th} = 4 m_p^2$), \tcs($s$) decreases monotonically up to $\sim$ 20 GeV and then starts to rise.
As it is well known, the region of the smooth decrease is expected to be described by the Reggeon
exchanges. Therefore, the region $s < 4m_p^2$ corresponds to an unphysical region for scattering
states. 

\item{2.} The usual dataset for amplitude analyses starts at $\sqrt{s_{\mathrm{min}}}$ = 5 GeV
(the energy cutoff), which is not far above the threshold $\sqrt{s_{th}} \sim $ 2 GeV.
Or, in other words, for fits with $\sqrt{s_{\mathrm{min}}}$ = 5 GeV (as those by COMPETE
and PDG) it seems unreasonable to consider $\sqrt{s_{th}} \sim $ 2 GeV as zero.

\item{3.} In what concerns our parameterizations (FMS models, Sect. 3.2) and in all the
data reductions here developed (Sect. 4.1), the energy scale is assumed at $s_0 = s_{th} = 4 m_p^2$
and therefore this scale cannot be considered null. 

\item{4.} Most importantly, as we shall show in what follows, in data reductions the \textit{leading effect} 
of a finite (not zero)
lower limit can be analytically absorbed by the subtraction constant. As a practical consequence
the use of the subtraction constant as a free fit parameter takes account, at least approximately,
of the finite lower limit. On the other hand, as we shall also show, if the subtraction
constant is assumed zero (or not considered) the high-energy approximation is present.

\end{description}

In order to treat in detail an specific analytic example, let us consider the parametrization
for the total cross section based on non-degenerated Regge trajectories and the simple pole case
to represent the pomeron contribution (we derive this model in Appendix B, in the
context of the Regge-Gribov formalism). Here, we use as analytic
inputs the expressions for the total cross section, Eq. (\ref{absigrp}). To avoid confusion
between the $s_{th}$ and the energy scale, we set here $s_0 = 1$ GeV$^2$:
\begin{eqnarray}
\sigma_{\mathrm{tot}}(s) = a_1 s^{-b_1} + \tau a_2 s^{-b_2} + \delta s^{\epsilon},
\label{acstex}
\end{eqnarray}
where $\tau = -1$ for $pp$ and $\tau = +1$ for $\bar{p}p$. The point is to determine the real parts
(and eventually $\rho(s)$) by means of the crossing relations, Eq. (\ref{abph}) and the IDR (\ref{acidre}) and (\ref{acidro}),
with $s_{th}$ fixed and not zero.

From (\ref{acstex}), with the crossing relations (\ref{abph}) and the optical theorem Eq. (\ref{ot}), we obtain
\begin{eqnarray}
\frac{ \mathrm{Im} A_{+}}{s} = a_1 s^{-b_1} + \delta s^{\epsilon},
\qquad
\frac{ \mathrm{Im} A_{-}}{s} = - a_2 s^{-b_2}.
\label{aceoa}
\end{eqnarray}
Substituting in (\ref{acidre}) and (\ref{acidro}) and separating the integrals in the form
\begin{eqnarray}
\int_{s_{th}}^{\infty} ....\ ds' = \int_{0}^{\infty} ....\ ds' - \int_{0}^{s_{th}} ....\ ds',
\label{acistep}
\end{eqnarray}
the first integral in the RHS led to trigonometric functions and the second one can be
put in series expansion \cite{grads,abram}. We obtain

\begin{eqnarray}
\frac{\mathrm{Re} A_{+}(s)}{s} = \frac{K}{s} - a_1 \tan \left(\frac{\pi b_1}{2}\right) s^{- b_1} +
\delta \tan \left(\frac{\pi \epsilon}{2}\right) s^{\epsilon} + \Delta^{+},
\nonumber
\end{eqnarray}
where
\begin{eqnarray}
\Delta^{+} = \frac{2}{\pi} \sum_{j=0}^{\infty} \frac{a_1 s_{th}^{-b_1}}{2j + 1 - b_1}\left[\frac{s_{th}}{s}\right]^{2j + 1}
+
\frac{2}{\pi} \sum_{j=0}^{\infty} \frac{\delta s_{th}^{\epsilon}}{2j + 1 + \epsilon}\left[\frac{s_{th}}{s}\right]^{2j + 1}
\nonumber
\end{eqnarray}
and 
\begin{eqnarray}
\frac{\mathrm{Re} A_{-}(s)}{s} =  - a_2 \cot \left(\frac{\pi b_2}{2}\right) s^{- b_2}  - \Delta^{-},
\nonumber
\end{eqnarray}
where
\begin{eqnarray}
\Delta^{-} = \frac{2}{\pi s} \sum_{j=0}^{\infty} 
\frac{a_2 s_{th}^{1 -b_2}}{2j + 2 - b_2}\left[\frac{s_{th}}{s}\right]^{2j + 1}.
\nonumber
\end{eqnarray}
Therefore, denoting
\begin{eqnarray}
\Delta \equiv \Delta^{+} + \tau \Delta^{-},
\nonumber
\end{eqnarray}
the real parts of the $pp$ and $\bar{p}p$ amplitudes can be expressed by
\begin{eqnarray}
\! \! \! \! \! \! \! \!
\frac{\mathrm{Re} A(s)}{s} = \frac{K}{s} - a_1 \tan \left(\frac{\pi b_1}{2}\right) s^{- b_1} + \tau
a_2 \cot \left(\frac{\pi b_2}{2}\right) s^{- b_2} + 
\delta \tan \left(\frac{\pi \epsilon}{2}\right) s^{\epsilon} + \Delta,
\label{acidrr}
\end{eqnarray}
where
\begin{eqnarray}
\Delta = 
\frac{2}{\pi} \sum_{j=0}^{\infty} \left[\frac{a_1 s_{th}^{-b_1}}{2j + 1 - b_1} +
\frac{\delta s_{th}^{\epsilon}}{2j + 1 + \epsilon} + \tau \frac{1}{s} \frac{a_2 s_{th}^{1 -b_2}}{2j + 2 - b_2}\right]
\left[\frac{s_{th}}{s}\right]^{2j + 1}, 
\nonumber
\end{eqnarray}
with $\tau = -1$ for $pp$ and $\tau = +1$ for $\bar{p}p$.

We see that if $s_{th} \rightarrow 0$ then $\Delta \rightarrow 0$ and we obtain
the Regge-Gribov result for $\rho$, Eq. (\ref{abrhorp}), except for the presence here of the subtraction constant in the form $K/s$
and that is the point we are interested in here. By expanding the delta term we obtain
\begin{eqnarray}
\Delta = 
\frac{2}{\pi} \left\{ \left[\frac{a_1 s_{th}^{1 -b_1}}{1 - b_1} + \frac{\delta s_{th}^{1 + \epsilon}}{1 + \epsilon}\right]\frac{1}{s} +
\tau \left[\frac{a_2 s_{th}^{2 -b_2}}{2 - b_2}\right]\frac{1}{s^2} + 
\left[\frac{a_1 s_{th}^{3 -b_1}}{3 - b_1} + \frac{\delta s_{th}^{3 + \epsilon}}{3 + \epsilon}\right]\frac{1}{s^3} +
\tau \left[\frac{a_2 s_{th}^{4 -b_2}}{4 - b_2}\right]\frac{1}{s^4} + ...
\right\}. 
\nonumber
\end{eqnarray}

Now, denoting the coefficient of the leading contribution by
\begin{eqnarray}
\frac{2}{\pi} \left[\frac{a_1 s_{th}^{1 -b_1}}{1 - b_1} + \frac{\delta s_{th}^{1 + \epsilon}}{1 + \epsilon}\right]
\equiv f(s_{th}, a_1, b_1, \delta, \epsilon),
\nonumber
\end{eqnarray}
we have 
\begin{eqnarray}
\Delta = \frac{f(s_{th}, a_1, b_1, \delta, \epsilon)}{s} + \mathcal{O} 
(1/s^2).
\nonumber
\end{eqnarray}

Therefore, in Eq. (\ref{acidrr}), this leading term can be put together with the subtraction constant, defining 
an \textit{effective subtraction constant}:
\begin{eqnarray}
\frac{K + f(s_{th}, a_1, b_1, \delta, \epsilon)}{s} \equiv \frac{K_{eff}}{s},
\label{ackeff}
\end{eqnarray}
which is the same for $pp$ and $\bar{p}p$ scattering.

With this concept and definition, we can re-express the IDR in the form (note the lower limits):
\begin{eqnarray} 
\frac{\textrm{Re}\,A_{+}(s)}{s}  =
\frac{K_{eff}}{s} + \frac{2s}{\pi}\,P\int_{0}^{\infty} 
ds' \left[\frac{1}{s'^2-s^2}\right]  
\frac{\textrm{Im}\,A_{+}(s')}{s'} , 
\label{acidrekeff}
\end{eqnarray}
\begin{eqnarray} 
\frac{\textrm{Re}\,A_{-}(s)}{s} =
  \frac{2}{\pi}\,P\int_{0}^{\infty} 
ds' \left[\frac{s'}{s'^2-s^2}\right]  
\frac{\textrm{Im}\,A_{-}(s')}{s'}.
\label{acidrokeff}
\end{eqnarray}

This result deserves some comments and explanations as follows.

\begin{description}

\item{1.} If $K_{eff} = 0$, the IDR correspond to the high-energy approximation,
namely $s_{th} \rightarrow 0$. It is important to stress this point: if in the data reductions
with IDR (or DDR as we shall see), the subtraction constant is omitted (which means to be assumed zero) then
the high-energy approximation is implicit and therefore the unphysical region
from 0 to $s_{th}$.

\item{2.} Once used as a free fit parameter in data reductions the $K_{eff}$ has a
clear and important physical meaning as a first order contribution related to
the finite value of the lower limit. 
As a consequence \textit{it improves the applicability of the formalism in the regions of lower 
and intermediate energies}.

\item{3.} As we have shown in the example with the simple poles,  Eq. (\ref{ackeff}), $K_{eff}$ is connected with
the other free parameters present in the analytic input for the total cross section.
Moreover, this connection involves not only the Reggeon parameters ($a_1, b_1$) but also the Pomeron
parameters ($\delta, \epsilon$). As a consequence,
in data reductions it is expected that the subtraction constant as a free fit parameter
is correlated with all the other parameters, including those associated
with any form of the leading Pomeron contribution (high-energy region). In fact, we have already demonstrated
this effect in our previous analyses \cite{fms2} (see Table A1) and \cite{ms1} (see Table 6). 

\item{4.}
As we shall discuss in what follows, the same interpretation of $K_{eff}$
is present in the replacements of IDR by DDR.

\end{description}

We recall that amplitude analyses with fixed lower limit in IDR
were discussed in some detail by Bertini, Giffon, Jenkovszky and Paccanoni, already in 1996
\cite{bertini}, with the results expressed in terms of hyper-geometric functions.
A simplified case with the results expressed either through hyper-geometric functions
or their series expansions were also discussed by \'Avila and Menon in 2004 \cite{am04}.

\subsection{Derivative Dispersion Relations with the Effective Subtraction Constant}

\subsubsection{Basic Concepts and Results}

The IDR (\ref{acidre}) and (\ref{acidro}) have a non-local character: in order to obtain the
real part of the amplitude, the imaginary part must be known for all values of the energy.
Moreover, depending on the input, the integration can demand numerical techniques,
as was the case in the analyses by Amaldi et al., UA4/2 Collaboration and Bueno-Velasco, with
the leading contribution in the form $\ln^{\gamma}(s/s_{th})$ and real $\gamma$ 
\cite{amaldi,ua42,bv}.

On the other hand, for functions of interest in amplitude analyses (as the aforementioned one),
the IDR can be replaced by DDR, which beyond the nearly
local-character, provide analytic results for all these functions since the
variable involved is just $\ln s$. 

Essentially, in this replacement, with a change of variable $s = s_{th}\, e^{\xi}$, the integrands in 
(\ref{acidre}) and (\ref{acidro}) are expanded 
in power series which are then integrated 
by parts and the primitives are evaluated at the upper and lower integration limits. Detailed
treatment can be found in several works, but for our purposes it is sufficient to display and
discuss some specific authors and results (complete references to other contributions on
the subject can be found in the quoted papers that follows).

The well known result by Bronzan, Kane, and Sukhatme is based on the \textit{high-energy approximation},
namely $s_{th} \rightarrow 0$ in Eqs. (\ref{acidre}) and (\ref{acidro}) and without reference to the subtraction constant
\cite{bks}. In that work, the authors consider also an additional parameter, which, in fact,
is not necessary (this and other aspects related with the DDR have been critically discussed by
\'Avila and Menon in \cite{am04}). The results can be expressed as differentiation with respect to
the logarithm of the energy in the arguments of trigonometric operators. For crossing even and odd
amplitudes these singly subtracted DDR can be expressed by

\begin{eqnarray} 
\frac{\textrm{Re}\,A_{+}(s)}{s} =
\tan\left[\frac{\pi}{2}\frac{d}{d\ln s} \right] 
\frac{\textrm{Im}\,A_{+}(s)}{s}, 
\label{acddrbkse}
\end{eqnarray} 

\begin{eqnarray} 
\frac{\textrm{Re}\,A_{-}(s)}{s} =
\tan\left[\frac{\pi}{2}\left( 1 + \frac{d}{d\ln s}\right) \right]
\frac{\textrm{Im}\,A_{-}(s)}{s}.
\label{acddrbkso}
\end{eqnarray} 

In practice, with analytic inputs for $\textrm{Im}\,A_{+}(s)$, $\textrm{Im}\,A_{-}(s)$ the operators are expanded and the derivatives
performed term by term providing the corresponding real parts by summing
the series. It is easy to show that
the Reggeons and simple pole Pomeron parameterizations for the total cross section discussed in Sect. B.2 led
to exactly the same results for $\rho(s)$ as obtained in Regge-Gribov context (Appendix B). These relations were used by 
the COMPETE Collaboration in their amplitude analyses,
including the cases of double and triple poles ($\ln(s)$ and $\ln^2(s)$ contributions, respectively) \cite{compete1}
and also by the COMPAS Group \cite{pdg12}. 

Up to our knowledge, the first results for the DDR taking into account not only the finite
lower limit (namely without the high-energy approximation) but also the effect of
the primitive at both upper and lower limits were obtained by \'Avila and Menon in 2005 \cite{am06,am07a}.
The correction term can be expressed as a double infinite series or as a single series using
sum rules and the incomplete Gamma function, as subsequently demonstrated and
discussed by Ferreira and Sesma \cite{fs1,fs2}. However, for our purposes, the main point concerns the fact
that for simples poles, this correction term can be also expressed as inverse powers of $s$ so that the
leading contribution can be absorbed by the subtraction constant \cite{teseregina}. As discussed in the last section
this corresponds to the introduction of the \textit{effective subtraction constant} as a free
fit parameter in data reductions.
The complete practical equivalence in data reductions
between the IDR without the high energy approximation and the DDR with the subtraction constant as
a free fit parameter is demonstrated in \cite{am04,am07b,amwc} and in more detail in \cite{teseregina}.
The replacement of IDR by DDR has been also discussed by Cudell, Martynov and Selyugin
\cite{cms1, cms2} and more recently (2017) by Ferreira, Kohara and Sesma \cite{fks1,fks2}.

\subsubsection{DDR Approach}

Let us now focus on the DDR used in the FMS analyses. 
As already comment, for a leading contribution in the form $\ln^{\gamma}(s/s_{th})$
the IDR can not provide an analytic result, but that is not the case for the DDR.
Beyond the concept of the effective subtraction constant
(free fit parameter in amplitude analyses), we consider the operator expansion
introduced by Kang and Nicolescu \cite{kn} in 1975 and discussed in \cite{am04,am07a}.
The even and odd relations are given by

\begin{eqnarray} 
\! \! \! \! \! \! \! \! \! \! \! \! \! \! \! \!
\frac{\textrm{Re}\ A_{+}(s)}{s} = \frac{K_{eff}}{s} +
\left[ \frac{\pi}{2} \frac{d}{d\ln s} + 
\frac{1}{3} \left(\frac{\pi}{2}\frac{d}{d \ln s}\right)^3 +
\frac{2}{15} \left(\frac{\pi}{2}\frac{d}{d \ln s}\right)^5 + \dots
 \right] \frac{\textrm{Im}\ A_{+}(s)}{s},
\label{acddre}
\end{eqnarray}

\begin{eqnarray}  
\! \! \! \! \! \! \! \! \! \! \! \! \! \! \! \!
\frac{\textrm{Re}\ A_{-}(s)}{s}  &=& 
- \int \left\{ \frac{d}{d\ln s} \left[\cot \left( \frac{\pi}{2} 
\frac{d}{d\ln s} \right)\right]\frac{\textrm{Im}\ F_{-}(s)}{s} \right\} d\ln s \nonumber \\
&=&
- \frac{2}{\pi}\int \left\{ \left[ 1 - \frac{1}{3} \left(\frac{\pi}{2}\frac{d}{d \ln
s}\right)^2   \right. \right. \nonumber \\
&-& \left. \left. \frac{1}{45} \left(\frac{\pi}{2}\frac{d}{d \ln s}\right)^4
 - \dots \right] \frac{\textrm{Im}\ A_{-}(s)}{s} \right\} \, d \ln s.
\label{acddro}
\end{eqnarray}

It is easy to show that all Reggeons and simple pole Pomeron inputs for \tcs($s$)
led to the same previous results for $\rho(s)$ in Appendix B (Regge-Gribov
formalism). Moreover, for a leading \textit{even} input
in the form (equal for $pp$ and $\bar{p}p$ scattering),
\begin{eqnarray} 
\sigma^P(s) = \alpha + \beta \ln^{\gamma}(s/s_{0}),
\label{acsigp}
\end{eqnarray} 
from Eq. (\ref{acddre}), the corresponding ratio can be expressed as
\begin{eqnarray} 
\rho^P(s) = \frac{1}{\sigma_P(s)}
\left[
\mathcal{A}\,\ln^{\gamma - 1} \left(\frac{s}{s_0}\right) +
\mathcal{B}\,\ln^{\gamma - 3} \left(\frac{s}{s_0}\right) +
\mathcal{C}\,\ln^{\gamma - 5} \left(\frac{s}{s_0}\right) + \dots
\right]
\label{acrhop}
\end{eqnarray} 
where
\begin{eqnarray} 
\mathcal{A} = \frac{\pi}{2} \, \beta\, \gamma,  
\quad 
\mathcal{B} = \frac{1}{3} \left[\frac{\pi}{2}\right]^3 \, \beta\, \gamma\, [\gamma - 1][ \gamma - 2], 
 \nonumber \\
\mathcal{C} = \frac{2}{15} \left[\frac{\pi}{2}\right]^5 \, \beta\, \gamma\, [\gamma - 1][ \gamma - 2]
[\gamma - 3][ \gamma - 4], \dots
\label{acparrho}
\end{eqnarray} 

We see that for a double pole ($D$), $\gamma$ = 1, $\mathcal{A} = \pi\,\beta/2$, $\mathcal{B} = \mathcal{C} = \dots =0$
and
\begin{eqnarray} 
\rho^P_{D}(s) = \frac{\pi/2}{\ln(s/s_{0})}
\label{acrhodp}
\end{eqnarray}
and for a triple pole ($T$), $\gamma$ = 2, $\mathcal{A} = \pi\,\beta$, $\mathcal{B} = \mathcal{C} = \dots =0$
and
\begin{eqnarray} 
\rho^P_{T}(s) = \frac{\pi}{\ln(s/s_{0})}, 
\label{acrhotp}
\end{eqnarray}
as obtained in the Regge-Gribov formalism, Eqs. (\ref{abdoublep}) and (\ref{abtriplep}), through first order DDR.

In all our analyses, the data reductions with $\gamma$ as a real free fit parameter have shown that its
value does not exceed $\sim$ 2.5 \cite{fms2,ms1}. Therefore, the above third-order
expansion is sufficient to ensure the convergence of the fit, justifying, therefore, Eqs. (\ref{fmsrholg})-(\ref{fmsrhocoeflg})
in Sect. 3.2.1.

It is important to stress two advantages of this approach in amplitude analyses,
as follows. 

\begin{description}

\item{1.}
It provides analytic results in all cases of interest,  which are adequate for data reductions
and allow standard statistical determination of the uncertainties in all free fit
parameters involved (and consequently, analytic propagation of the uncertainties
to the physical quantities). 

\item{2.} 
With the subtraction constant as an additional free fit parameter, related to its \textit{effective role},
the approach is not constrained by the high-energy approximation: its applicability covers, in principle, all the
energies above the physical threshold, without reference to the unphysical region.

\end{description}

 Further discussions on the role and practical applicability of the subtraction
constant as a free fit parameter can be found in \cite{fms2}, Sect. 2.3.2,
\cite{ms2}, Sect. A.2 and \cite{am04}, Sect. 4.4.

\section{Asymptotic Uniqueness and the Phragm\'en-Lindel\"off Theorems}
\label{sad}

Beyond dispersion relations, asymptotic uniqueness associated with 
Phragm\'en-Lindel\"off theorems, constitute another analytic way
for the determination of the real part of the forward amplitude.
The formal ideas and basic theorems related to these asymptotic relations are
pedagogically presented in Sect. 7.1 of the Eden book \cite{eden},
in which we based this review on the main ideas and results.
With a different approach, the subject is also treated in \cite{block} (Sect. 10.3)
and \cite{bc} (Sect. IV.D).

\subsection{Basic Concepts}

Asymptotic Uniqueness is based on the concepts of crossing symmetry
and analyticity, associated with the \textit{forward} scattering amplitude in
the complex-$s$ plane. We re-write here some formulas already presented,
omitting in the argument $t=0$. Denoting 
by an asterisk the operation
of complex conjugation, the symmetric ($+$) and antisymmetric ($-$) scattering
amplitudes (under crossing) are defined by:

\begin{eqnarray}
A_{+}(-s) = A_{+}(se^{i \pi}) = A_{+}^{*}(s),
\qquad
A_{-}(-s) = A_{-}(se^{i \pi}) = - A_{-}^{*}(s),
\label{adevenodd}
\end{eqnarray}
from which the hadronic amplitudes are constructed,
\begin{eqnarray}
A_{pp}(s) = A_{+}(s) + A_{-}(s),
\qquad
A_{\bar{p}p}(s) = A_{+}(s) - A_{-}(s),
\label{adhadevenodd}
\end{eqnarray}
and the physical observables 
$\sigma_{\mathrm{tot}}(s)$
and
$\rho(s)$ are determined through Eqs. (\ref{ot}) and (\ref{rho}).
In the polar form, $A(s) = |A(s)| e^{i \theta}$, the phase of the amplitude
is given by
\begin{eqnarray}
\theta = \tan^{-1} \left\{ \frac{\mathrm{Im} A(s)}{\mathrm{Re} A(s)}\right\}.
\label{adphase}
\end{eqnarray}

The asymptotic uniqueness constitutes a way to determine
the phase of the amplitudes, once given an real input related to its imaginary
part. As we shall shown, these asymptotic results provide the crossing even
and odd amplitudes to within a $\pm$ factor, the sign being determined by 
physical conditions involved \cite{eden}. Specifically, from Eqs. (\ref{adhadevenodd}) and the optical
theorem Eq. (\ref{ot}),

\begin{eqnarray}
\frac{\mathrm{Im} A_+(s)}{s} = \frac{1}{2} 
\left\{\sigma_{pp} + \sigma_{\bar{p}p} \right\},
\qquad
\frac{\mathrm{Im} A_-(s)}{s} = \frac{1}{2} 
\left\{\sigma_{pp} - \sigma_{\bar{p}p} \right\}.
\nonumber
\end{eqnarray}
Since $\sigma_{pp}, \sigma_{\bar{p}p} > 0$, we have always $\mathrm{Im} A_+(s) > 0$.
For the Reggeons, once associated with the region where 
$\sigma_{\bar{p}p} > \sigma_{pp}$ we have $\mathrm{Im} A_-(s) < 0$.
For the Pomerons, dominating the region where $\sigma_{\bar{p}p} = \sigma_{pp}$,
$\mathrm{Im} A_-(s) = 0$. These results have been already obtained in
Appendix B, through the Regge-Gribov formalism.

Following \cite{eden}, we recall a fundamental corollary which, not only explain
the concept of the asymptotic uniqueness but also provide the essential
role for the determination of the phase of the amplitude\footnote{In the statement, 
$z = x + i0$ denotes the limit from the upper half plane.} \cite{eden}:

\vspace{0.3cm}

\noindent
\textbf{Corollary}

\begin{quote}
``If $f(z)$ is bounded by a polynomial, and $f(z)$ tends to the limits $L_1$ and $L_2$
along the rays $z = x + i0$ as $x \rightarrow + \infty$ and $- \infty$, then we must
have $L_1 = L_2$."
\end{quote}

As in Appendices B and C, we first discuss in Sect. D.2 the power laws (simple poles), followed 
in the next two Subsections
by the logarithmic laws, with focus on the log-squared (D.3) and log-raised-to-$\gamma$ laws (D.4).

\subsection{Power Law (Simple Poles)}

Let \tcs($s$) = $\beta s^{\alpha - 1}$ so that
\begin{eqnarray}
\mathrm{Im} A = \beta s^{\alpha}
\label{adaimpower}
\end{eqnarray}
and consider the complex function (corresponding to $f(z)$),
\begin{eqnarray}
\frac{A(s)}{s^{\alpha}}.
\nonumber
\end{eqnarray}

\noindent
- Asymptotic Behavior

%\vspace{0.3cm}

From the corollary, if for $s \rightarrow \infty$,
\begin{eqnarray}
\frac{A(s)}{s^{\alpha}} \rightarrow L_1 \equiv M e^{i \theta},
\nonumber
\end{eqnarray}
and
\begin{eqnarray}
\frac{A(se^{i\pi})}{[se^{i \pi}]^{\alpha}} \rightarrow L_2,
\nonumber
\end{eqnarray}
then $L_2 = L_1$ and therefore,
\begin{eqnarray}
A(s) = Ms^{\alpha} e^{i \theta}
\quad
\mathrm{and}
\quad
A(se^{i\pi}) = Ms^{\alpha} e^{i (\theta + \pi \alpha)}.
\nonumber
\end{eqnarray}

\noindent
- Crossing symmetry

%\vspace{0.3cm}

For a \textit{symmetric} ($+$) amplitude, from Eq. (\ref{adevenodd}) and the above result,
\begin{eqnarray}
Ms^{\alpha_+} e^{i (\theta_+ + \pi \alpha_+)} = + Ms^{\alpha_+} e^{- i \theta_+},
\nonumber
\end{eqnarray}
so that, for $n = 0, \pm 1, \pm 2,...$, the phase is given, \textit{explicitly}, by
\begin{eqnarray}
\theta_{+} = n\pi - \frac{\pi \alpha_+}{2}.
\nonumber
\end{eqnarray}

From the optical theorem Eq. (\ref{ot}) and Eqs. (\ref{adphase}) and (\ref{adaimpower}), denoting 
$\mathrm{Im} A_{+}(s) =  \beta_{+} s^{\alpha_{+}}$, $\beta_+ > 0$, 
the real part of the amplitude reads:

\begin{eqnarray}
\mathrm{Re} A_{+}(s) = \frac{\mathrm{Im} A_+}{\tan \theta_+} = 
\frac{\beta_{+} s^{\alpha_{+}}}{- \tan (\pi \alpha_+/2)}
= - \beta_{+} \cot (\pi \alpha_+/2) s^{\alpha_{+}},
\nonumber
\end{eqnarray}
leading to the \textit{complex symmetric amplitude}:
\begin{eqnarray}
 A_{+}(s) = \beta_{+} \left[ i - \cot \frac{\pi \alpha}{2} \right] s^{\alpha_{+}}.
\label{adevena}
\end{eqnarray}

In the same way, from (\ref{adevenodd}), for an \textit{antisymmetric} ($-$) amplitude,
\begin{eqnarray}
Ms^{\alpha_-} e^{i (\theta_- + \pi \alpha_-)} = - Ms^{\alpha_-} e^{- i \theta_-}.
\nonumber
\end{eqnarray}
With $-1 = e^{i\pi}$ and for $n = 0, \pm 1, \pm 2,...$, the phase is now
given, \textit{explicitly}, by
\begin{eqnarray}
\theta_{-} = n\pi + \frac{\pi}{2} (1 - \alpha_{-}).
\nonumber
\end{eqnarray}

As commented in Sect. B.2.1 (see Eq. (\ref{absigns})), in the odd case, from Eq. (\ref{adaimpower}) 
we consider $\mathrm{Im} A_- = - \beta_- s^{-\alpha_-}$,
with $\beta_- > 0$. From Eq. (\ref{adphase}), we obtain
\begin{eqnarray}
\mathrm{Re} A_{-}(s) = \frac{\mathrm{Im} A_-}{\tan \theta_-} = - \frac{\beta_{-} s^{\alpha_{-}}}{\tan (\pi/2 - \pi \alpha/2)}
= - \beta_{-} \tan (\pi \alpha/2) s^{\alpha_{-}},
\nonumber
\end{eqnarray}
and the \textit{complex antisymmetric amplitude}:
\begin{eqnarray}
 A_{-}(s) = - \beta_{-} \left[ i + \tan \frac{\pi \alpha_-}{2} \right] s^{\alpha_{-}}.
\label{adodda}
\end{eqnarray}

We note that the asymptotic results Eqs. (\ref{adevena}) and (\ref{adodda}), are exactly the same as those obtained in the
Regge-Gribov formalism for the even and odd amplitudes (Appendix B) and also through dispersion relations
(Appendix C), without the subtraction constant (therefore, corresponding to a high-energy approximation result,
Sect. C.1).

\vspace{0.3cm}

\noindent
- Reggeons and Simple Pole Pomeron

%\vspace{0.3cm}

Therefore, for $pp$ and $\bar{p}p$ scattering, associating even and odd contributions
for Reggeons and the simple pole (even) Pomeron, the predictions for the forward physical observables
are the same as Eqs. (\ref{absigrp}) and (\ref{abrhorp}) (including the energy scale):
\begin{eqnarray}
\sigma_{\mathrm{tot}}(s) &=& a_1 s^{-b_1} + \tau a_2 s^{-b_2} + \delta s^{\epsilon}, \nonumber \\ 
                      \nonumber     \\ 
\rho(s) &=& \frac{1}{\sigma_{\mathrm{tot}}(s)} \left\{- a_1 \tan \left(\frac{\pi b_1}{2}\right) s^{- b_1} + 
\tau a_2 \cot \left(\frac{\pi b_2}{2} \right) s^{-b_2}  + \delta \tan \left(\frac{\pi \epsilon}{2}\right) s^{\epsilon}    \right\},
\nonumber
\end{eqnarray}
with $\tau = -1$ for $pp$ and $\tau = +1$ for $\bar{p}p$.

\subsection{Log-squared Law (Triple Pole)} 

In order to stress some important differences in case of $\gamma = 2$, 
and $\gamma$ as a real parameter, we treat first the log-squared law 
and in the next Subsection the log-raised-to-$\gamma$
law. As we shall show, the former is not a particular case of the latter
for $\gamma = 2$. Once representing the Pomeron (even signature) we consider here only
the \textit{symmetric} relation in Eq. (\ref{adevenodd}). Also, we use the index $T$ standing
for triple pole leading contribution at high energies.

We shall follow here the same steps of Sect. D.2 (power laws) in the case of
$\sigma^{T} = \beta \ln^2{s}$, so that
$\mathrm{Im} A^{T}(s) = \beta s \ln^2(s)$.

\vspace{0.3cm}

\noindent
- Asymptotic Behavior

%\vspace{0.3cm}

From the corollary,

\begin{eqnarray}
A^{T}(s) = Ms\ln^{2}(s)e^{i \theta}
\quad
\mathrm{and}
\quad
A^{T}(se^{i\pi}) = Ms\ln^{2}(se^{i\pi}) e^{i (\theta + \pi)}.
\label{adstepcor}
\end{eqnarray}

\noindent
- Crossing Symmetric Amplitude

%\vspace{0.3cm}

Omitting the index $+$, from the above expressions
and the symmetric relation (\ref{adevenodd}), 

\begin{eqnarray}
\ln^{2}(se^{i\pi}) e^{i (\theta + \pi)} =
\ln^{2}(s)e^{-\, i \theta}.
\nonumber
\end{eqnarray}
By substituting
\begin{eqnarray}
\ln^{2}(se^{i\pi}) =
\ln^{2}(s)\,\left[1 + i\frac{\pi}{\ln(s)}\right]^{2},
\nonumber
\end{eqnarray}
extracting the square root from both sides and taking the complex conjugate,
\begin{eqnarray}
e^{i\theta} = \pm \left[\frac{\pi}{\ln s} + i\right].
\nonumber
\end{eqnarray}

We note that due to the $e^{i \pi}$, here in the \textit{argument} of the logarithm,
the phase cannot be determined \textit{explicitly}, as in the case of the power law.
In this case, from $A^{T}(s)$ in Eq. (\ref{adstepcor}), denoting $M = \beta$ and taking
the $+$ sign for $\mathrm{Im} A^{T}(s) >0$, we obtain the complex amplitude
\begin{eqnarray}
\frac{A^{T}(s)}{s} = \beta [ \pi \ln(s) + i \ln^2(s) ].
\label{adatriple}
\end{eqnarray}
 
Therefore, as obtained in the Regge-Gribov and dispersion formalisms (Appendices B and C),
the Pomeron contribution associated with a triple pole reads
\begin{eqnarray}
\sigma^{T} = \beta \ln^2(s),
\qquad
\rho^{T} = \frac{1}{\sigma^{T}_{P}}[\beta \pi \ln(s)] = \frac{\pi}{\ln(s)}.
\label{adsigrhot}
\end{eqnarray}

Equation (\ref{adatriple}) with the Reggeon components derived in Sect. D.2,
defines the AU-L2 model introduced in Sect. 3.3.1.

\subsection{Log-raised-to-$\gamma$ Law}

Let us now treat the main focus of this work, namely
$\mathrm{Im} A(s) = \beta s \ln^{\gamma}(s)$,
with $\gamma$ a real number. We shall consider two methods,
one related to the phase of the amplitude (Sect. D.4.1) and
another one by considering a binomial expansion (Sect. D.4.2).
As a matter of notation we shall ommit any index in the amplitude.

\subsubsection{Phase of the Amplitude} We first derive an exact
result, to be used in our data reductions (Subsect. a),
followed by an high-energy approximate result (Subsect. b)
and relations with the AU-L2 and FMS-L$\gamma$ models (Subsect. c).

\vspace{0.2cm}

\noindent
\textit{a. Exact Result}

As before, following the same steps:

\noindent
- Asymptotic Behavior

From the corollary,
\begin{eqnarray}
A(s) = Ms\ln^{\gamma}(s)e^{i \theta}
\quad
\mathrm{and}
\quad
A(se^{i\pi}) = Ms\ln^{\gamma}(se^{i\pi}) e^{i (\theta + \pi)}.
\label{adlgstep1}
\end{eqnarray}

\noindent
- Crossing Symmetric Amplitude

From the symmetric relation in Eq. (\ref{adevenodd}), 
\begin{eqnarray}
\ln^{\gamma}(se^{i\pi}) e^{i (\theta + \pi)} =
\ln^{\gamma}(s)e^{-\, i \theta}.
\label{adlgstep2}
\end{eqnarray}

By expressing as before,
\begin{eqnarray}
\ln^{\gamma}(se^{i\pi}) =
\ln^{\gamma}(s)\,\left[1 + i\frac{\pi}{\ln(s)}\right]^{\gamma},
\nonumber
\end{eqnarray}
we obtain
\begin{eqnarray}
e^{-i(2\theta + \pi)} = \left[1 + i\frac{\pi}{\ln(s)}\right]^{\gamma},
\label{adlgstep3}
\end{eqnarray}
or
\begin{eqnarray}
e^{-i(2\theta + \pi)} = \left[1 + \frac{\pi^2}{\ln^2(s)}\right]^{\gamma/2} e^{i \gamma \phi(s)},
\nonumber
\end{eqnarray}
where
\begin{eqnarray}
\phi(s) = \tan^{-1}\left\{ \frac{\pi}{\ln(s)}\right\}.
\label{adlgphi}
\end{eqnarray}
By extracting the square root,
\begin{eqnarray}
e^{-i(\theta + \pi/2)} = \pm \frac{1}{\ln^{\gamma/2}(s)}
 \left[\ln^2(s) + \pi^2 \right]^{\gamma/4} e^{i \gamma \phi /2},
\nonumber
\end{eqnarray}
we obtain
\begin{eqnarray}
\ln^{\gamma}(s) e^{i\theta} = \pm \ln^{\gamma/2}(s)
 \left[\ln^2(s) + \pi^2 \right]^{\gamma/4} e^{- i [\gamma \phi + \pi]/2}.
\nonumber
\end{eqnarray}

Now, from (\ref{adlgstep1}), asymptotically, $A(s) = Ms\ln^{\gamma}(s)e^{i \theta}$.
Denoting $\beta > 0$ the coefficient, and under the condition $\mathrm{Im} A(s) > 0$, the above
equation provides the exact result for the complex amplitude:

\begin{eqnarray}
\boxed{
\frac{A(s)}{s} = 
\beta \ln^{\gamma/2}(s)\,
\left[ \ln^2(s) + \pi^2 \right]^{\gamma/4} 
\left[ \sin \left(\frac{\gamma \phi}{2} \right) + i \cos \left(\frac{\gamma \phi}{2} \right) \right]},
\label{adamplitudeaulg}
\end{eqnarray}
where $\phi = \phi(s)$ is given by Eq. (\ref{adlgphi}).

Equation (\ref{adamplitudeaulg}) with the Reggeon components derived in Sect. D.2,
defines the AU-L$\gamma$ model introduced in Sect. 3.3.2.

\vspace{0.2cm}

\noindent
\textit{b. High-energy Approximate Result}

At sufficiently high energies and since from the data reductions $\gamma < 3$,
we can approximate

\begin{equation}
\tan \phi = \frac{\pi}{\ln(s)} \approx \phi, 
\qquad
\sin\left(\frac{\gamma\phi}{2}\right)\approx \frac{\gamma\phi}{2} =
\frac{\gamma \pi}{2 \ln(s)},
\qquad
\cos\left(\frac{\gamma\phi}{2}\right)\approx 1
\nonumber
\end{equation}
and 
\begin{equation}
\left[ \ln^2(s) + \pi^2 \right]^{\gamma/4} = \ln^{\gamma/2}(s) 
\left[1 +\frac{\pi^2}{\ln^2 (s)}\right]^{\gamma/4}\approx \ln^{\gamma/2}(s).
\nonumber
\end{equation}
Substituting in Eq. (\ref{adamplitudeaulg}) we obtain the approximate result
\begin{eqnarray}
\frac{A(s)}{s} \approx \beta \ln^{\gamma}(s) \left[ \frac{\gamma \pi}{2\ln(s)} + i\ \right]. 
\label{adapprox}
\end{eqnarray}

\vspace{0.2cm}

\noindent
\textit{c. Relations with the AU-L2 and FMS-L$\gamma$ models}

Let us first consider the exact result given by Eq. (\ref{adamplitudeaulg}). For $\gamma$ = 2, we obtain
\begin{eqnarray}
\boxed{
\frac{A^{\gamma=2}(s)}{s} = 
\beta \ln(s)\,
\left[ \ln^2(s) + \pi^2 \right]^{1/2} 
\left[ \sin \left(\phi \right) + i \cos \left(\phi \right) \right]},
\label{adamplitudeaug=2}
\end{eqnarray}
which does not correspond to the triple pole contribution, Eq. (\ref{adatriple}).
Equation (\ref{adamplitudeaug=2}) with the Reggeon components derived in Sect. D.2,
defines the AU-L$\gamma$=2 model introduced in Sect. 3.3.3.

On the other hand, from the high-energy approximate result (\ref{adapprox}),
for $\gamma$ = 2, we obtain the triple pole contribution (\ref{adatriple}).
Moreover, from (\ref{adapprox}), in the general case,
\begin{eqnarray}
\sigma(s) \approx \beta\,\ln^{\gamma}(s),
\qquad
\rho(s) \approx \frac{\gamma \pi}{2 \ln(s)}.
\nonumber
\end{eqnarray}
It is interesting to note that this result correspond to the first order expansion
for $\rho(s)$ in the DDR approach, Eqs. (\ref{acrhop}) and (\ref{acparrho}):
\begin{eqnarray}
\rho^P_{1st} = \frac{\mathcal{A} \ln^{\gamma - 1}(s)}{\sigma_P} 
= \frac{\gamma \pi}{2 \ln(s)}.
\nonumber
\end{eqnarray}

\subsubsection{Binomial Expansion} 

Here we present another result for the amplitude, without the 
explicit determination of the phase, but by using a binomial
expansion (Subsection a). We also compare the results with the AU-L2 
and FMS-L$\gamma$ models (Subsection b).

\vspace{0.2cm}

\noindent
\textit{a. General Result}

Returning to Eq. (\ref{adlgstep3}), by extracting the square root, we can express
\begin{eqnarray}
e^{i\theta} = \pm \left[1 - i\frac{\pi}{\ln(s)}\right]^{\gamma/2}.
\nonumber
\end{eqnarray}
By considering the binomial expansion,
\begin{eqnarray}
(1 + x)^p = 1 + \sum_{k=1} \frac{1}{k!} p (p-1) (p-2) \cdots (p-[k-1]) x^k
\nonumber
\end{eqnarray}
in the variable       
\begin{eqnarray}
x = -i\,\frac{\pi}{\ln s},
\nonumber
\end{eqnarray}
since from Eq. (\ref{adlgstep1}), $A(s)/s = \beta \ln(s) e^{i \theta}$, for
$\mathrm{Im} A(s) > 0$, we obtain the complex amplitude in the form
of an expansion series:

\begin{eqnarray}
\frac{A(s)}{s} &=& \beta \left\{\frac{\gamma}{1!} \left[\frac{\pi}{2}\right] \ln^{\gamma-1}(s) -
\frac{\gamma (\gamma-2) (\gamma-4)}{3!}  \left[\frac{\pi}{2}\right]^3  \ln^{\gamma-3}(s)  \right. \nonumber \\
&+& \left. \frac{\gamma (\gamma-2) (\gamma-4) (\gamma - 6) (\gamma -8)}{5!}  \left[\frac{\pi}{2}\right]^5  \ln^{\gamma-5}(s)
+ \cdots \right\} \nonumber \\
 &+& i\, \beta \left\{\ln^{\gamma}(s) -
\frac{\gamma (\gamma-2)}{2!}  \left[\frac{\pi}{2}\right]^2  \ln^{\gamma-2}(s)  \right. \nonumber \\
&+& \left. \frac{\gamma (\gamma-2) (\gamma-4) (\gamma - 6)}{4!}  \left[\frac{\pi}{2}\right]^4  \ln^{\gamma-4}(s)
+ \cdots \right\}.
\label{adamplitudebe}
\end{eqnarray}

\vspace{0.2cm}

\noindent
\textit{b. Relations with AU-L2 and FMS-L$\gamma$ Models}
 
 For $\gamma$ = 2, we obtain the L2 model (AU or FMS) for the leading contribution,
\begin{eqnarray}
\sigma^P(s) = \beta \ln^2(s),
\qquad
\rho^P(s) = \frac{1}{\sigma^P(s)} \{ \beta \pi \ln(s) \}.
\nonumber
\end{eqnarray}

Comparing Eq. (\ref{adamplitudebe}) with the leading contribution in the FMS-L$\gamma$ model,
Eqs. (\ref{acsigp}) - (\ref{acparrho}), the results are the same only in first order, namely
\begin{eqnarray}
\sigma^P(s) \sim \beta \ln^{\gamma}(s),
\qquad
\rho^P(s) = \frac{1}{\sigma^P(s)} \{ \beta \gamma \frac{\pi}{2} \ln^{\gamma - 1}(s) \}.
\nonumber
\end{eqnarray}

We have derived three results using the AU approach for the L$\gamma$ law:
the exact result (\ref{adamplitudeaulg}), the particular case for $\gamma$=2, 
Eq. (\ref{adamplitudeaug=2}) and the binomial expansion (\ref{adamplitudebe}).
It is important to note that, in all cases, the AU approach introduces terms (functions of the energy) 
in the expression of \tcs($s$) 
that are not present in the results obtained through the DDR approach, namely the Amaldi et al. 
parametrization, Eq. (\ref{fmsstlg}), which also defines
the FMS-L$\gamma$ model.

%\section*{References}

\end{document}